%% file: placco.tex
\documentclass[twocolumn]{aastex63} 

\usepackage{hyperref,enumerate}
\input{colordvi}

\hypersetup{
    unicode=false,                  
    pdftoolbar=true,                
    pdfmenubar=true,                
    pdffitwindow=true,              
    pdfstartview={FitH},            
    pdftitle={J1830-4555},         
    pdfauthor={vplacco},            
    pdfsubject={Astronomy},         
    pdfcreator={dvipdf},            
    pdfproducer={dvipdf},           
    pdfkeywords={metal-poor stars}, 
    pdfnewwindow=true,              
    colorlinks=true,                
    linkcolor=red,                  
    citecolor=blue,                 
    filecolor=magenta,              
    urlcolor=cyan,                  
    breaklinks=true,
    linktocpage
}

\usepackage{hyperref,enumerate,numdef}

\newcommand{\kmsec}{\mbox{km~s$^{\rm -1}$}}
\newcommand{\eps}[1]{\ensuremath{\log\epsilon\,(\mathrm{#1})}}
\newcommand{\xx}{{\tablenotemark{\footnotesize{a}}}}

\newcommand{\abund}[2]{\ensuremath{[\mathrm{#1}/\mathrm{#2}]}}
\newcommand{\cfe}{\abund{C}{Fe}}
\newcommand{\nfe}{\abund{N}{Fe}}
\newcommand{\xfe}[1]{\abund{#1}{Fe}}

\newcommand{\metal}{\abund{Fe}{H}}

\newcommand{\teff}{\ensuremath{T_\mathrm{eff}}}
\newcommand{\logg}{\ensuremath{\log\,g}}

\newcommand{\A}[1]{\ensuremath{A(\mathrm{#1})}}

\newcommand{\K}{\ensuremath{\mathrm{K}}}

\newcommand{\rave}{\object{RAVE~J1830$-$4555}}
\newcommand{\ravel}{\object{RAVE~J183013.5$-$455510}}

\shorttitle{\rave}
\shortauthors{Placco et al.}

\begin{document}

\title{The \emph{R}-Process Alliance: The Peculiar Chemical Abundance Pattern of \ravel
\footnote{
Based on observations gathered with the $6.5\,$m Magellan
Telescopes located at Las Campanas Observatory, Chile.
Based on observations collected at the European Organisation for Astronomical
Research in the Southern Hemisphere - 099.D-0428(A).
%
}}

\author[0000-0003-4479-1265]{Vinicius M.\ Placco}
\affiliation{Department of Physics, University of Notre Dame, Notre Dame, IN 46556, USA}
\affiliation{JINA Center for the Evolution of the Elements, USA}

\author[0000-0002-7529-1442]{Rafael M.\ Santucci}
\affiliation{Instituto de Estudos S\'ocio-Ambientais, Planet\'ario, 
Universidade Federal de Goi\'as, Goi\^ania, GO 74055-140, Brazil}
\affiliation{Instituto de F\'isica, Universidade Federal de Goi\'as, Campus
Samambaia, Goi\^ania, GO 74001-970, Brazil}

\author[0000-0002-8129-5415]{Zhen Yuan}
\affiliation{Key Laboratory for Research in Galaxies and Cosmology, Shanghai
Astronomical Observatory, Chinese Academy of Sciences, Shanghai 200030, China}

\author[0000-0001-9178-3992]{Mohammad K.\ Mardini}
\affiliation{Key Laboratory of Optical Astronomy, National Astronomical Observatories,
Chinese Academy of Sciences, Beijing, China}
\affiliation{School of Astronomy and Space Science, University of Chinese
Academy of Sciences,  Beijing, China}	

\author[0000-0002-5463-6800]{Erika M.\ Holmbeck}
\affiliation{Department of Physics, University of Notre Dame, Notre Dame, IN 46556, USA}
\affiliation{JINA Center for the Evolution of the Elements, USA}

\author[0000-0002-5901-9879]{Xilu Wang}
\affiliation{Department of Physics, University of Notre Dame, Notre Dame, IN 46556, USA}
\affiliation{JINA Center for the Evolution of the Elements, USA}
\affiliation{Department of Physics, University of California, Berkeley, CA 94720, USA}

\author[0000-0002-4729-8823]{Rebecca Surman}
\affiliation{Department of Physics, University of Notre Dame, Notre Dame, IN 46556, USA}
\affiliation{JINA Center for the Evolution of the Elements, USA}

\author[0000-0001-6154-8983]{Terese T.\ Hansen}
\affiliation{Mitchell Institute for Fundamental Physics and Astronomy and Department 
of Physics and Astronomy, Texas A\&M University, College Station, TX 77843, USA}

\author[0000-0001-5107-8930]{Ian U.\ Roederer}
\affiliation{Department of Astronomy, University of Michigan, Ann Arbor, MI 48109, USA}
\affiliation{JINA Center for the Evolution of the Elements, USA}

\author[0000-0003-4573-6233]{Timothy C.\ Beers}
\affiliation{Department of Physics, University of Notre Dame, Notre Dame, IN 46556, USA}
\affiliation{JINA Center for the Evolution of the Elements, USA}

\author[0000-0001-6159-8470]{Arthur Choplin}
\affiliation{Department of Physics, Faculty of Science and Engineering, Konan
University, 8-9-1 Okamoto, Kobe, Hyogo 658-8501, Japan}
\affiliation{Geneva Observatory, University of Geneva, Maillettes 51, CH-1290
Sauverny, Switzerland}

\author[0000-0001-6154-8983]{Alexander P.\ Ji}
\altaffiliation{Hubble Fellow}
\affiliation{Observatories of the Carnegie Institution for Science, 
Pasadena, CA 91101, USA}

\author[0000-0002-8504-8470]{Rana Ezzeddine}
\affiliation{Department of Astronomy, University of Florida, Bryant Space 
Science Center, Gainesville, FL 32611, USA}

\author[0000-0002-2139-7145]{Anna Frebel}
\affiliation{Department of Physics and Kavli Institute for Astrophysics and
Space Research, \\ Massachusetts Institute of Technology, Cambridge, MA 02139, USA}
\affiliation{JINA Center for the Evolution of the Elements, USA}

\author[0000-0002-5095-4000]{Charli M.\ Sakari}
\affiliation{Department of Physics \& Astronomy, San Francisco State University, San Francisco, CA 94132, USA}

\author[0000-0002-9594-6143]{Devin D.\ Whitten}
\affiliation{Department of Physics, University of Notre Dame, Notre Dame, IN 46556, USA}
\affiliation{JINA Center for the Evolution of the Elements, USA}

\author[0000-0003-0998-2744]{Joseph Zepeda}
\affiliation{Department of Physics, University of Notre Dame, Notre Dame, IN 46556, USA}
\affiliation{JINA Center for the Evolution of the Elements, USA}

\correspondingauthor{Vinicius M.\ Placco}
\email{vplacco@nd.edu}

\begin{abstract}

We report on the spectroscopic analysis of \ravel, an extremely metal-poor star, highly
enhanced in CNO, and with discernible contributions from the rapid
neutron-capture process. There is no evidence of binarity for this object.  At
\metal=$-3.57$, this is one of the lowest metallicity stars currently observed,
with 18 measured abundances of neutron-capture elements. The presence of Ba, La,
and Ce abundances above the Solar System $r$-process predictions suggest that
there must have been a non-standard source of $r$-process elements operating at
such low metallicities.  
%
%
One plausible explanation is that this enhancement originates from material
ejected at unusually fast velocities in a neutron star merger event.
We also explore the possibility that the neutron-capture elements were produced
during the evolution and explosion of a rotating massive star.
In addition, based on comparisons with yields from zero-metallicity faint
supernova, we speculate that \rave\ was formed from a gas cloud pre-enriched by
both progenitor types. From analysis based on Gaia DR2 measurements, we show
that this star has orbital properties similar to the Galactic metal-weak
thick-disk stellar population.

\vspace{1.0cm}

\end{abstract}


\section{Introduction}
\label{intro}

One of the most intriguing challenges in stellar astrophysics today is to paint
a compelling picture of how the Universe chemically evolved from hydrogen and
helium (with traces of lithium) to the wealthy diversity of elements we observe
today in the atmosphere of the Sun and other stars.
Nucleosynthesis taking place during the evolution of stars, either in burning or
explosive stages, is the culprit for such diversity
\citep{merrill1952,hoyle1954,arnett1996}. The
underlying physical processes by which chemical elements, from carbon to
uranium, are formed has a reasonably well-established framework
\citep[e.g.,][]{b2fh,cameron1957}. The
next steps are to identify possible astrophysical sites where such
nucleosynthesis events could occur and describe the mixing processes that seed the
formation of subsequent stellar generations.

Therefore, a star such as the Sun (with a main-sequence age of over 4 Gyr) was
formed from a gas cloud that carried over 9 Gyr of chemical evolution from
previous stellar generations. As a result, and the intrinsic stochasticity
associated with star formation, it is impossible to pinpoint a single genealogy
record for such relatively young stars.
However, by observing stars formed from gas clouds enriched by a single (or a
handful of) nucleosynthesis episode(s), as is expected to be the case for the most
metal-deficient stars in the Galaxy, it is possible to characterize and study
the progenitor population(s) of these stars.

The field of stellar archaeology was built upon the premise that old,
slow-evolving, low-mass, low-metallicity stars can preserve in their atmospheres
the chemical imprint of primordial stellar populations in the Galaxy and the
Universe \citep{bromm2004,bromm2009,nomoto2013}.
More importantly, it is believed that a subset of these objects are indeed
``true'' second-generation stars, also known as Extremely Metal-Poor 
(EMP; [Fe/H]\footnote{\abund{A}{B} = $log(N_X/{}N_Y)_{\star} -
\log(N_X/{}N_Y) _{\odot}$, where $N$ is the number density of atoms of
elements $X$ and $Y$ in the star ($\star$) and the Sun ($\odot$),
respectively.} $< -3.0$) stars \citep{beers2005}. The chemical abundance
patterns of these EMP stars can place direct constraints on the nature
of the first (Population III) stars to be formed in the Universe
\citep[e.g.][]{christlieb2002,frebel2006,norris2007,caffau2011b,
ito2013,hansen2014,keller2014,placco2014b,starkenburg2014,frebel2015b,melendez2016,
caffau2016,roederer2016,placco2016b,aguado2018,starkenburg2018,ezzeddine2019,mardini2019b,mardini2019}
and possible astrophysical site(s) for their occurence, such as dwarf
galaxies \citep{salvadori2015,hansen2017,longeard2018,nagasawa2018,marshall2019} and damped Ly-$\alpha$ systems
\citep{cooke2011,cooke2014,welsh2020}.

A large fraction of EMP stars exhibit enhancements in carbon (and similarly
nitrogen and oxygen -- 43\% according to \citealt{placco2014c})
and are classified as
carbon-enhanced metal-poor (CEMP; \cfe$>+0.7$, \citealt{aoki2007}). These objects are
further classified by their paucity or enhancement in neutron-capture elements
(CEMP-no and CEMP-$s/r/i$, respectively - \citealt{beers2005,frebel2018}) and have very
distinct nucleosynthetic pathways and enrichment processes
\citep{yoon2016,frebel2018}.

The light-element (from C to Zn) abundance pattern found in EMP stars (mostly
CEMP-no) is believed to be the result of the evolution of massive Pop III stars in
the early Universe. Candidates for CEMP-no progenitor population are (i)
metal-free massive stars \citep{heger2010}, (ii) mixing and fallback  ``faint
supernovae'' \citep{umeda2005,nomoto2006,tominaga2014}, and (iii) rapidly
rotating, near-zero-metallicity, massive stars
\citep[{\emph{spinstars;}}][]{meynet2010,chiappini2013,cescutti2013,cescutti2014}.
A subset of these EMP stars, also known as ``mono-enriched,'' are thought to be
the direct descendents of the first stars \citep{hartwig2018}.
\citet{placco2016b} provide a brief explanation on the main
characteristics of these progenitor types and the possible metallicity regimes where
their occurence appears to  have better agreement with observations.

For the heavy elements (from Ga to U), formed by the {\emph{slow}}, {\emph{intermediate}}, and
{\emph{rapid}} neutron-capture processes ($s$-, $i$-, and $r$-process;
\citealt{frebel2018,hansen2019,prantzos2020}), there are a number of possible astrophysical sites
responsible for their production. The observed $s$-process abundances in
CEMP-$s$, CEMP-$r/s$, and CEMP-$i$ stars\footnote{
CEMP-$s$: \cfe$>+0.7$, \xfe{Ba}$>+1.0$, \abund{Ba}{Eu}$>+0.5$, \abund{Ba}{Pb}$>-1.5$;
CEMP-$r/s$: \cfe$>+0.7$, $0.0<$\abund{Ba}{Eu}$<+0.5$ and $-1.0<$\abund{Ba}{Pb}$<-0.5$;
CEMP-$i$: \cfe$>+0.7$, $0.0<$\abund{La}{Eu}$<+0.6$ and \abund{Hf}{Ir}$\sim +1.0$. See Table 1
in \citet{frebel2018} for further details.
}
are thought to be a result of the evolution of low- to
intermediate-mass, low-metallicity, asymptotic giant branch stars
\citep{herwig2005,hampel2016}. The newly synthesized elements are then moved to
the atmosphere of the less-evolved low-metallicity companion via mass transfer in a
binary system \citep{starkenburg2014,hansen2015b,hansen2016c,cseh2018}.

The onset and operation of the $r$-process require high neutron fluxes and
densities. Possible astrophysical sites that would sustain these conditions
include (i) the aftermath of events such as merging neutron stars
\citep{abbott2017,drout2017,shappee2017}, (ii) supernova-triggering collapse of rapidly
rotating massive stars \citep[{\emph{collapsars}};][]{siegel2019}, and (iii) 
common-envelope jet supernovae \citep{grichener2019}.
Observational evidence suggests that these events could have occured early in
the history of the Universe \citep{roederer2014b} in environments such as dwarf galaxies, which were
recently found to harbor low-metallicity, $r$-process enhanced stars
\citep{ishimaru2015,vincenzo2015,ji2016,hansen2017,roederer2017,roederer2018b}. However, there is still no consensus in
the literature as to which $r$-process nucleosynthesis channel (or a combination
of) can successfully reproduce observations and be incorporated in Galactic
chemical-evolution models
\citep{matteucci2014,cescutti2015,shen2015,vandevoort2015,wehmeyer2015,cote2019,haynes2019,holmbeck2019}.

\begin{deluxetable*}{lcccl}
\tablecaption{Observational Data for \protect\ravel \label{candlist}}
\tablewidth{0pt}
\tabletypesize{\scriptsize}
\tabletypesize{\small}
\tablehead{
    \colhead{Quantity} &
    \colhead{Symbol} &
    \colhead{Value} &
    \colhead{Units} &
    \colhead{Reference}}
\startdata
Right ascension           & $\alpha$ (J2000)    & 18:30:13.54            & hh:mm:ss.ss   & Simbad\xx                  \\
Declination               & $\delta$ (J2000)    & $-$45:55:10.1          & dd:mm:ss.s    & Simbad\xx                  \\
Galactic longitude        & $\ell$              & 348.9                  & degrees       & Simbad\xx                  \\
Galactic latitude         & $b$                 & $-$15.7                & degrees       & Simbad\xx                  \\
Gaia DR2 source ID        &                     & 6708532208165979392    &               & \citet{gaia2018}           \\
Parallax                  & $\varpi$            & 0.3214 $\pm$ 0.0429    & mas           & \citet{gaiadr2}            \\
Inverse parallax distance & $1/\varpi$          & 3.11$^{+0.48}_{-0.37}$ & kpc           & this work                  \\
Distance                  & $D$                 & 2.88$^{+0.43}_{-0.33}$ & kpc           & \citet{bailer-jones2018}   \\
Distance                  & $D$                 & 2.75$^{+0.58}_{-0.40}$ & kpc           & \citet{starhorse19}        \\
Proper motion ($\alpha$)  & PMRA                & 7.949 $\pm$ 0.084      & mas yr$^{-1}$ & \citet{gaiadr2}            \\
Proper motion ($\delta$)  & PMDec               & $-$6.712 $\pm$ 0.081   & mas yr$^{-1}$ & \citet{gaiadr2}            \\
$G$ magnitude             & $G$                 & 11.8125 $\pm$ 0.0002   & mag           & \citet{gaia2018}           \\
$G_{\rm BP}$ magnitude    & $G_{\rm BP}$        & 12.2984 $\pm$ 0.0015   & mag           & \citet{gaia2018}           \\
$G_{\rm RP}$ magnitude    & $G_{\rm RP}$        & 11.1767 $\pm$ 0.0009   & mag           & \citet{gaia2018}           \\
$B$ magnitude             & $B$                 & 12.915 $\pm$ 0.010     & mag           & \citet{henden2014}         \\
$V$ magnitude             & $V$                 & 12.059 $\pm$ 0.010     & mag           & \citet{henden2014}         \\
$J$ magnitude             & $J$                 & 10.393 $\pm$ 0.023     & mag           & \citet{skrutskie2006}      \\
$H$ magnitude             & $H$                 &  9.852 $\pm$ 0.022     & mag           & \citet{skrutskie2006}      \\
$K$ magnitude             & $K$                 &  9.744 $\pm$ 0.020     & mag           & \citet{skrutskie2006}      \\
Color excess              & $E(B-V)$            & 0.0486                 & mag           & \citet{schlafly2011}       \\
Radial velocities         & RV                  & 61.3 $\pm$ 1.9         & \kmsec        & Gaia DR2 (4 epochs)        \\
                          &                     & 63.5 $\pm$ 1.6         & \kmsec        & RAVE DR5 (MJD: 55743.622)  \\
                          &                     & 61.1 $\pm$ 1.0         & \kmsec        & du Pont  (MJD: 57894.294)  \\
                          &                     & 63.2 $\pm$ 0.5         & \kmsec        & Magellan (MJD: 57979.596)  \\
                          &                     & 62.4 $\pm$ 1.0         & \kmsec        & du Pont  (MJD: 58734.518)  \\
\enddata
\tablenotetext{a}{\href{http://simbad.u-strasbg.fr/simbad/sim-id?Ident=RAVE+J183013.5-455510}
{http://simbad.u-strasbg.fr/simbad/sim-id?Ident=RAVE$+$J183013.5$-$455510}}
\end{deluxetable*}

From an observational perspective, the {\emph{R}}-Process Alliance (RPA) has
been instrumental in the quest to increase the number of known $r$-process
enhanced stars in the Galaxy. In its two years of existence, the RPA has already
identified 26 new $r$-II (\xfe{Eu}$> +1.0$) and 146 new $r$-I ($+0.3 \leq$
\xfe{Eu}$ \leq +1.0$) stars \citep{hansen2018,sakari2018b,ezzeddine2019b}, an increase of 87\%
and 130\%, respectively, from all previous literature
studies\footnote{\citet{holmbeck2020} proposes a new dividing line
between the $r$-I and $r$-II classes at \xfe{Eu}$=+0.7$, 
which changes the number of identified $r$-I and $r$-II stars by the RPA to 121
and 51, respectively.}.
This ongoing
effort has already provided in-depth analyses of a number of unique low-metallicity
stars in the Galaxy
\citep[e.g.i,][]{cain2018,gull2018,holmbeck2018,roederer2018,sakari2018} and will
continue to do so in its next stages.

In this work, we report on the identification and analysis of \ravel\ (hereafter
\rave), a CNO-enhanced extremely metal-poor (\metal=$-3.57$) star exhibiting
discernible $r$- and $s$-process patterns, with chemical abundances measured for
18 neutron-capture elements. This ancient star belongs to the metal-weak
thick-disk (MWTD) population of the Milky Way galaxy, and radial-velocity measurements
spanning more than eight years show no variations outside 1-$\sigma$.
This paper is outlined as follows: Section~\ref{secobs} describes the medium-
and high-resolution spectroscopic observations. The determinations of stellar
parameters and chemical abundances are presented, respectively, in
Sections~\ref{secatm} and \ref{absec}, including a comparison with data from the
literature. Analyses of radial-velocity variations, chemical abundance
pattern, and the kinematics of \rave\ are presented in Section~\ref{disc}. Our
conclusions are provided in Section~\ref{final}.

\begin{figure*}[!ht]
\epsscale{1.15}
\plotone{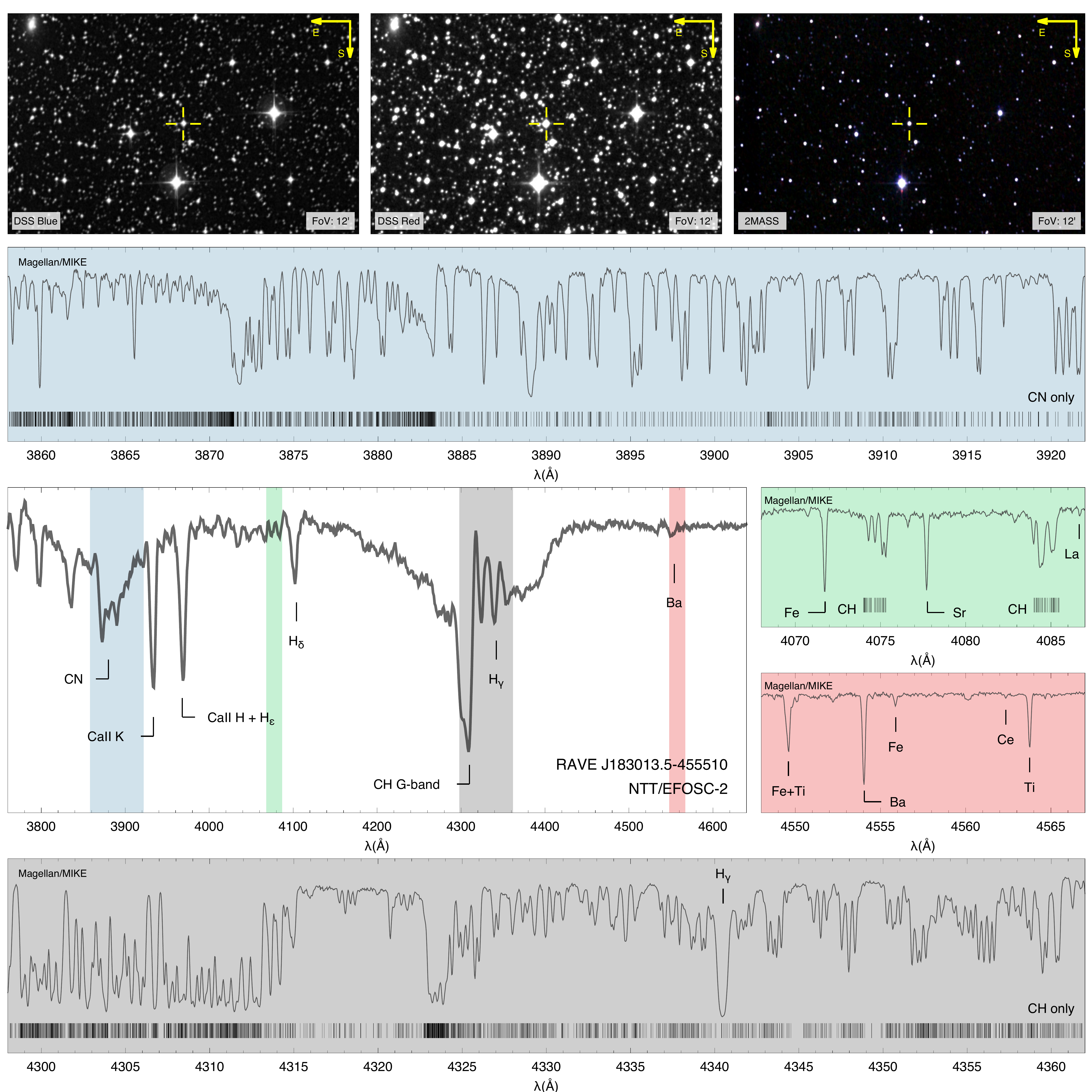}
\caption{Top row: DSS (blue/red) and 2MASS (combined) finding charts for
\protect{\rave}. Middle panel: medium-resolution NTT spectrum. Colored panels:
selected regions of the high-resolution Magellan spectrum. Atomic and molecular 
features of interest used in the analysis are highlighted. An interactive
version of this figure can be accessed at
\href{https://vmplacco.github.io/J1830-4555.html}{https://vmplacco.github.io/J1830-4555.html}.}
\label{spectra}
\end{figure*}

\section{Observations}
\label{secobs}

\rave\ is a relatively bright ($V = 12$) star in the Southern Hemisphere, in a
region not heavily obscured by dust. Table~\ref{candlist} lists basic
photometric and astrometric information for \rave; the top row of
Figure~\ref{spectra} shows the finding charts from the Digitized Sky Survey
\citep[DSS Blue and Red;][]{lasker1990} and from the Two Micron All Sky Survey
\citep[2MASS;][]{skrutskie2006}. Below we describe the medium-resolution and
high-resolution spectroscopic observations of \rave.

\subsection{Medium-resolution Spectroscopy}

\rave\ was first selected as a metal-poor star candidate from the fifth data
release of the RAdial Velocity Experiment \citep[RAVE
DR5;][]{steinmetz2006,kunder2017} and followed-up with medium-resolution
spectroscopy as part of the validation efforts described in
\citet{placco2018,placco2019}. 
Observations were carried out in semester 2017A with the 3.58m ESO New
Technology Telescope,  using the EFOSC-2 spectrograph \citep{efosc}.
The instrument setup included Grism7 (600~gr~mm$^{\rm{-1}}$) and a 1$\farcs$0
slit, yielding a wavelength coverage of 3500-5200\,{\AA}, resolving power of
$R\sim 2,000$ ($1 \times 1$ binning), and signal-to-noise ratio of S/N$\sim 50$
per pixel at 4000\,{\AA}. Calibration frames included FeAr exposures,
quartz-lamp flatfields, and bias frames. All reduction and extraction tasks were
performed using
IRAF\footnote{\href{http://iraf.noao.edu}{http://iraf.noao.edu}.} packages.
Figure
\ref{spectra} (middle panel) shows a portion of the NTT spectrum, indicating
absorption features and regions of interests for stellar parameter determination
and chemical abundance analysis.

\subsection{High-resolution Spectroscopy}

High-resolution spectroscopy for \rave\ was obtained on 2017 August 14 using the
Magellan Inamori Kyocera Echelle \citep[MIKE;][]{mike} spectrograph, mounted on
the 6.5m Magellan-Clay Telescope at Las Campanas Observatory.  The observing
setup included a $0\farcs7$ slit with $2\times2$ on-chip binning, yielding a
resolving power of $R\sim37,000$ (blue spectrum) and $R\sim30,000$ (red
spectrum). The S/N is $\sim 80$ per pixel at $3900\,${\AA} and $\sim 120$ at
$5200\,${\AA} after a total exposure of 2,000~s. The MIKE spectrum covers a wide
range of optical wavelengths ($\sim3300-9000\,${\AA}), making it ideal for
chemical abundance determinations, in particular for neutron-capture elements
(see Section~\ref{absec} for details).  The data were reduced using the routines
developed for MIKE spectra, described in
\citet{kelson2003}\footnote{\href{http://code.obs.carnegiescience.edu/python}
{http://code.obs.carnegiescience.edu/python}}.
The colored panels of Figure~\ref{spectra} show selected regions of the MIKE
spectrum, highlighting atomic and molecular features of interest for abundance
determination.
\rave\ was also observed in the 2017A and 2019B semesters with the Echelle
spectrograph on the du Pont 2.5~m telescope at the Las Campanas Observatory, as
part of the RPA snapshot campaign \citep[see][for further details]{hansen2018}.
These spectra were used to confirm the atmospheric parameters determined for the
MIKE spectrum and also for radial-velocity comparisons. The observational data
and radial velocities for \rave\ are listed in Table~\ref{candlist}.

\section{Stellar Atmospheric Parameters}
\label{secatm}


\subsection{Medium-resolution Spectrum}

Stellar atmospheric parameters (\teff, \logg, and \metal) were calculated from
the ESO/NTT spectrum using the n-SSPP \citep{beers2014,beers2017}, a modified
version of the SEGUE Stellar Parameter Pipeline
\citep[SSPP;][]{lee2008a,lee2008b,lee2013}. These were used to flag \rave\ as a
candidate for high-resolution spectroscopic follow-up. The parameters were also
estimated by the
CASPER\footnote{\href{https://github.com/DevinWhitten/CASPER}{https://github.com/DevinWhitten/CASPER}}
(Chemical Abundance and Stellar Parameter Estimation Routine) software,
described in \citet{yoon2019}.  CASPER also estimated the carbon abundance for
\rave\ as \eps{C}=$+6.82 \pm 0.24$, which is in excellent agreement with the
value determined from the high-resolution spectrum (\eps{C}=$+6.76 \pm 0.10$;
see Section~\ref{carbsec} for further details). The final parameters are listed in
Table~\ref{paramtab}, together with values from the literature and the
high-resolution spectra (see details below).

\begin{deluxetable}{@{}llccc@{}}[!ht]
\tablewidth{0pc}
\tabletypesize{\small}
\tabletypesize{\footnotesize}
\tablecaption{Derived Stellar Parameters \label{paramtab}}
\tablehead{
\colhead{              }&
\colhead{\teff{}(K)    }&
\colhead{\logg{}(cgs)  }&
\colhead{\metal{}      }&
\colhead{$\xi$(km/s)   }}
\startdata
\multicolumn{5}{c}{{Literature values}} \\
\hline
Gaia       & 4993 (100) & \nodata     & \nodata        & \nodata     \\  
RAVE       & 4984 (100) & 3.34 (0.47) & $-$3.51 (0.16) & \nodata     \\  
\hline
\multicolumn{5}{c}{{This work}} \\
\hline
ESO/NTT    & 4781 (150) & 0.95 (0.35) & $-$4.15 (0.20) & \nodata     \\  
CASPER     & 4905 (150) & 1.70 (0.40) & $-$3.84 (0.16) & \nodata     \\  
du Pont    & 4720 (100) & 1.40 (0.20) & $-$3.56 (0.10) & 2.00 (0.20) \\  
Magellan   & 4765 (100) & 1.20 (0.20) & $-$3.57 (0.10) & 1.95 (0.20) \\  
\enddata
\end{deluxetable}

\vspace{1.0cm}

\subsection{High-resolution Spectra}

The stellar parameters for the high-resolution data were determined
spectroscopically, using the latest version of the
MOOG\footnote{\href{https://github.com/alexji/moog17scat}{https://github.com/alexji/moog17scat}}
code \citep{sneden1973}, employing one-dimensional plane-parallel model
atmospheres with no overshooting \citep{castelli2004}, computed under the
assumption of local thermodynamic equilibrium (LTE). 
The effective temperature was determined by minimizing the trend between the
abundances of individual \ion{Fe}{1} lines and their excitation potential
($\chi$). After
that, the temperature is corrected to the ``photometric scale'' using the
calibration described in \citet{frebel2013}.  With the temperature fixed, the
microturbulent velocity ($\xi$) was determined by removing the trend in the \ion{Fe}{1}
abundances and the reduced equivalent width (REW$ = \log(\rm{EW}/\lambda)$), and
the surface gravity determined by forcing the agreement between the \ion{Fe}{1}
and \ion{Fe}{2} average abundances.
The equivalent widths were obtained automatically by fitting Gaussian profiles to the observed
absorption lines and then visually inspected. Table~\ref{eqw} lists the lines
employed in this analysis, their measured equivalent widths, and the derived
chemical abundances.
This procedure was used to determine the parameters using both the du
Pont/Echelle and Magellan/Mike spectra. The resulting parameters are listed in
Table~\ref{paramtab}.

\startlongtable

\begin{deluxetable}{lrrrrr}
\tabletypesize{\tiny}
\tabletypesize{\footnotesize}
\tablewidth{0pc}
\tablecaption{\label{eqw} Equivalent-Width Measurements}
\tablehead{
\colhead{Ion}&
\colhead{$\lambda$}&
\colhead{$\chi$} &
\colhead{$\log\,gf$}&
\colhead{$EW$}&
\colhead{$\log\epsilon$\,(X)}\\
\colhead{}&
\colhead{({\AA})}&
\colhead{(eV)} &
\colhead{}&
\colhead{(m{\AA})}&
\colhead{}}
\startdata
\input{line.tex}
\enddata
\end{deluxetable}

\section{Chemical Abundances}
\label{absec}

Elemental-abundance ratios, \xfe{X}, were calculated adopting the Solar photospheric
abundances from \citet{asplund2009}.  The average measurements (or
upper limits) for $36$ elements, derived from the Magellan/MIKE spectrum, are
listed in Table~\ref{abund}.  The $\sigma$ values are the standard
error of the mean.  Abundances were calculated by both equivalent-width analysis
and spectral synthesis. 

\begin{deluxetable}{@{}lcrrrcr@{}}
\tabletypesize{\small}
\tabletypesize{\footnotesize}
\tablewidth{0pc}
\tablecaption{Abundances for Individual Species \label{abund}}
\tablehead{
\colhead{Species}                     & 
\colhead{$\log\epsilon_{\odot}$\,(X)} & 
\colhead{$\log\epsilon$\,(X)}         & 
\colhead{$\mbox{[X/H]}$}              & 
\colhead{$\mbox{[X/Fe]}$}             & 
\colhead{$\sigma$}                    & 
\colhead{$N$}}
\startdata
\input{abund.tex}
\enddata
\tablenotetext{a}{Using the carbon evolutionary corrections of \citet{placco2014c}.}
\end{deluxetable}

Uncertainties in the elemental-abundance determinations, as well as the systematic
uncertainties due to changes in the atmospheric parameters, were treated in the
same way as described in \citet{placco2013,placco2015b}.  Table~\ref{sys} shows
how variations within the quoted uncertainties in each atmospheric parameter
affect the derived chemical abundances.  Also listed is the total uncertainty
for each element, which is calculated from the quadratic sum of the individual
error estimates. For this purpose, we used spectral features with abundances
determined by equivalent-width analysis only.  The adopted variations for
the parameters are $+$150~K for \teff, $+$0.3~dex for \logg, and $+$0.3
km\,s$^{-1}$ for $\xi$.  

\begin{deluxetable}{@{}lrrrrr@{}}
\tabletypesize{\small}
\tabletypesize{\footnotesize}
\tablewidth{0pc}
\tablecaption{Example Systematic Abundance Uncertainties for \protect\rave \label{sys}}
\tablehead{
\colhead{Elem}&
\colhead{$\Delta$\teff}&
\colhead{$\Delta$\logg}&
\colhead{$\Delta\xi$}&
\colhead{$\sigma/\sqrt{n}$}&
\colhead{$\sigma_{\rm tot}$}\\
\colhead{}&
\colhead{$+$150\,K}&
\colhead{$+$0.3 dex}&
\colhead{$+$0.3 km/s}&
\colhead{}&
\colhead{}}
\startdata
\input{error.tex}
\enddata
\end{deluxetable}

\subsection{Carbon, Nitrogen, and Oxygen}
\label{carbsec}

\begin{figure*}[!ht]
\epsscale{1.15}
\plotone{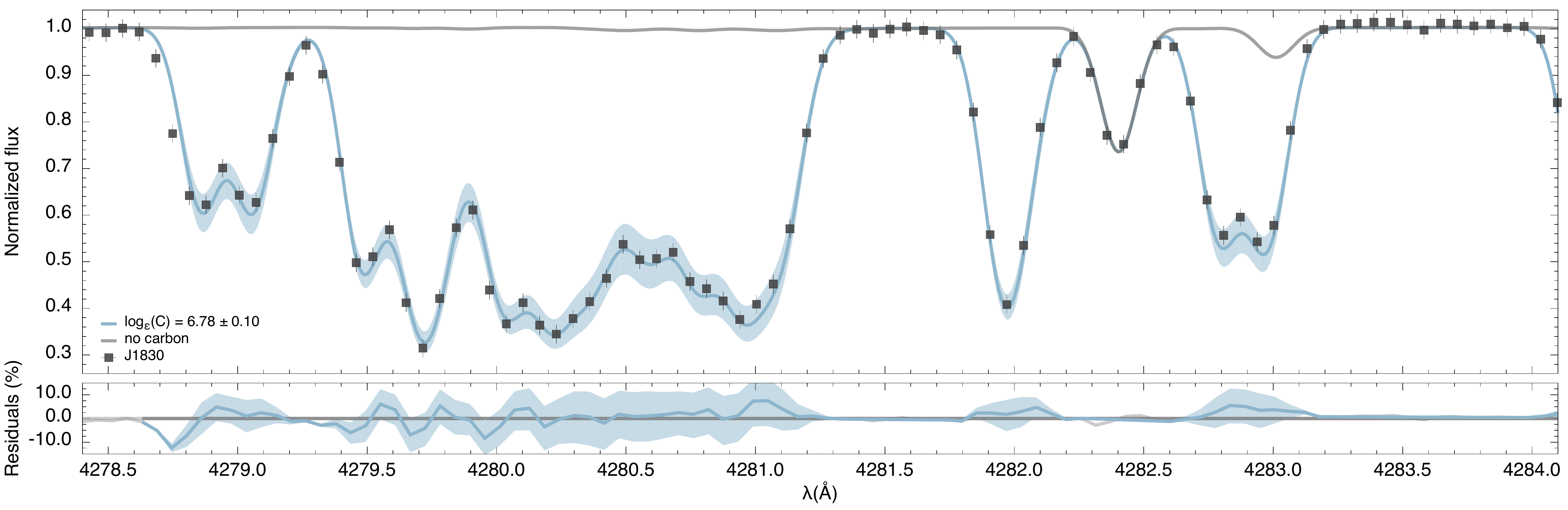}
\epsscale{1.15}
\plotone{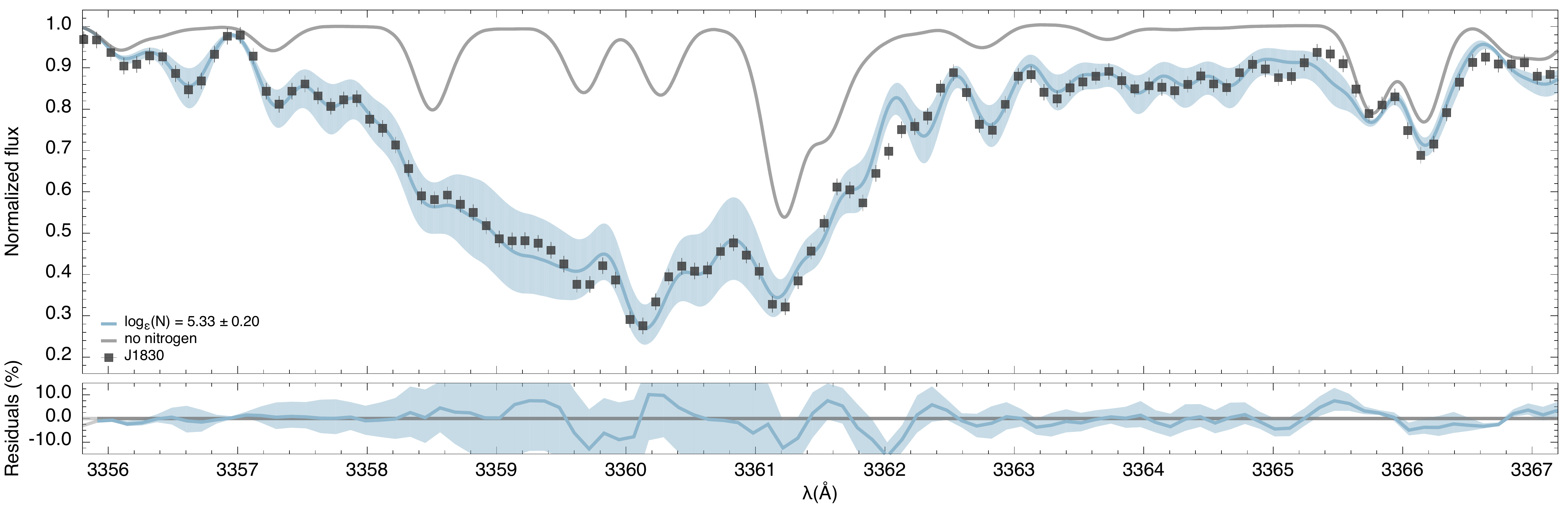}
\epsscale{0.575}
\plotone{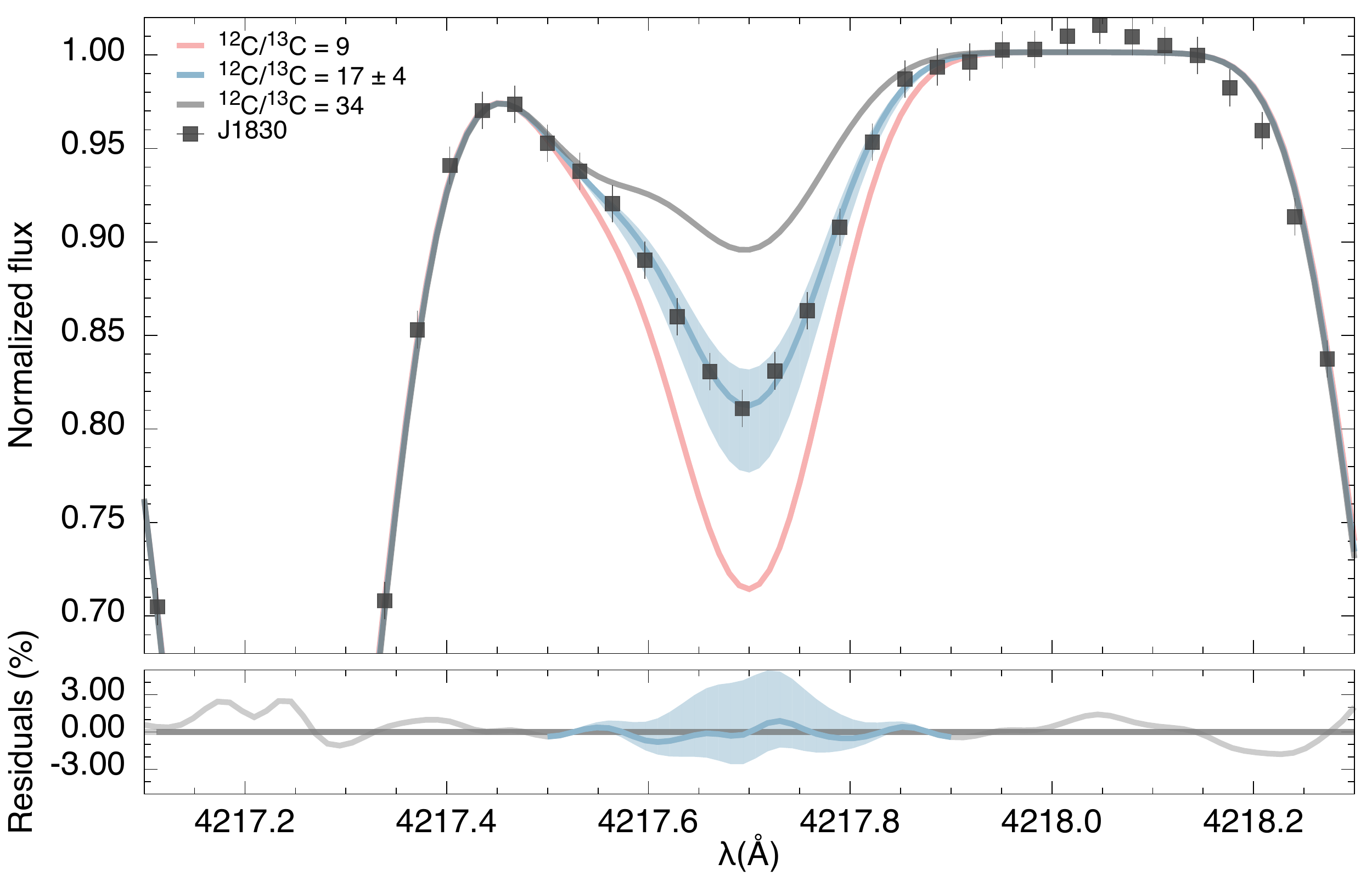}
\epsscale{0.575}
\plotone{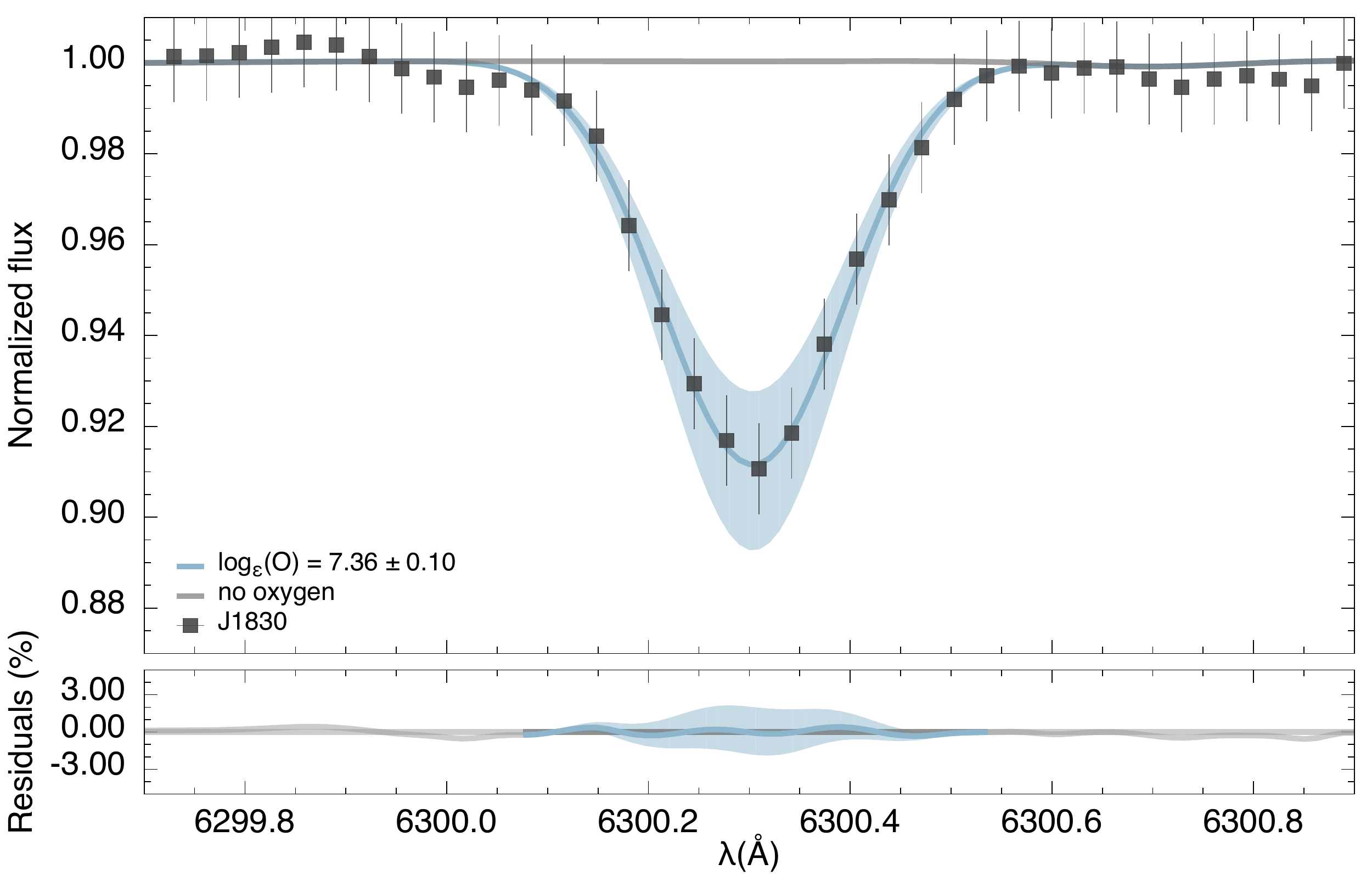}
\caption{Spectral syntheses for the determination of carbon (upper panel),
nitrogen (middle panel), and oxygen (lower right panel) abundances. The top
panel of each plot shows the best-fit syntheses (blue lines) and uncertainties
(shaded regions) compared to the observed spectra (points). Also shown are
syntheses after removing all the contributions from specific elements (gray lines).The
bottom panels show the residuals between the observed spectra and the syntheses.
The lower left panel shows the determination of the $^{12}$C/$^{13}$C ratio (see
text for details).}
\label{cn_syn}
\end{figure*}

The carbon abundance for \rave\ was derived from nine different regions of the
MIKE spectrum, including CH/C$_2$ molecules and a \ion{C}{1} atomic feature. All of
the individual abundances (listed in Table~\ref{eqw}) are within 0.12~dex;
the average value found is \eps{C}=6.76 (\cfe $= +1.90$). 
The top panel of Figure~\ref{cn_syn} shows the spectral synthesis of the CH
$G$-band at $\lambda$4280\,{\AA} for \rave. The points represent the observed
spectrum, the solid blue line is the best abundance fit, and the shaded area
represents a variation of $\pm 0.1$~dex in abundance, used to estimate the
uncertainty.  The gray line shows the synthesized spectrum in the absence of
carbon. The lower panel shows the residuals (in \%) between the observed data
and the best fit, which are all below 10\% for the synthesized region. As
\rave\ is on the upper red-giant branch, the observed carbon abundance does not
reflect the chemical composition of its natal cloud. We determined the carbon
depletion due to CN processing for \rave\ to be 0.44~dex, by using the online
calculator\footnote{\href{http://vplacco.pythonanywhere.com/}{http://vplacco.pythonanywhere.com/}}
described in \citet{placco2014}.

\begin{figure*}[!ht]
\epsscale{1.15}
\plotone{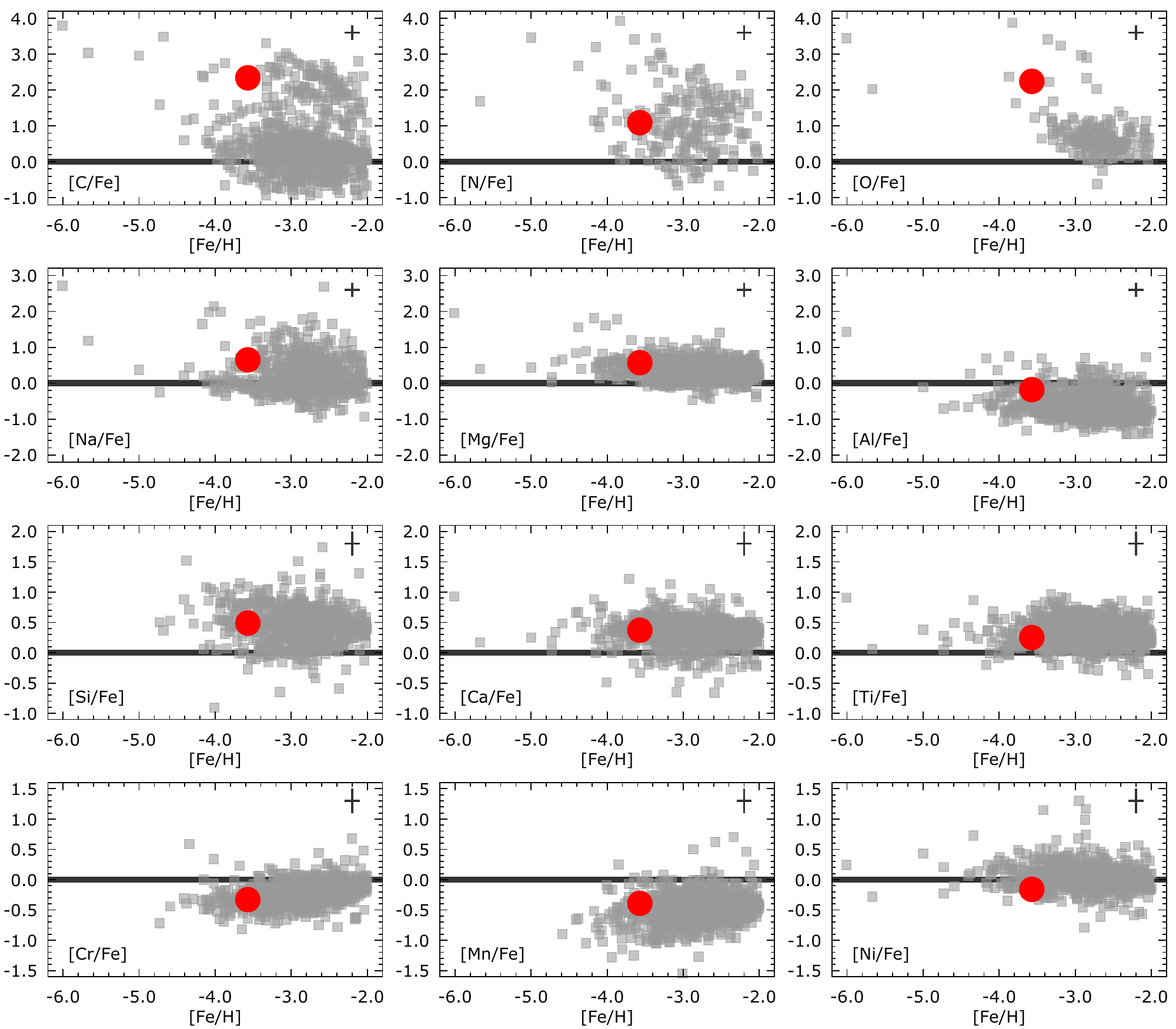}
\caption{Light-element abundance ratios, as a function of the metallicity, for
\protect{\rave}\ (red filled circle) and the JINAbase literature compilation
\citep{jinabase}.}
\label{ab_light}
\end{figure*}

The $^{12}$C/$^{13}$C isotopic ratio is an indicator of the extent of mixing
processes on the red-giant branch stage of evolution.  Using a fixed carbon
abundance of \eps{C}=6.76 for the CH features at $\lambda$4217\,{\AA}, we
derived $^{12}$C/$^{13}$C = $17 \pm 4$, suggesting that substantial
processing of $^{12}$C into $^{13}$C has taken place in \rave.
The lower left panel of Figure~\ref{cn_syn} show the determination of the
$^{12}$C/$^{13}$C isotopic ratio and its uncertainty. Note that the residuals
between the observed data and $^{12}$C/$^{13}$C = 17 are all within 3\%. For the
remainder of the analysis, we have fixed the carbon abundance (\eps{C}=6.76) and
isotopic ratios ($^{12}$C/$^{13}$C = 17), which is of particular importance for
stars with such high levels of carbon as \rave.

The nitrogen abundance was determined from spectral synthesis of the NH band at
$\lambda$3360\,{\AA} (\eps{N}=5.33) and the CN band at $\lambda$3883\,{\AA}
(\eps{N}=5.38). For the CN band, we used a fixed carbon abundance as explained
above. Individual determinations agree within 0.05~dex and the final average
abundance is \eps{N}=5.35 (\nfe $= +1.09$). The middle panel of
Figure~\ref{cn_syn} shows the spectral synthesis for the NH region at
$\lambda$3360\,{\AA}. Similar to carbon, the shaded area (encompassing $\pm
0.2$~dex from the best fit) successfully describes the behavior of this region.

\begin{figure*}[!ht]
\epsscale{1.15}
\plotone{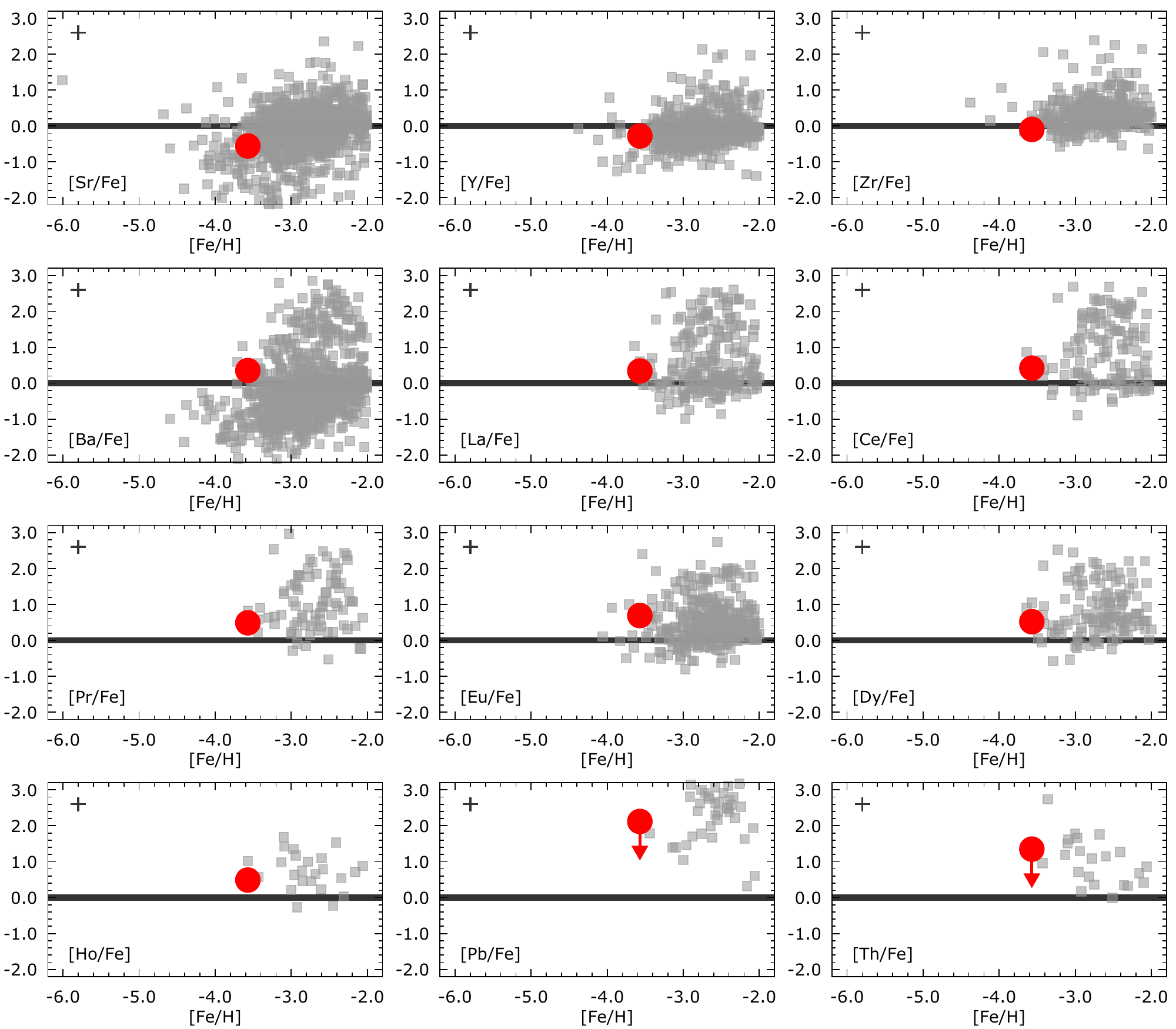}
\caption{Neutron-capture element abundance ratios, as a function of the metallicity, for
\protect{\rave}\ (red filled circle) and the JINAbase literature compilation
\citep{jinabase}.}
\label{ab_heavy}
\end{figure*}

\begin{figure*}[!ht]
\epsscale{0.575}
\plotone{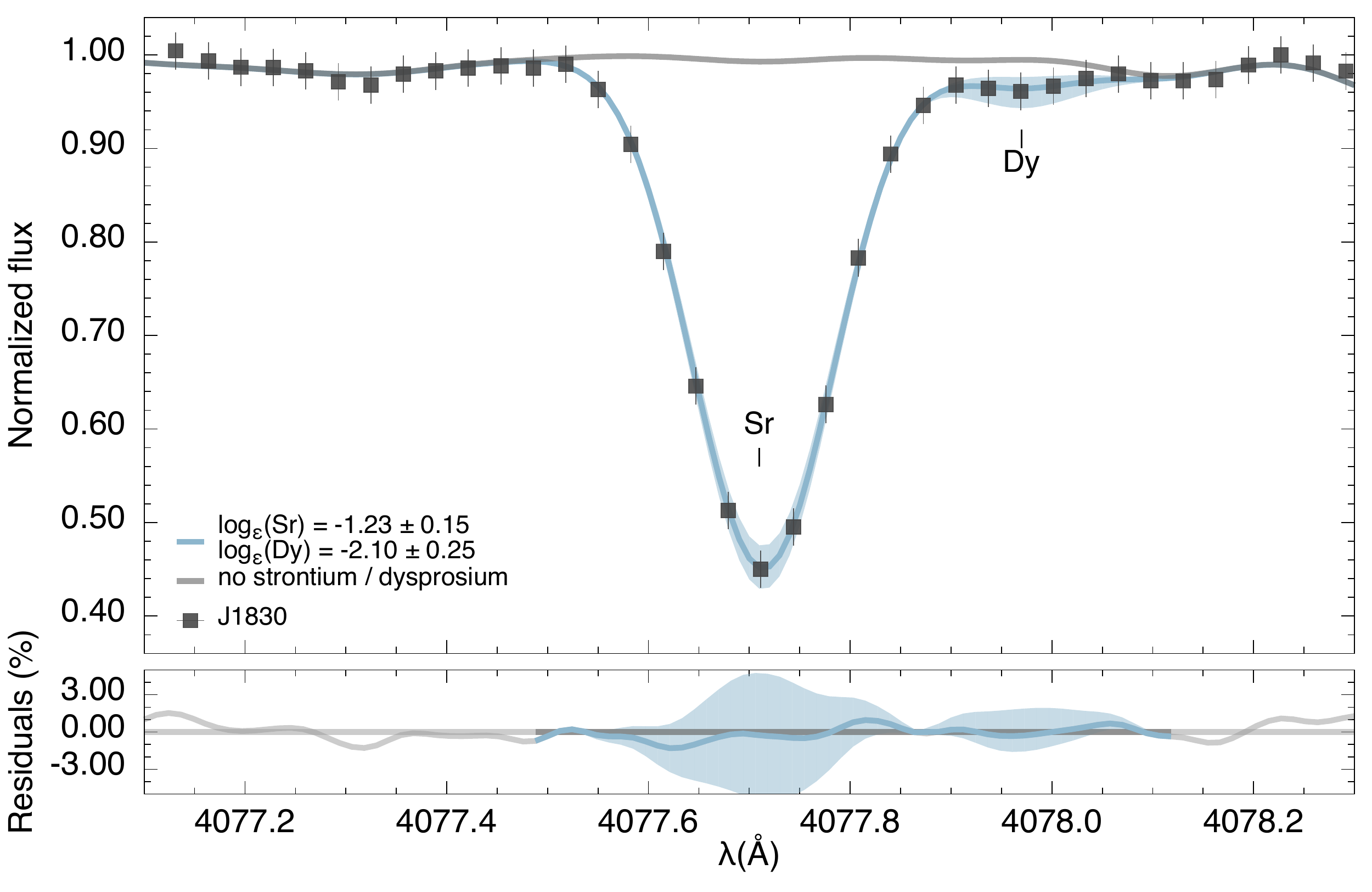}
\epsscale{0.575}
\plotone{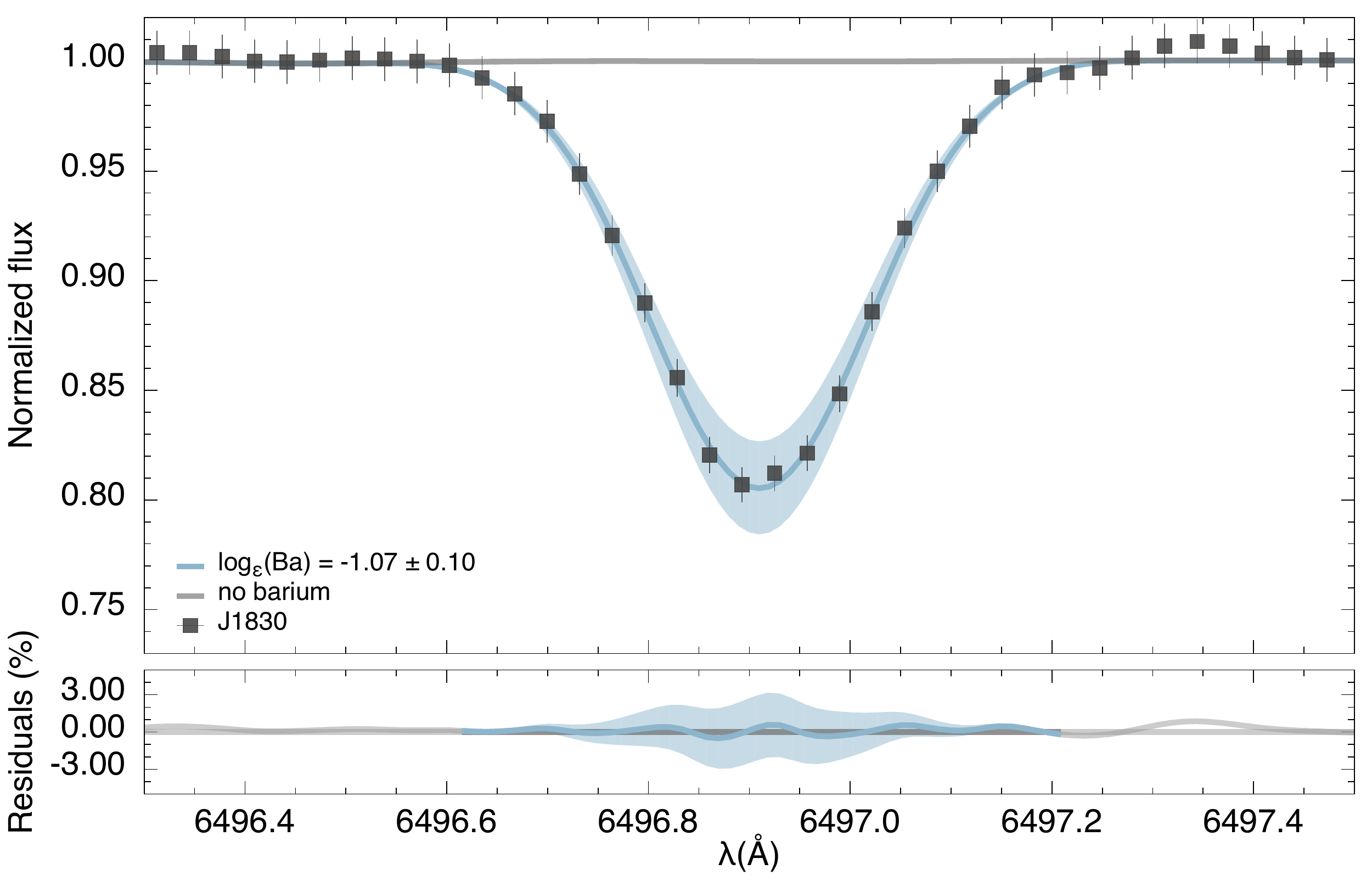}
\epsscale{0.575}
\plotone{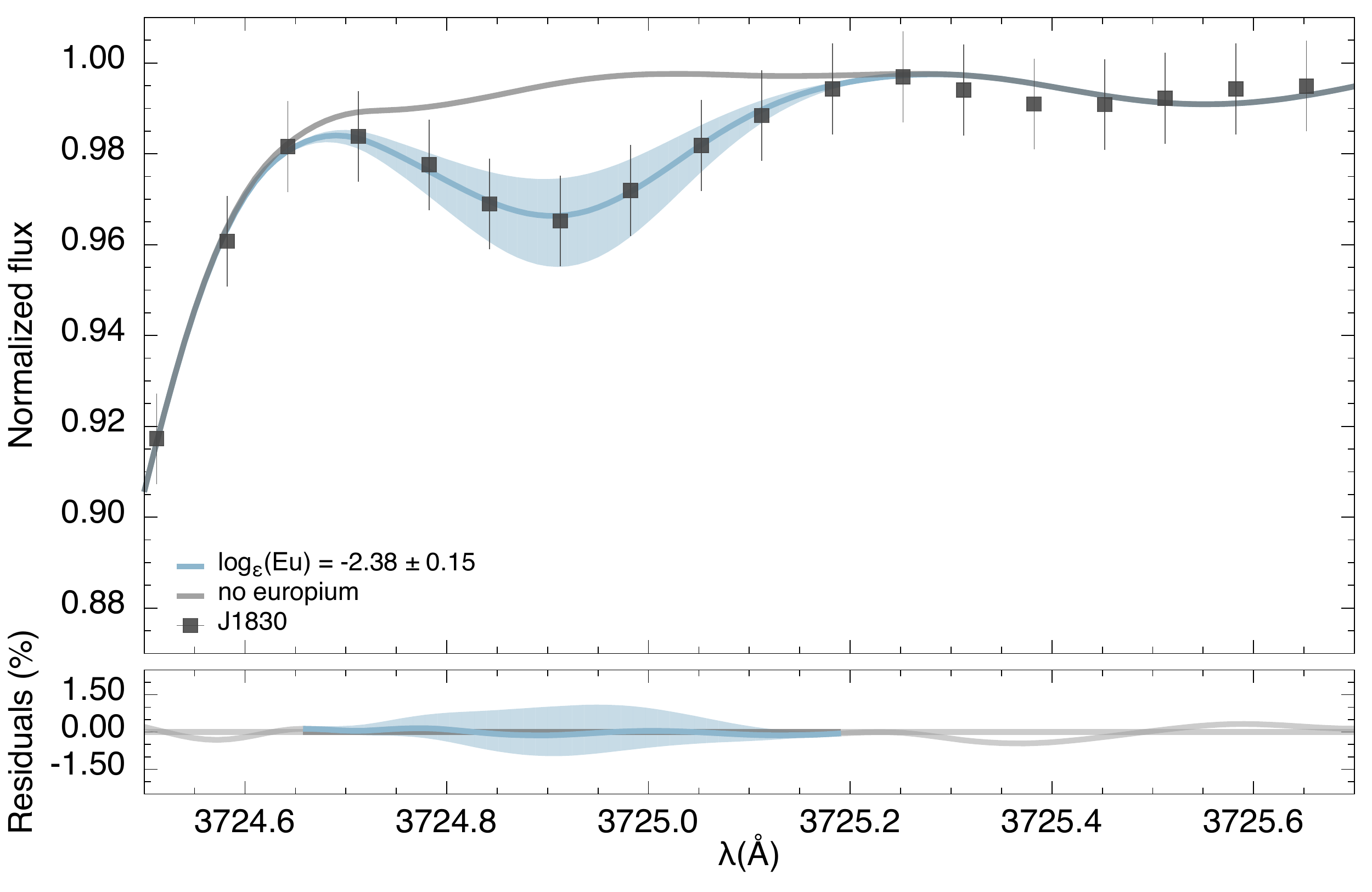}
\epsscale{0.575}
\plotone{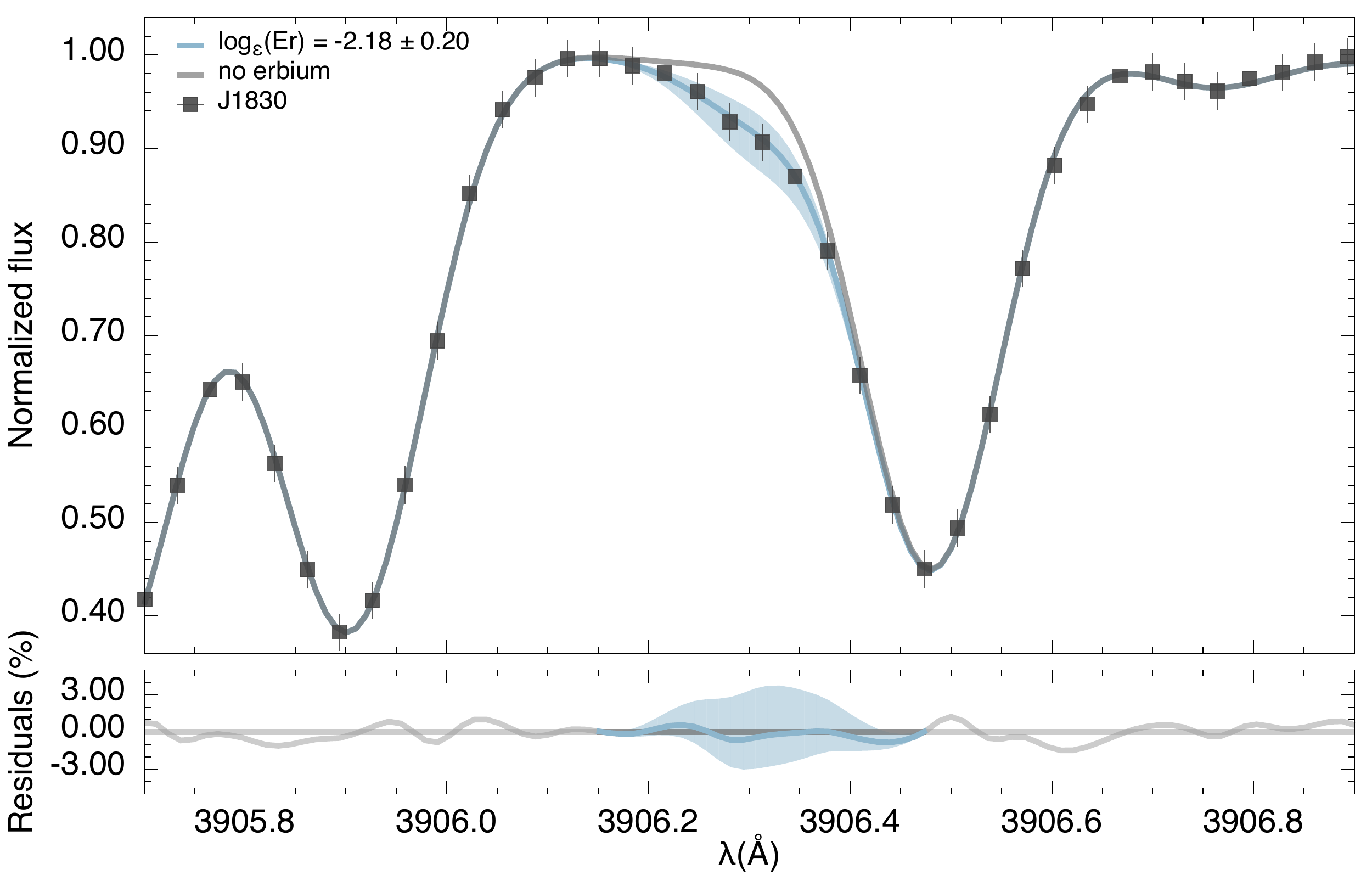}
\epsscale{0.575}
\plotone{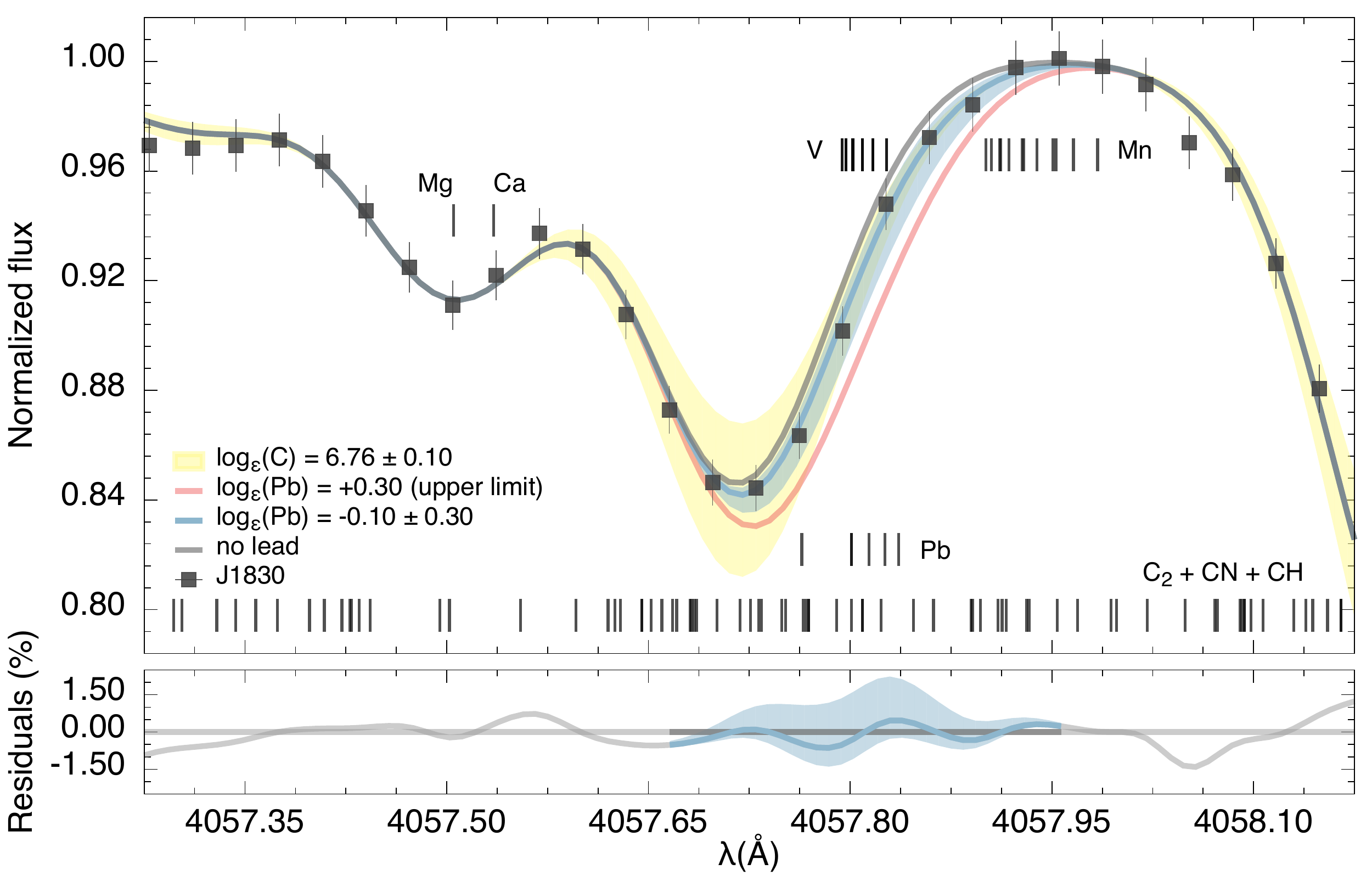}
\epsscale{0.575}
\plotone{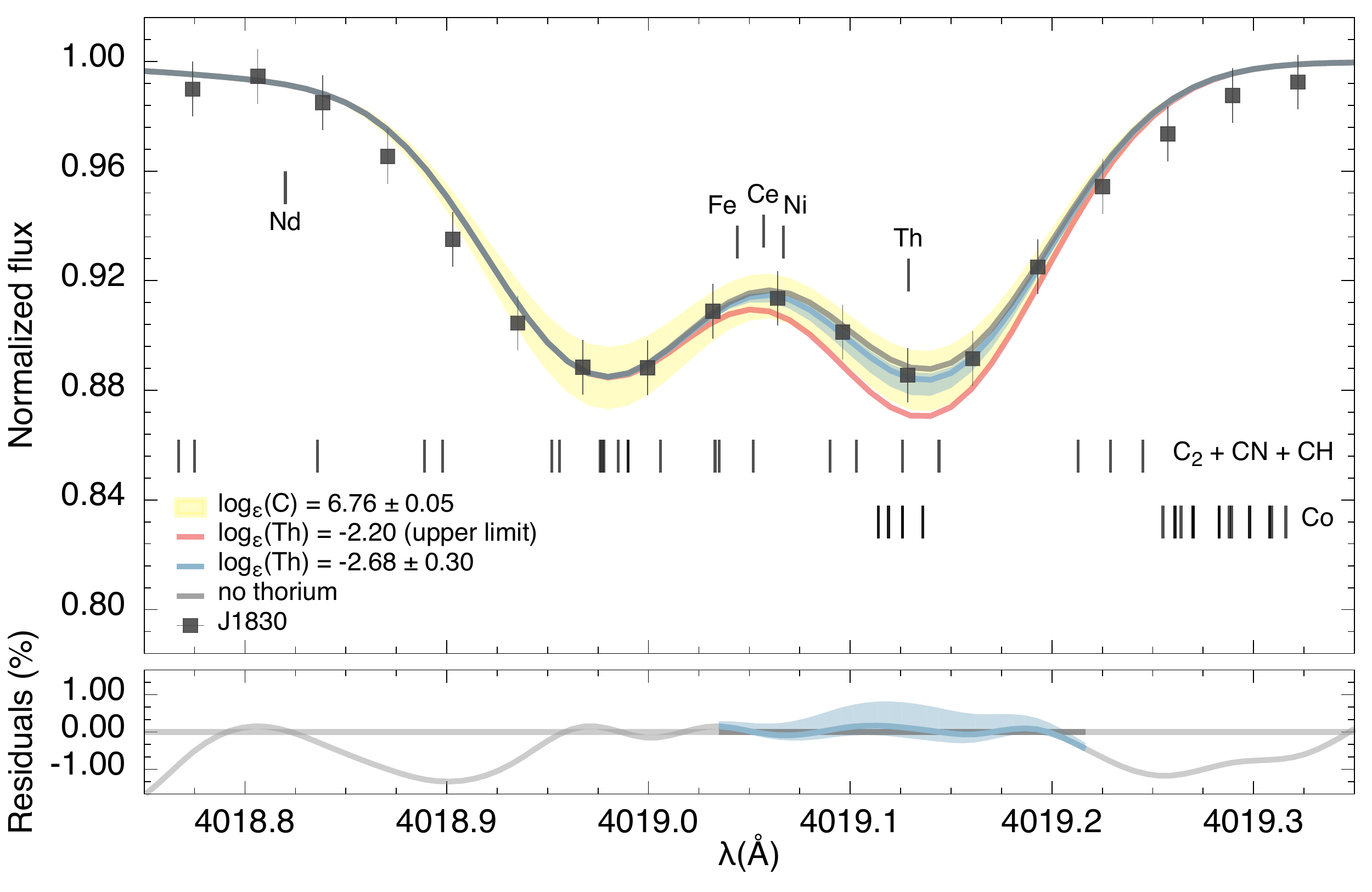}
\caption{Same as Figure~\ref{cn_syn}, but for neutron-capture elemental abundance
determinations.}
\label{ncap_syn}
\end{figure*}

The lower right panel of Figure~\ref{cn_syn} shows the synthesis for the
forbidden oxygen
transition at $\lambda$6300\,{\AA}. Due to the relatively high signal-to-noise
ratio of the MIKE spectrum on this region, we were able to determine an
abundance of \eps{O}=7.36 (\xfe{O} $= +2.24$) with an uncertainty of 0.1~dex. 
Both carbon and nitrogen abundances were determined
assuming \eps{O}=7.36 for the synthetic spectra.

\subsection{From Sodium to Nickel}

Abundances of Na, Mg, K, Ca, Sc, Ti, Cr, Co, and Ni were determined by
equivalent-width analysis only. For Ti, we were able to measure transitions from
two different ionization stages; the abundances agree within 0.03~dex.  For
Al, Si, and Mn, spectral synthesis was used to determine the abundances.  Individual
line measurements are listed in Table~\ref{eqw} and final average abundances are
listed in Table~\ref{abund}. Figure~\ref{ab_light} shows the comparison between
the light-element abundances (from C to Ni), as a function of the metallicity,
for \rave\ and stars in the JINAbase compilation
\citep{jinabase}\footnote{The
values were scaled using the \citet{asplund2009} Solar photospheric abundances.}.
Apart from
the notably high carbon and oxygen abundances, the measurements for \rave\ are 
within the general trends for its metallicity range. In Section~\ref{lightsec} we 
explore the main characteristics of the possible progenitor population of \rave,
which could help explain its light-element abundance pattern.

\subsection{Neutron-capture Elements}

The chemical abundances for 20 neutron-capture elements (from Sr to Th) were
measured in the spectrum of \rave\ through spectral synthesis. We have used the
atomic and molecular line lists generated by the \texttt{linemake}
code\footnote{\href{https://github.com/vmplacco/linemake}{https://github.com/vmplacco/linemake}}.
Individual references for transitions are given in their README file. Where
appropriate, we accounted for line broadening by hyperfine splitting structure
and isotopic shifts. As stated above, for all syntheses we fixed the abundances
of carbon, nitrogen, and the $^{12}$C/$^{13}$C ratio. We also used the
$r$-process isotopic fractions from \citet{sneden2008} for specific elements, as
described below.  Individual line measurements can be found in Table~\ref{eqw}
and final average abundances in Table~\ref{abund}. 

Figure~\ref{ab_heavy} shows the comparison between the neutron-capture element
abundances (from Sr to Th), as a function of the metallicity, for \rave\ and
stars in the JINAbase compilation \citep{jinabase}. Similar to the light
elements, \rave\ appears to follow the general trends shown by the literature
data. However, it is interesting to note that \rave\ is among the lowest
metallicity stars for which elements heavier than barium ($Z=56$) have ever been
measured\footnote{Other examples of stars with Eu detected include
CS~22891$-$200 (\metal = $-3.9$; \citealt{roederer2014b}) and SMSS~J0248$-$6843
(\metal = $-3.7$; \citealt{jacobson2015}).}. 
%
In the following, we provide details on the determinations of neutron-capture
element abundances and upper limits.

\paragraph{Strontium, Yttrium, Zirconium}

These elements belong to the first $r$-process peak and are believed to be
formed by the limited $r$-process \citep{frebel2018}. The upper left panel of
Figure~\ref{ncap_syn} shows the spectral synthesis for the Sr $\lambda$4077
absorption feature. The abundance found for this line (\eps{Sr} = $-$1.23)
agrees well with the value found for the $\lambda$4215 line (\eps{Sr} =
$-$1.28). For Y, we were only able to measure one feature ($\lambda$5205  --
\eps{Y} =  $-$1.64), which lies on the blue wing of a well-modeled Cr feature.
For Zr, two lines were measured, $\lambda$3998 (\eps{Zr} = $-$1.20) and
$\lambda$4045 (\eps{Zr} = $-$0.97), with abundances agreeing within 1-$\sigma$.

\paragraph{Ruthenium and Palladium}

These elements are part of the less explored region within $41 \leq Z \leq 55$.
Abundances for these elements are challenging to measure, and successful
determinations are mostly made in the near ultra-violet range \citep[e.g.][among
others]{roederer2012d}. From the MIKE spectrum we were able to measure one Ru
($\lambda$3728 - \eps{Ru} = $-$1.20) and one Pd ($\lambda$3404 - \eps{Pd} =
$-$1.43) features.

\paragraph{Barium, Lanthanum}

These elements are the main representatives of the second-peak of the
$s$-process. For Ba, abundances were determined from five lines, with an average
abundance of \eps{Ba} = $-$1.04, agreeing within 0.15~dex. In all cases, we have
accounted for the Ba isotopic fractions, following \citet{sneden2008}. The upper
right panel of Figure~\ref{ncap_syn} shows the synthesis for the Ba
$\lambda$6496 line, where the shaded area represents an abundance variation of
$\pm0.10$~dex. For La, we were able to identify two lines ($\lambda$3995 and
$\lambda$4086) in regions less affected by blends and carbon features, with an
average abundance of \eps{La} = $-$2.13.


\paragraph{Cerium, Praseodymium, Neodymium, Samarium}

A total of nine lines were used to determine the abundances of these elements.
The agreement is within 1-$\sigma$ for Pr (two lines - 0.15~dex) and Nd (three lines -
0.10~dex), and 2-$\sigma$ for Sm (three lines - 0.30~dex). Only one line was
used to determine the abundance of Ce, with an uncertainty of 0.20~dex.


\paragraph{Europium}

Europium is an important indicator of the $r$-process and has been widely used
to distinguish between the various subclasses of neutron-capture enhanced
metal-poor stars. In the Solar-System, Eu is mainly formed by the $r$-process
(97\%, according to \citealt{burris2000}), and there are many strong absorption
features that can be measured in the optical wavelength regime. Unfortunately,
most Eu lines (notably $\lambda4129$ and $\lambda4205$) were within regions with
strong molecular carbon absorption features, and hence not possible to be properly
synthesized. We were able to measure abundances for two lines ($\lambda3724$ and
$\lambda4435$) with an average abundance of \eps{Eu} = $-$2.36. The middle-left
panel of Figure~\ref{ncap_syn} shows the synthesis of the Eu $\lambda3724$ line.
It is possible to see that the neighboring Fe line on the blue side of the Eu
feature is well-modeled, and the residuals confirm the good agreement between
synthesis and observations.

\paragraph{Gadolinium, Dysprosium, Holmium, Erbium, Thulium, Ytterbium}

These elements are in the $64 \leq Z \leq 70$ range and have $r$-process
fractions of at least 70\% \citep{burris2000}. All of the absorption features
measured in the MIKE spectrum are in the blue region ($\lambda \leq
4200$\,{\AA}). The agreement between individual line measurements are within
1-$\sigma$ for Gd (two lines - 0.15~dex), Ho (two lines - 0.15~dex), and Er
(three lines - 0.10~dex), and within 2-$\sigma$ for Dy (two lines - 0.30~dex).
Only one line was measured for both Tm and Yb, with uncertainties of 0.25~dex.
The top-left and middle-right panels of Figure~\ref{ncap_syn} show,
respectively, the spectral synthesis for Dy and Er. In both cases, there is a
good agreement between the observations and best-fit abundances, with residuals
within 3\%.


\paragraph{Lead}

Pb is a third-peak element typically produced by the $s$-process. However, at
metallicities of \metal$\leq -3.5$, Pb is expected to be produced by the
$r$-process by the same $\alpha$-decay chains as thorium and uranium
\citep{wanajo2002}. Even the high S/N ratio of the MIKE spectrum did not allow for an
abundance determination from the weak Pb line at $\lambda$4057, which is blended
with a strong carbon feature (see below). We were able to place an upper limit
on the Pb abundance. The lower left
panel of Figure~\ref{ncap_syn} shows the synthesis of the lead feature and also
identifies other species that contribute to the observed absorption. For the
best-fit and uncertainty determinations (blue solid line and blue shaded
region), the abundances of C, N, Mg, Ca, V, and Mn were fixed based on results
shown in Table~\ref{abund}, and the isotopic fractions were taken from
\citet{sneden2008}. The gray and red solid lines show, respectively, syntheses
without contribution of lead and with a lead abundance that is 0.4~dex higher
than the best-fit, which we consider to be our upper limit. 
To further assess possible sources of contamination, we varied the
abundance of carbon by 1-$\sigma$ (yellow shaded region). It is possible to see
that any change in carbon would directly affect the blue side of the absorption feature 
centered at $\sim 4057.7$\,{\AA}. Our final value for the lead abundance is \eps{Pb} $<+0.30$.


\paragraph{Thorium}

Th is a radioactive actinide, and the second heaviest element observable in
stellar spectra. 
We were able to determine an upper limit for the Th abundance from one absorption feature at
$\lambda$4019. Results are shown in the lower right panel of
Figure~\ref{ncap_syn}. Similar to the Pb determination, we fixed the
abundances of C, N, Fe, Co, Ni, Ce, and Nd before attempting to fit the Th
feature. We also explored how changes in the carbon abundance affect the line
strengths; results suggest that, even though C can be well constrained by the feature to the
blue side of the Th line, only an upper limit can be determined. Our final value
is \eps{Th} $<-2.20$.

\section{Analysis}
\label{disc}

\rave\ has an intriguing chemical abundance pattern. The light
elements present similar behavior to those of CEMP-no stars, while the heavy elements
resemble the abundance pattern of an $r$-I star. 
In addition, a binary scenario does not appear to be a possibility due to
the lack of appreciable radial-velocity variations. As a consequence, the
chemical makeup of \rave\ requires an interstellar cloud pre-enriched with elements
ranging from carbon to thorium. Below we speculate on the possible pathways
that may have led to this scenario.

\subsection{Radial-Velocity Variations}

The binary fractions among the different sub-classes of low-metallicity stars
have been subject to extensive study in the literature. \citet{hansen2011} found
that only 18\% of their sample of 17 $r$-process enhanced stars were in binary
systems. A follow-up study with additional radial-velocity data
\citep{hansen2015b} confirmed the conclusion that the chemical peculiarities of
$r$-I and $r$-II stars are not caused by binary companions.
In addition, about 83\% of the CEMP-no stars do not present radial-velocity
variations consistent with a binary system \citep{hansen2016}. 

\begin{figure}[!ht]
\epsscale{1.15}
\plotone{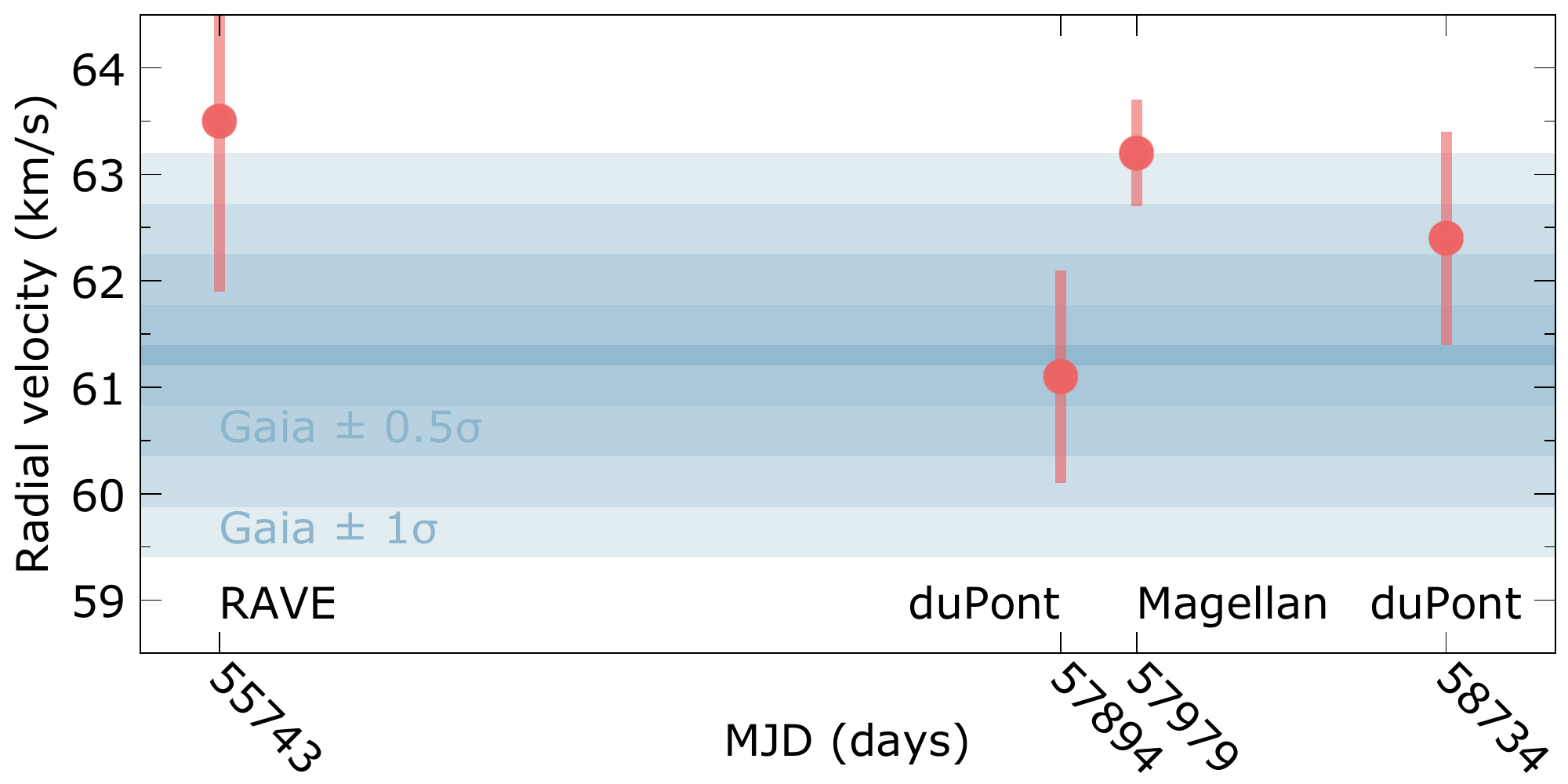}
\caption{Radial velocities for \protect\rave, as a function of observation
date. The solid line and shaded areas show, respectively, the Gaia average value
(based on 4 measurements) and its uncertainty.}
\label{rv}
\end{figure}


Due to its low metallicity and chemical-abundance pattern (light elements
comensurate with the CEMP-no class and neutron-capture elements with the $r$-I
class), \rave\ would not be expected to belong to a binary system.  The star
HE~1012$-$1540 \citep{cohen2008} presents a similar case. It is a single EMP
star (\metal$\sim -3.5$) with CNO enhancement and mildly enhanced in
neutron-capture elements 
(\xfe{Ba}=$+0.20$; \citealt{cohen2013} and \xfe{Ba}=$+0.07$; \citealt{roederer2014}).

The expectation that \rave\ is a single star is supported\footnote{We also
acknowledge the possibility that a lack of observed orbital motion for \rave\ could be
due to the system's orientation (e.g. face-on) or a very long period (P$_{\rm
orb} \gtrsim 10$~years).} by the radial-velocity measurements listed in
Table~\ref{candlist} .  Furthermore, the uncertainty from the Gaia value (based
on 4 epochs -- Gaia DR2 does not provide individual RV measurements) is similar to
the uncertainties for the other spectroscopic values. Figure~\ref{rv} shows the
individual measurements as a function of the observation date. The Gaia DR2 average
value is shown as a solid line, with its uncertainty given by the shaded areas.

\begin{figure}[!ht]
\epsscale{1.15}
\plotone{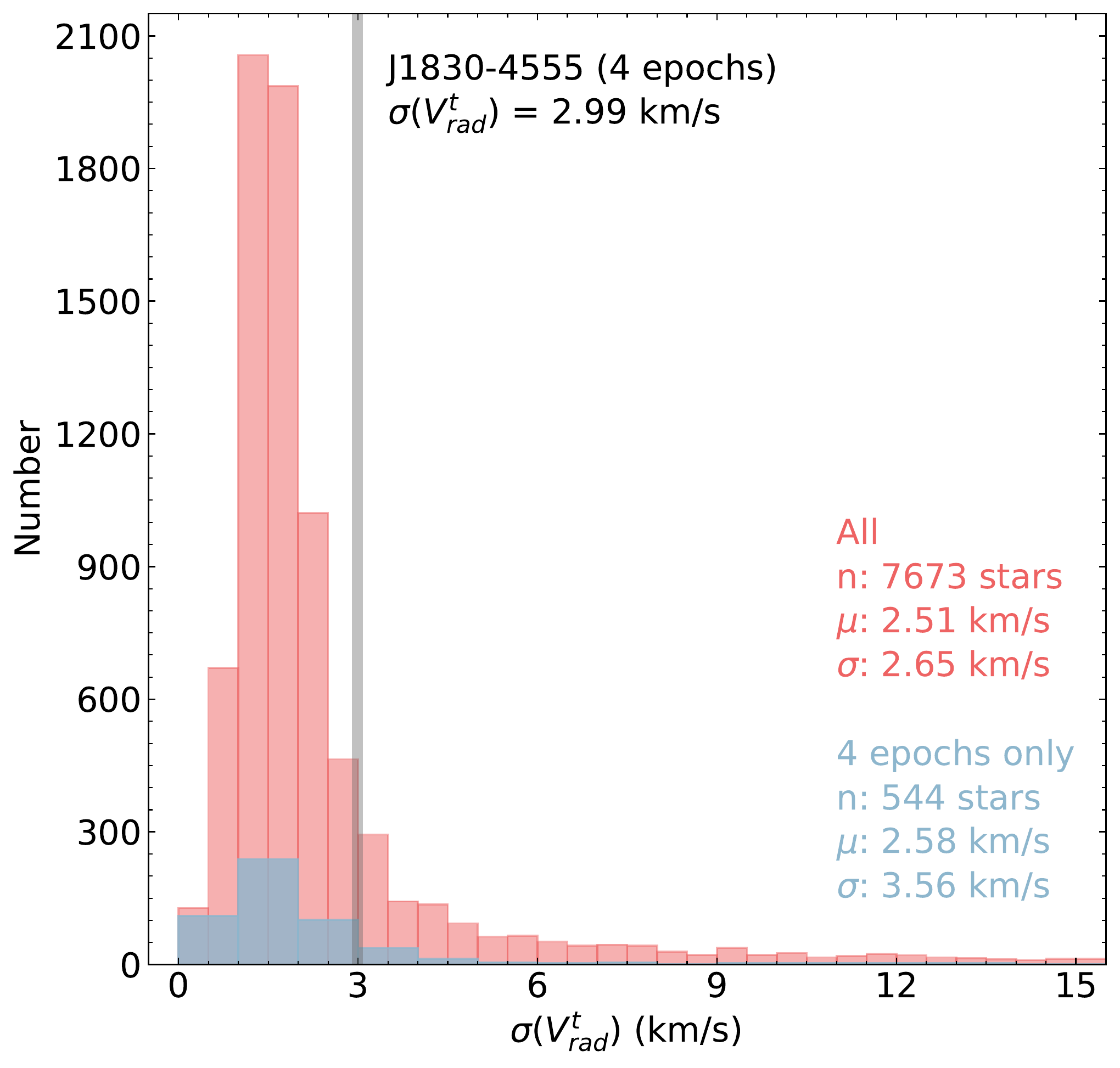}
\caption{Distribution of radial-velocity standard deviations for stars in the
Gaia DR2 database with similar temperatures and magnitudes as \protect\rave\
(see text for details). The number of objects, mean, and standard deviation of
the distribution are listed.  The vertical solid line marks the
$\sigma(V_{rad}^t)$ value for \protect\rave.}
\label{rvar}
\end{figure}

To further test the non-binary hypothesis, we made a comparison between the
standard deviation of the Gaia measurement for \rave\ and stars with similar
magnitudes ($G \pm 0.05$~mag, $G_{\rm BP} \pm 0.05$~mag, and $G_{\rm RP} \pm
0.05$~mag) and temperatures (T$_{\rm eff} \pm 100$~K) present in the Gaia DR2
database. By inspecting the RV standard deviations for this subsample of stars,
it is expected that binary stars would show characteristically larger values,
while single stars (with similar parameters) would have similar RV dispersions.

In total, we found 7,673 stars in the Gaia DR2 database that fulfill the
magnitude and temperature constraints presented above. To calculate the standard
deviation of the epoch radial velocities $\sigma(V_{rad}^t)$, we used the
following relation\footnote{Derived from the equations provided in Table 14.1.1
of \href{https://gea.esac.esa.int/archive/documentation/GDR2/}
{https://gea.esac.esa.int/archive/documentation/GDR2/}}:

\begin{equation}
\sigma(V_{rad}^t) = \sqrt{\left(\frac{2\cdot
\mathtt{o\_RV}}{\pi}\right)\cdot(\mathtt{e\_RV}^2 - 0.11^2)},
\end{equation}

\noindent where \texttt{o\_RV} is the number of epochs used to compute the radial
velocity (\texttt{rv\_nb\_transits}) and \texttt{e\_RV} is the radial-velocity
error (\texttt{radial\_velocity\_error}). A constant noise floor of 0.11 km/s
is added in quadrature to take into account calibration contributions.

Figure~\ref{rvar} shows the distribution of RV standard deviations
for all stars (red histogram) and also for stars with 4 radial-velocity epochs
measured (blue histogram), which is the case for \rave. The labels on the figure
show the number of stars, average, and standard deviation for
$\sigma(V_{rad}^t)$ in both cases. The extended tails of the distributions are
strong evidence of the presence of binaries. \rave\ has a value consistent
with the average for both distribution, which adds
confidence to the assertion of its non-binary status.

\subsection{The Light-Element Abundance Pattern} 
\label{lightsec}

At \metal=$- 3.57$, \rave\ is well within the realm of the so-called
mono-enriched stars, for which interstellar clouds were polluted by a single
progenitor population \citep{hartwig2018}. One of the main diagnostics to
identify mono-enriched stars is, along with \metal, the \abund{Mg}{C} abundance
ratio. In the case of \rave, both the observed and ``natal'' values
(\abund{Mg}{C}=$-1.32$ and $-1.76$, respectively) are consistent with that
classification \citep[Figure 11 of][]{hartwig2018}.
Even though we argue in later sections that \rave\ could have been formed from a
gas cloud polluted by more than one progenitor, here we speculate on the
possible origin of the light elements, from carbon to nickel, from a single progenitor.

\begin{figure*}[!ht]
\epsscale{1.15}
\plotone{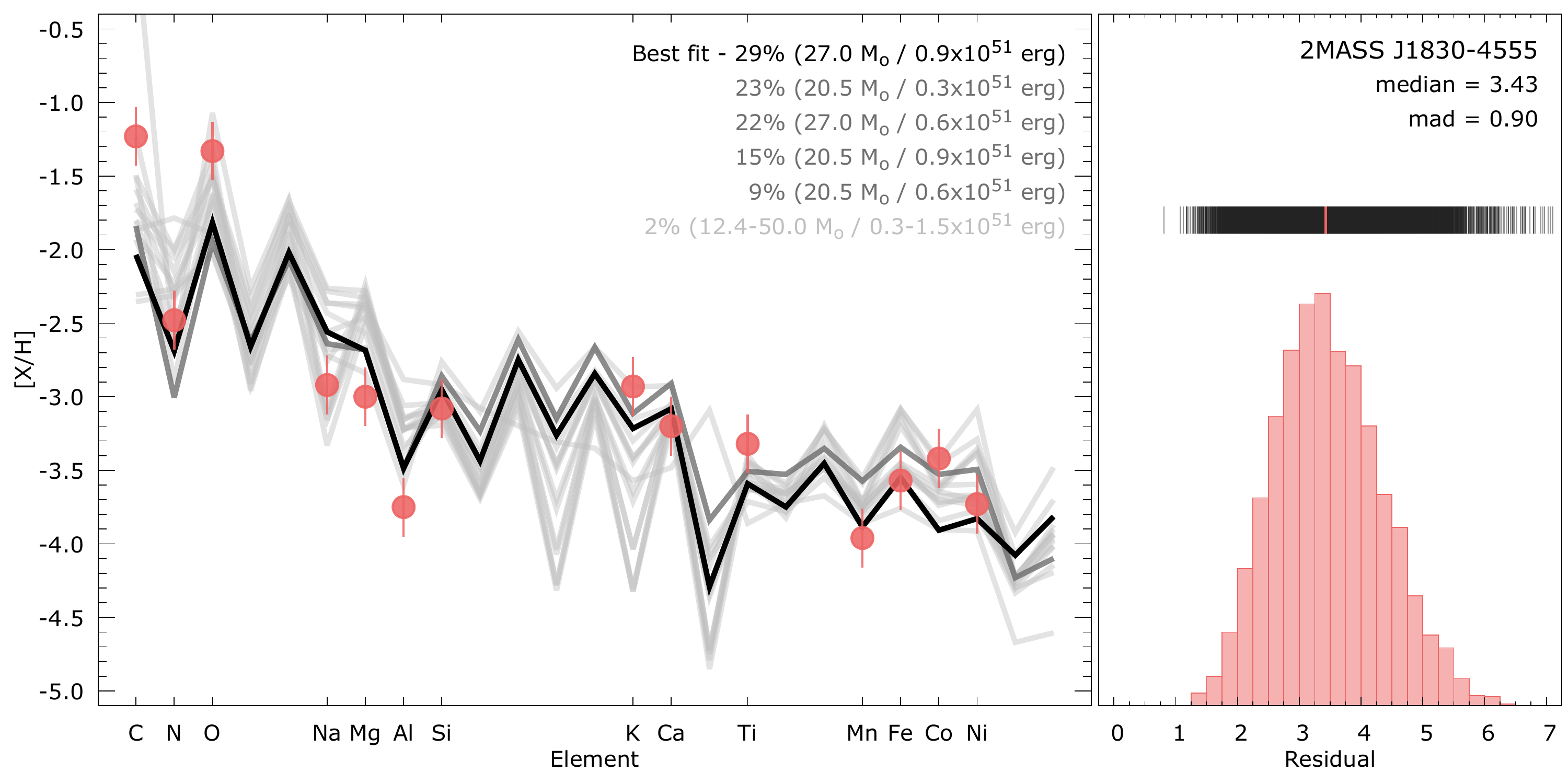}
\caption{Best model fits for \protect\rave, ordered by increasing median
residuals.  The left panel shows the observed [X/H] abundance ratios (red filled
circles), along with the yields (solid lines) from the S4 models described in
\citet{heger2010}. The masses and explosion energies are provided in the legend
at the upper right, color-coded by their fractional occurrence. The right panel
shows the distribution of the residuals (in dex) for the 10,000 simulations 
(see text for details). The colored bar overlaying the upper density
distribution in the right panel marks the median value, shown in the legend at
the top right, along with the median absolute deviation (MAD).}
\label{starfit}
\end{figure*}

\begin{figure*}[!ht]
\epsscale{1.15}
\plotone{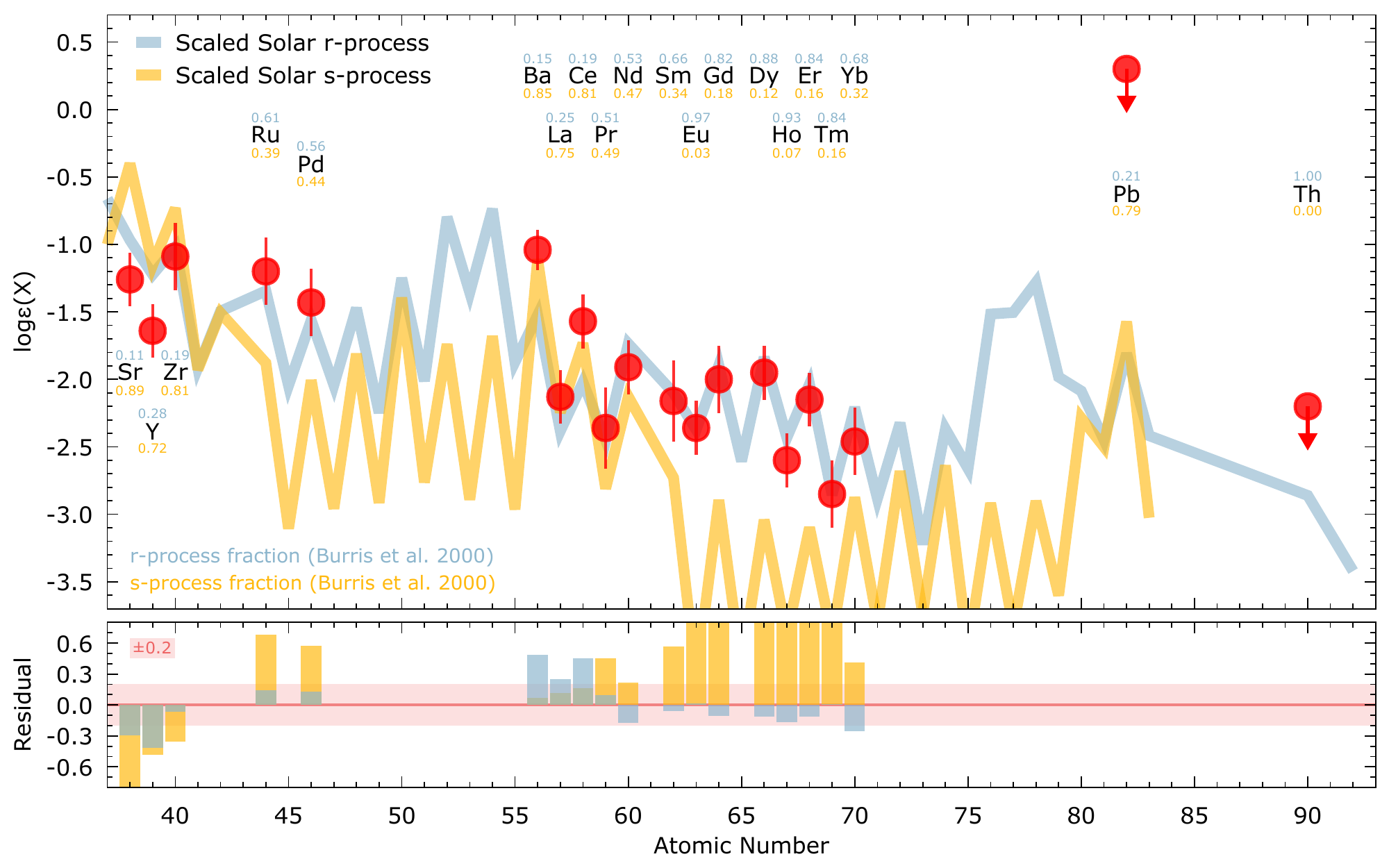}
\caption{Upper panel: Heavy-element chemical abundance pattern of \protect\rave,
compared with the scaled Solar System abundances. The $r$- and $s$-process
contributions in the Sun are calculated based on the fractions given in
\citet{burris2000} and scaled to match the observed abundances of Eu and Ba,
respectively. Lower panel: residuals between observations and the scaled Solar
System abundance patterns.}
\label{pattern}
\end{figure*}

We attempted to model the light-element abundance signature of \rave\ with
theoretical Pop\,III supernova nucleosynthesis
yields\footnote{\href{http://starfit.org}{http://starfit.org}} by
\citet{heger2010}. 
These models follow the evolution and explosion of metal-free stars,
where the initial composition is pristine big bang nucleosynthesis and both mass
loss and rotation are neglected throughout the evolution.
The (S4) fallback models have masses from 10 to
100\,M$_\odot$ and explosion energies from $0.3 \times 10^{51}$\,erg to $10
\times 10^{51}$\,erg. 
Details of their $\chi^2$ matching algorithm can be found in
\citet{placco2015b}, \citet{frebel2015}, \citet{placco2016}, and
\citet{placco2016b}, where this procedure is applied to EMP stars in the
literature. 

The fitting results are shown in Figure~\ref{starfit}. Similar to
\citet{placco2016b}, we generated 10,000 abundance patterns for \rave, by
re-sampling the $\log\epsilon (\mbox{X})$ and $\sigma$ values from
Table~\ref{abund}. By running the {\sc{starfit}} code for each re-sampled
pattern (and determining its respective best-fit model), we found that only 16
different models were used. Their fractions can be seen on the left panel of
Figure~\ref{starfit}. The ``best-fit'' result found in 29\% of the re-samples is a
model with 27.0\,M$_\odot$ and $0.9\times10^{51}$\,erg.  
In general, Figure~\ref{starfit} shows that a possible progenitor for \rave
could have had masses of 20.5-27 M$_{\odot}$ and explosion energies of $0.3-0.9
\times 10^{51}$\,erg. This support the conclusion presented in
\citet{mardini2019b}, which suggests that stellar masses $\sim$20\,M$_{\odot}$
may reflect the initial mass function of the first stars.
The right panel of Figure~\ref{starfit} shows the distribution of the residuals
for each best-fit model.  Comparing the median residual\footnote{The residual
for each iteration is taken as the sum of the absolute values of the differences
between predicted yields and measurements.} (3.43) with the values
for the UMP stars presented in Figure~3 of \citet{placco2016b} confirms that
\rave\ likely belongs to the group of stars for which the faint-SN
models of \citet{heger2010} can explain the observed light-element
abundances.

\begin{figure*}[!ht]
\epsscale{1.15}
\plotone{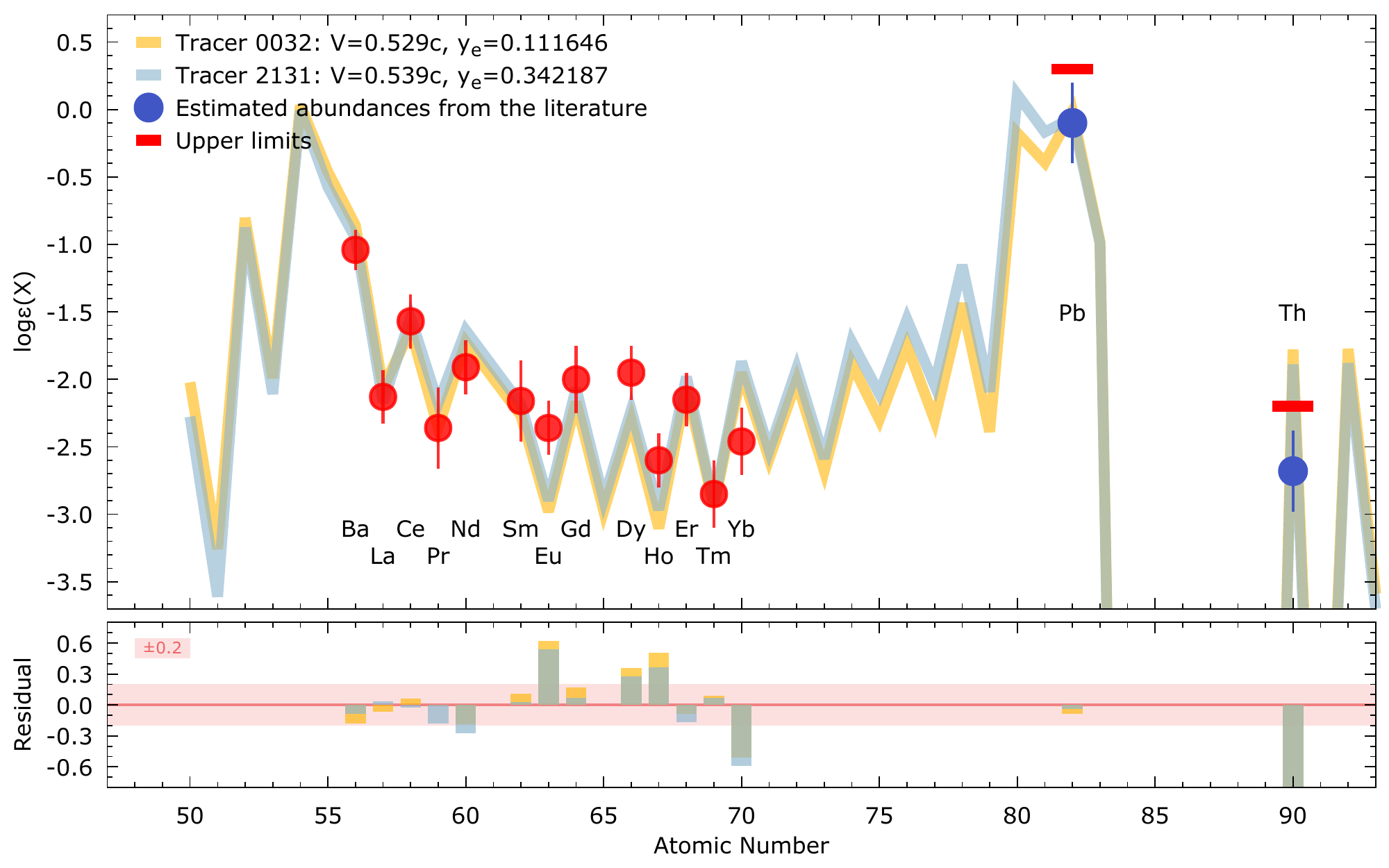}
\caption{Upper panel: Heavy-element chemical abundance pattern of \protect\rave\
(see text for explanations on the choices for the Pb and Th values -- upper
limits also shown for reference), compared with the $r$-process nucleosynthesis
yields for two very fast ($v>0.5c$) example tracers 0032 and 2131 from a neutron
star merger simulation \citep{bovard2017}. The abundance pattern is scaled to
match the observed abundance of Ce.  Lower panel: residuals between observations
and the scaled fast $r$-process ejecta abundance patterns.}
\label{nsm}
\end{figure*}

\subsection{The Heavy-Element Abundance Pattern}
\label{heavysec}

\subsubsection{Comparison with Solar System $r$ and $s$ Fractions}

According to the classifications proposed by \citet{beers2005} and
\citet{frebel2018}, the heavy-element abundance pattern of \rave\ has a
signature of the {\emph{main r-process}} and it is classified as a $r$-I star
(\xfe{Eu}=$+0.69$ and \abund{Ba}{Eu}=$-0.34$).
The upper panel of Figure~\ref{pattern} shows the heavy-element abundance
pattern of \rave, compared to the Solar System $r$-process (scaled to Eu) and
$s$-process (scaled to Ba), using the fractions of \citet{burris2000}. Each
label shows the element name and its $r$ and $s$ fractions in the Sun. The lower panel shows
the residuals between observations and the $r$ and $s$ scaled patterns. The red
shaded area denotes the typical uncertainty ($\sim0.2$~dex) in the abundance
measurements.

It is possible to see that the elements from Pr to Yb reproduce the normalized
$r$-process pattern quite well, within 1-$\sigma$. The same applies to the
lighter elements Ru and Pd. In contrast, the first peak elements (Sr, Y, and Zr)
appear to be under-produced when compared with the scaled patterns. 
Under the assumption that all the neutron-capture elements in J1830-4555 were
produced by a single neutron star merger, we can estimate the lanthanide
fraction that would be measured if we were to observe the neutron star merger's
kilonova directly. Following the methods in \citet{ji2019}, we estimate the
lanthanide fraction to be $\log X_{\rm La} = -1.47$. This lanthanide fraction is
in the $\sim 80$~th percentile of lanthanide fractions for all stars and is
higher than the lanthanide fraction observed in GW170817.
For Ba, La, and Ce, there is a clear over-production when compared to the scaled
$r$-process pattern\footnote{The $r$-process abundance pattern is derived by
subtracting the $s$-process contributions from the Solar System values
\citep{roederer2010b}.}, which could suggest a contribution of the $s$-process
to the observed abundance pattern of \rave. 

The operation of the $s$-process can also be traced by abundance ratios such as
\abund{Ba}{Eu}, \abund{La}{Eu}, and \abund{Pb}{Eu}. The traditional limit for
$r$-process enhanced stars set by \citet{beers2005} is \abund{Ba}{Eu}$<0$, which
is met in the case of \rave. In addition, \citet{roederer2010b} sets approximate
minimum ratios expected from AGB pollution to be \abund{La}{Eu}$\approx 0.0$ and
\abund{Pb}{Eu}$\approx +0.3$ \citep[cf. their Figure 3 and also Figure 15
of][]{placco2013}. For \rave, the \abund{La}{Eu} ratio is consistent with the
$r$-process expectation (\abund{La}{Eu}=$-0.35$), yet the upper limit for the
lead-to-europium ratio exceeds the threshold, at \abund{Pb}{Eu}$<+1.43$,
potentially making it consistent with the $s$-process expectation. 
However, even though the operation of the $s$-process at \metal$< -3.5$ is
possible, the case of \rave\  would require both a high-mass AGB (donor) star
and a binary system signature. \citet{choplin2017} speculates that
some CEMP-$s$ stars, which appear to be single according to radial-velocity
monitoring program presented in \citet{hansen2016c}, could have been formed from the
ejecta of low-metallicity {\emph{spinstars}}. However, their [Fe/H] regime is
somewhat higher than the value found for \rave\ and the models are also not able
to reproduce high Pb abundances.

\subsubsection{Comparison with Ejecta from a Neutron Star Merger Event}

Given that the standard $s$- and $r$-process patterns are not a clear match to
the chemical abundances of \rave, we consider the possibility that a
non-standard $r$-process could be responsible for the observed Ba, La, Ce, and
Pb over-production. For this exercise, we assume that \rave\ would have
{\emph{measured}} abundances of Pb and Th, set to be 0.4 and 0.5~dex lower than
the estimated upper limits (\eps{Pb}$=-0.10$ and \eps{Th}$=-2.70$),
respectively. These values would be consistent with measurements reported in the
literature for stars in the same metallicity range: 
CS~30322-023 (\metal=$-3.44$ and \eps{Pb}$=+0.10$; \citealt{masseron2006}) and 
CS~30315-029 (\metal=$-3.43$ and \eps{Th}$=-2.45$; \citealt{siqueira2014}).
Figure~\ref{nsm} compares the chemical abundances of \rave\
(for $Z\geq56$) with theoretical predictions for very fast ($v>0.5c$) ejecta
from a neutron star merger event. For the example calculations shown here, we
take two tracers from the SFHO-M1.35 model of \citet{bovard2017} that simulates
the merger of two 1.35\,M$_\odot$ neutron stars. 
This model considers binary neutron star systems on quasicircular orbits
and initial configurations built from three different equations of state. The
flow of ejected material is followed by tracer particles or measured in
spherical surfaces at fixed distances from the center of the event
\citep[see][for further details]{bovard2017}.
The nucleosynthesis
calculations are from \citet{wang2019}, where details of the calculation appear
and the outcome of the full model set is shown. Here we focus on example tracers
0032 and 2131, which have high speeds, $v=0.529c$ and $v=0.539c$, and distinct
initial electron fractions, $Y_{e}=0.112$ and $Y_{e}=0.342$, respectively. 

\begin{figure*}[!ht]
\epsscale{1.15}
\plotone{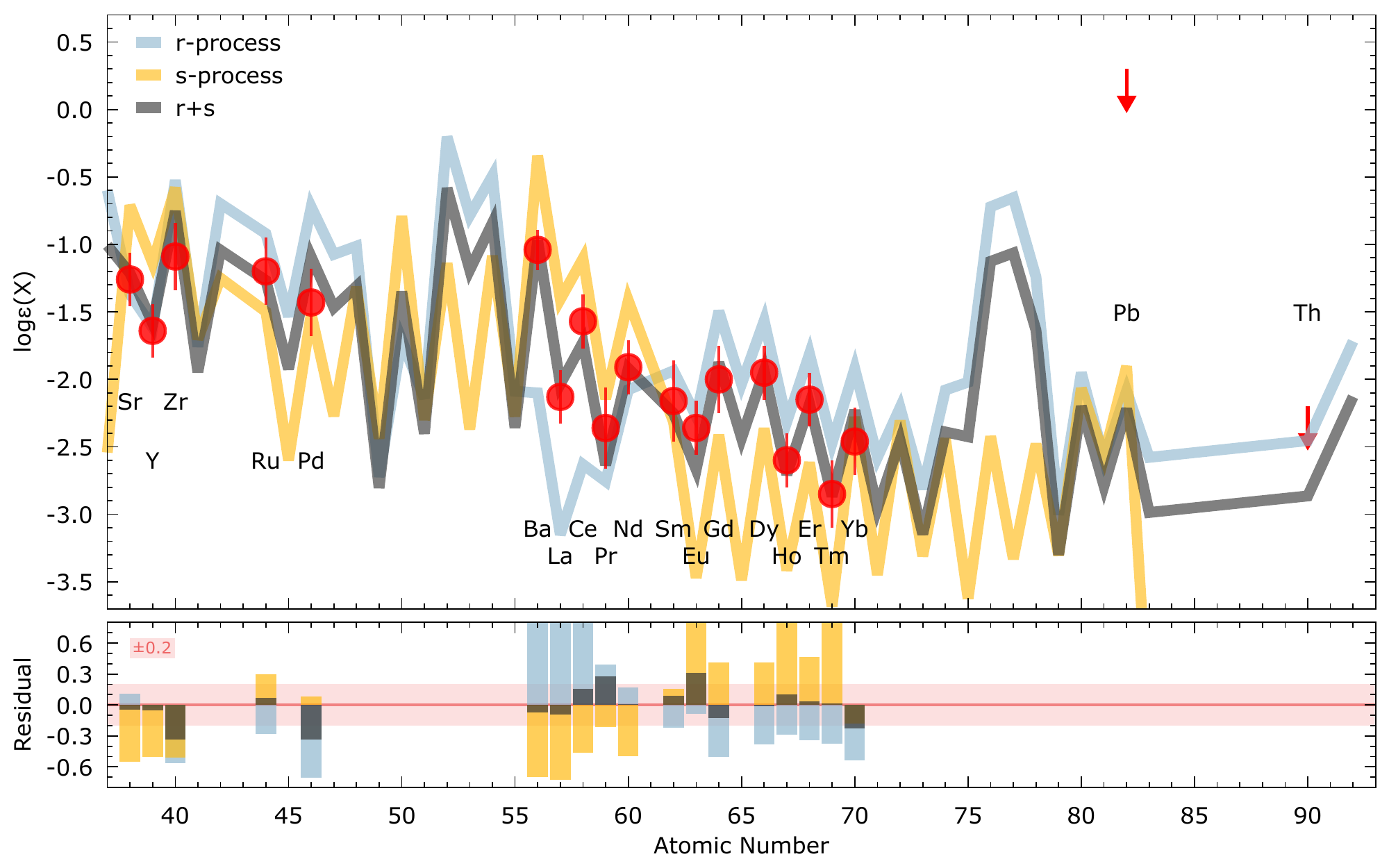}
\caption{Best fits for trans-iron elements, if considering $s$-process only
(yellow), $r$-process only (blue) and a combination of $s$- and $r$-process (black,
see text for details).}
\label{iprocess1}
\end{figure*}

\begin{figure*}[!ht]
\epsscale{1.15}
\plotone{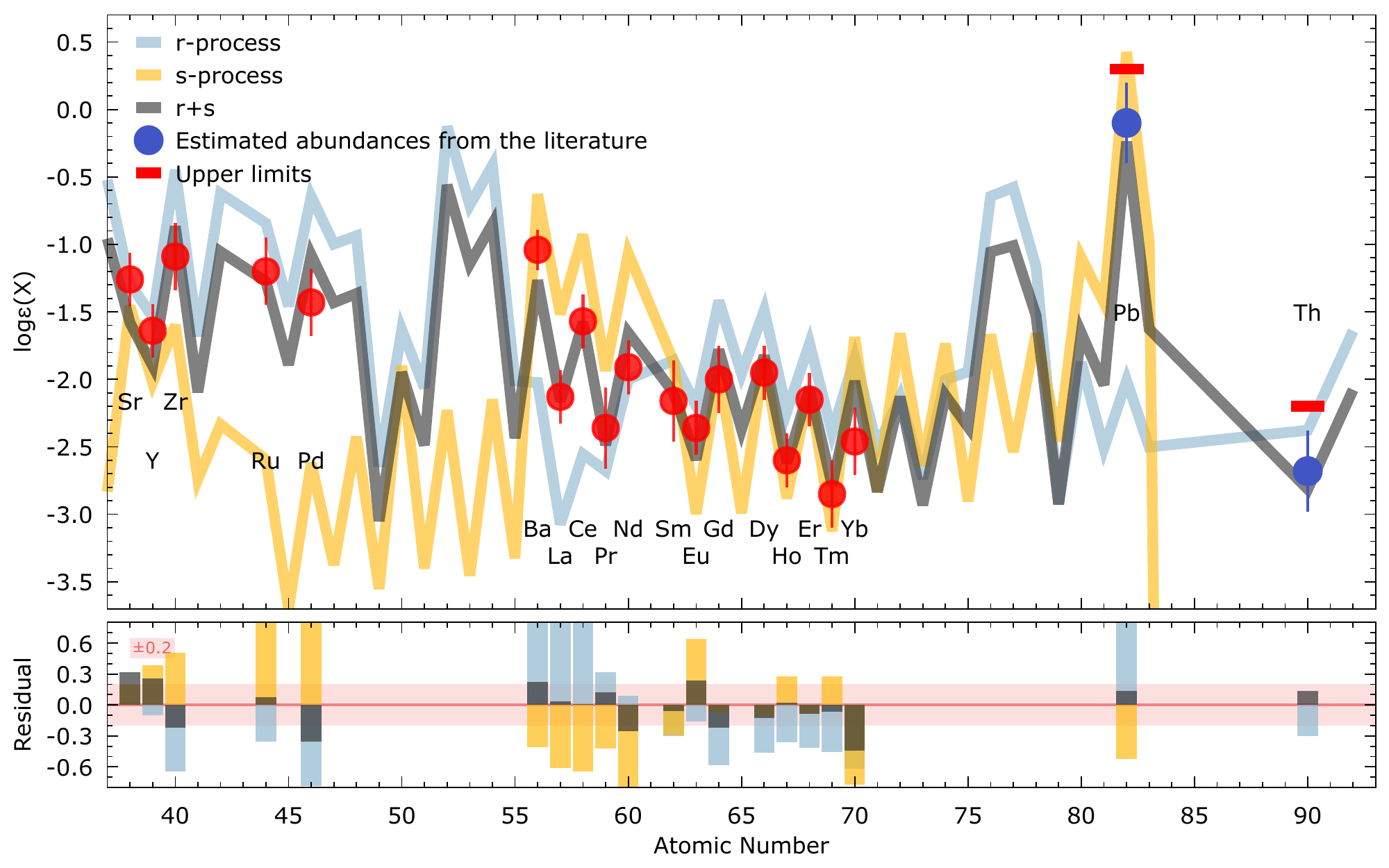}
\caption{Same as Figure~\ref{iprocess1}, but considering Pb and Th as
measurements (see text for details).  In this case, a stronger $s$-process is
required to reproduce the Pb abundance and the $r$-process pattern is the same as
in Figure~\ref{iprocess1}.}
\label{iprocess2}
\end{figure*}

\begin{figure*}
\epsscale{1.10}
\plotone{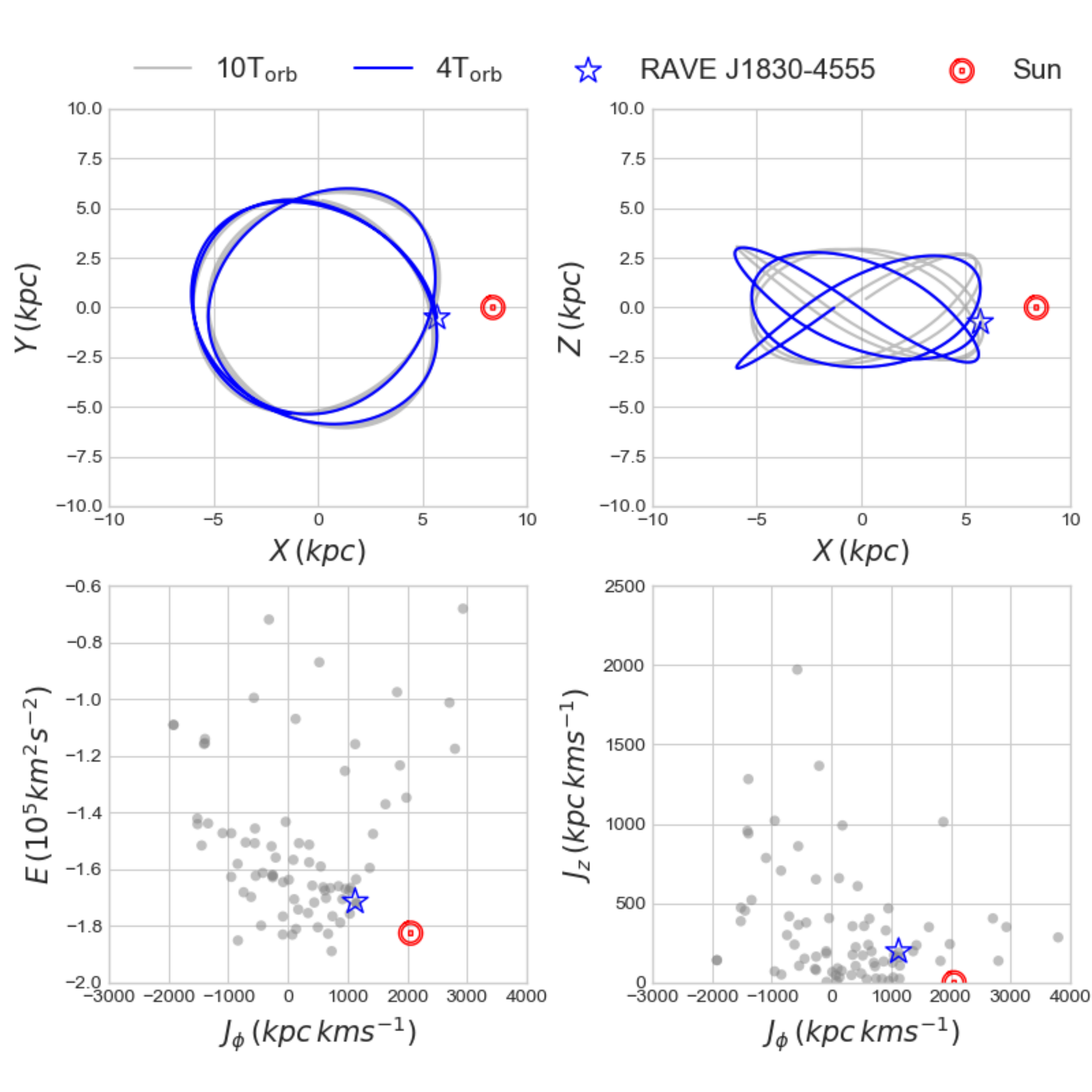}
\caption{Upper panels: Integrated orbit for \protect{\rave} (current position
marked as a blue star) over its last 10 periods (gray line - 1.0~Gyr
look-back time) and 4 periods (blue line - 0.4~Gyr look-back time) in
the $xy$ and $xz$ planes. Lower panels: Behavior of \protect{\rave} in two
different orbital-energy vs. action planes, compared with data from
\citet{roederer2018b} for $r$-process enhanced stars. The current position of the
Sun in all panels is marked with a red circle.}
\label{fig:action}
\end{figure*}

When neutron-rich material is ejected from an $r$-process event at very high
speeds, the temperature and density evolve so quickly that a large neutron
excess persists through the late stages of the $r$-process. In this scenario,
neutron capture can continue throughout the decay to stability, shifting the
second peak into the barium region and the third peak to the lead region. The
resulting final elemental patterns are a good fit to \rave\ abundances for $Z\geq
56$, especially for Ba, La, and Ce. 
Though speculative, this type of fast
$r$-process provides a consistent explanation for the unusual abundance features
of \rave.
One possible way of testing the hypothesis of shifting the third peak to the
lead region would be by comparing upper limits for Os ($Z$=76) or Ir ($Z$=77)
that are either at, or lower than, the scaled solar $r$-process pattern.


\subsubsection{Comparison with $r$ and $s$-process in Massive Stars}

Non-rotating massive stars are known to experience a weak $s$-process, mainly
during the core helium-burning phase \citep[e.g.][]{langer1989, prantzos1990,
raiteri1991a}.  Rotational mixing can significantly boost the $s$-process in
massive stars \citep{pignatari2008, frischknecht2016, choplin2018, limongi2018,
banerjee2019}.  Indeed, during the core helium-burning phase, rotation-induced mixing
progressively transports $^{12}$C and $^{16}$O from the helium core to the
hydrogen shell. It boosts the CNO cycle and produces extra $^{13}$C and
$^{14}$N. These newly synthesized elements are engulfed back by the convective
helium core. Some $^{22}$Ne is synthesized through the chain
$^{14}$N($\alpha,\gamma$)$^{18}$F($\beta^+$)$^{18}$O($\alpha,\gamma$)$^{22}$Ne.  Neutrons are then released through
the $^{13}$C($\alpha,n$) and $^{22}$Ne($\alpha,n$) reactions.
A massive star dying as a jet-like magnetorotational supernova or as a collapsar
may also experience an $r$-process event \citep[e.g.][]{winteler2012,
nishimura2015, siegel2019}. 

We calculate the $s$-process patterns thanks to a one-zone nucleosynthesis code
mimicking the core helium-burning phase of a rotating massive star.  The code
follows the central temperature and density of a complete 25~$M_{\odot}$ stellar
model during core helium burning.  The initial chemical composition is taken
from a low-metallicity stellar model at core helium-burning ignition. The
initial abundances of elements heavier than Fe are set equal to zero. The
initial Fe mass fraction is $2.03 \times 10^{-7}$ which corresponds to [Fe/H] $=
-3.5$ \citep[using the solar abundances of][]{asplund2009}.  During the
nucleosynthesis calculation, $^{13}$C and $^{14}$N are injected at a constant
rate in order to mimic the effect of rotational mixing during stellar evolution
\citep[see][for details about the injection method]{choplin2016}.
We also consider a $r$-process contribution from a magnetorotational supernova
model of \cite{nishimura2015}. We selected their \textsc{B11$\beta$1.00} model,
which has an initial magnetic field of $10^{11}$ G  and a ratio of rotational
energy to gravitational binding energy of $10^{-2}$ (cf. their Table~1).

The best fit when considering only an $s$-process ($r$-process) event is shown by
the yellow (blue) line in Figure~\ref{iprocess1}. The amount of added hydrogen is taken as
a free parameter to minimize the $\chi^2$ value (adding hydrogen has the effect
of shifting the pattern down).  The best fit when combining the two set of
yields is shown by the black line. In this case the dilution factor between the
$s$- and $r$-process material is also let as a free parameter.
If the Pb and Th abundances are set as measurements instead of upper limits, a
reasonable agreement can also be found if considering a stronger $s$-process
(Figure~\ref{iprocess2}).  The high Pb abundance can be reproduced if the
injection rate of $^{13}$C and $^{14}$N is increased by a factor of 30 during
the $s$-process calculation.  This would correspond to a stronger
rotation-induced mixing during stellar evolution.

Although we considered rotating massive stars as $s$-process sources in this
exercise, an $s$-process originating from AGB stars cannot be excluded.  If the
$s$-process pattern comes from an AGB star, two astrophysical sources are
required to explain the heavy-element abundance pattern of \rave\ (an AGB and an
$r$-process source). By contrast, if the $s$-process comes from a rotating
massive star, just one source could be sufficient since a rotating massive star could
produce both the $s$-process pattern during stellar evolution and the
$r$-process pattern at the time of the supernova (e.g. during magnetorotational
supernova or collapasar).

In this scenario, the massive star cannot be a Pop III star, since the
$s$-process is a secondary process that requires some heavy seeds (e.g. Fe)
during stellar evolution. Without these heavy seeds (i.e. in the case of a
Pop III star), the $s$-process would be too weak (or even non-existent) and the
$s$-process abundances (e.g. Ba) of \rave\ could not be reproduced. Thus, a
possible scenario would be that one (or more) Pop III massive stars first
exploded (similar to the progenitor suggested in Section~\ref{lightsec}) and
injected some Fe into the interstellar medium (up to [Fe/H] $\sim -3.5$) and then
a second generation rotating massive star formed, exploded as a
magnetorotational SN or collapsar and enriched the interstellar medium with a
mixture of $s$- and $r$-elements. Then \rave\ formed as a third generation star.

%

\subsection{Orbital Properties}

In this section, we investigate the orbital properties of \rave, using the proper
motion measurements from {\it Gaia} DR2 \citep[][]{gaiadr2}. We take the
distance determined from the \texttt{StarHorse} code \citep{starhorse19}, which
uses a Bayesian method combining the parallaxes and optical photometry to derive
the stellar parameters and distances, and sets the parallax zero-point offset to
be $+0.05$~mas for bright sources. The radial velocity is taken from the
Magellan spectrum. All of the kinematic parameters used in these calculations are
listed in Table~\ref{candlist}. The values of Solar motion used are ($U$, $V$,
$W$) = (11.10, 12.24, 7.25) kms$^{-1}$~\citep{sch19}, and the motion of the LSR
is $v_{\rm LSR}$ = 232.8 kms$^{-1}$~\citep{mc11}. We then calculate the orbital
energy of \rave\ and trace its orbit in the gravitational potential of
\citet{mc17} using \textsc{AGAMA}~\citep{agama}. In addition, we generate
$10^{4}$ realizations of the same set of parameters by sampling the distances
assuming Gaussian distributions according to its 16th and 84th percentile values, as
well as the proper motions by taking into account their observational errors and
covariance matrix.  

The integrated orbit of \rave\ over the last 10 periods (1.0~Gyr
look-back time) is indicated as a gray line in
the upper panels of Figure~\ref{fig:action}. The orbit over the last 4 periods
(0.4~Gyr look-back time) is
highlighted in blue. This orbit integration clearly shows that \rave\ possesses
a very circular motion in the Galactic plane and a small vertical motion in the
$Z$ direction, confined within $z_{\rm max}$ = 3.1$^{+0.5}_{-0.4}$ kpc. This is
consistent with its low eccentricity $e$ = 0.125$^{+0.12}_{-0.11}$, and may be
associated with the metal-weak thick disk.  This can be also be inferred from
inspection of the ($E$, $J_{\phi}$) and ($J_{\rm z}$, $J_{\phi}$) planes shown
in the lower panels of Figure~\ref{fig:action}, where the gray circles represent
the 89 $r$-process enhanced stars from \citet{roederer2018b}.  Compared to the
solar values, \rave\ has a lower orbital energy, which is consistent with its
smaller pericentric and apocentric distances ($r_{\rm peri}$ =
5.26$^{+0.15}_{-0.16}$ kpc, $r_{\rm apo}$ = 6.77$^{+0.35}_{-0.35}$ kpc). \rave\
also has a non-negligible action in the $Z$ direction with $J_{\rm z}$ =
202.5$^{+65.1}_{-61.9}$ kpc km s$^{-1}$. 

In summary, \rave\ has typical metal-weak thick-disk dynamics. It was very
likely born {\emph{in-situ}} and its orbit was heated during early merger
events. Alternatively, \rave\ could have come from accreted systems with very
prograde orbits.  \citet{sestito2020} argues that a star such as \rave\ could be
the by-product, at early times, of the assembly of the proto Galaxy, minor
mergers, or {\emph{in-situ}} formation. 

\section{Conclusions}
\label{final}

We have presented the first high-resolution spectroscopic study of the extremely
metal-poor, CNO-enhanced star \rave. This star shows an intriguing chemical
abundance pattern, combining a light-element abundance pattern that resembles
one of a ``mono-enriched'' star with a heavy-element pattern matching either the one
from the fast ejecta from a neutron-star merger event 
or the one from a rotating massive star experiencing an $r$-process event
during its explosion. Measurements of lead and thorium abundances in stars
similar to \rave\ would help distinguishing between these (and other) possible
formation scenarios.
The lack of radial velocity variations suggest that \rave\ is not
in a binary system, ruling out the possibility of chemical enrichment via mass
transfer from an evolved companion. Analysis of the orbital parameters derived
from Gaia DR2 data places \rave\ in the metal-weak thick-disk population of the
Milky Way, presenting interesting constraints on the population of its
progenitor(s). At \metal$< -3.5$, this peculiar third-generation star could have
been formed {\emph{in-situ}} or during the early stages of the assembly of the
Milky Way.

\vfill
\clearpage
\newpage

\acknowledgments

The authors would like to thank Fiorenzo Vincenzo for providing feedback
on the first version of this manuscript.
The authors acknowledge partial support for this work from the National Science
Foundation (NSF) under Grant No. PHY-1430152 (Physics Frontier Center / JINA
Center for the Evolution of the Elements).
R.M.S. acknowledges support from CNPq (Project 436696/2018-5) and
CAPES/PROAP-IF.
R.S. and X.W. acknowledge support from the Department of Energy Contract No.
DE-FG02-95-ER40934 and National Science Foundation Contract No. PHY-1630782
Focused Research Hub in Theoretical Physics: Network for Neutrinos, Nuclear
Astrophysics, and Symmetries (N3AS).
I.U.R. acknowledges financial support from grants AST 16- 13536 and AST-1815403,
awarded by the NSF.
A.C. acknowledges funding from the Swiss National Science Foundation under grant
P2GEP2\_184492.
A.P.J. is supported by NASA through Hubble Fellowship grant HST-HF2-51393.001
awarded by the Space Telescope Science Institute, which is operated by the
Association of Universities for Research in Astronomy, Inc., for NASA, under
contract NAS5-26555
A.F. is supported by NSF CAREER grant AST-1255160.  
This research has made use of NASA's Astrophysics Data System Bibliographic
Services; the arXiv pre-print server operated by Cornell University; the SIMBAD
database hosted by the Strasbourg Astronomical Data Center; the IRAF software
packages distributed by the National Optical Astronomy Observatories, which are
operated by AURA, under cooperative agreement with the NSF; 
the SAGA Database \citep[Stellar Abundances for Galactic
Archeology;][]{saga2008}; the JINAbase chemical-abundance database
\citep{jinabase}; and the online Q\&A platform {\texttt{stackoverflow}}
(\href{http://stackoverflow.com/}{http://stackoverflow.com/}).
This research was made possible through the use of the AAVSO Photometric All-Sky
Survey (APASS), funded by the Robert Martin Ayers Sciences Fund.
This publication makes use of data products from the Two Micron All Sky Survey,
which is a joint project of the University of Massachusetts and the Infrared
Processing and Analysis Center/California Institute of Technology, funded by the
National Aeronautics and Space Administration and the National Science
Foundation.
This work has made use of data from the European Space Agency (ESA) mission {\it
Gaia} (\url{https://www.cosmos.esa.int/gaia}), processed by the {\it Gaia} Data
Processing and Analysis Consortium (DPAC,
\url{https://www.cosmos.esa.int/wdpac/consortium}). Funding for the DPAC
has been provided by national institutions, in particular the institutions
participating in the {\it Gaia} Multilateral Agreement.

\vfill

\software{
{\texttt{awk}}\,\citep{awk}, 
{\texttt{gnuplot}}\,\citep{gnuplot}, 
{\texttt{IRAF}}\,\citep{tody1986,tody1993}, 
{\texttt{matplotlib}}\,\citep{matplotlib}, 
{\texttt{n-SSPP}}\,\citep{beers2014}, 
{\texttt{numpy}}\,\citep{numpy}, 
{\texttt{pandas}}\,\citep{pandas}, 
{\texttt{R-project}}\,\citep{rproject}, 
{\texttt{sed}}\,\citep{sed}.
}

\newpage

\bibliographystyle{aasjournal}

\input{placco.bbl}

\end{document}

%% file: line.tex
         CH  &  4049.000  &  \nodata &   \nodata &       syn &      6.73 \\
         CH  &  4246.000  &  \nodata &   \nodata &       syn &      6.78 \\
         CH  &  4261.000  &  \nodata &   \nodata &       syn &      6.73 \\
         CH  &  4280.000  &  \nodata &   \nodata &       syn &      6.78 \\
         CH  &  4313.000  &  \nodata &   \nodata &       syn &      6.71 \\
         C2  &  4737.000  &  \nodata &   \nodata &       syn &      6.83 \\
         C2  &  4940.000  &  \nodata &   \nodata &       syn &      6.78 \\
         C2  &  5165.000  &  \nodata &   \nodata &       syn &      6.78 \\
 \ion{C}{1}  &  8335.000  &  \nodata &   \nodata &       syn &      6.78 \\
         NH  &  3360.000  &  \nodata &   \nodata &       syn &      5.33 \\
         NH  &  3380.000  &  \nodata &   \nodata &       syn &      5.33 \\
         CN  &  3883.000  &  \nodata &   \nodata &       syn &      5.38 \\
$[$\ion{O}{1}$]$  &  6300.300  &     0.00 &  $-$9.820 &       syn &      7.36 \\
 \ion{Na}{1} &  5889.950  &     0.00 &     0.108 &    138.08 &      3.40 \\
 \ion{Na}{1} &  5895.924  &     0.00 &  $-$0.194 &    114.15 &      3.24 \\
 \ion{Mg}{1} &  3829.355  &     2.71 &  $-$0.208 &    132.26 &      4.61 \\
 \ion{Mg}{1} &  3832.304  &     2.71 &     0.270 &    161.32 &      4.59 \\
 \ion{Mg}{1} &  4702.990  &     4.33 &  $-$0.380 &     31.11 &      4.53 \\
 \ion{Mg}{1} &  5172.684  &     2.71 &  $-$0.450 &    142.09 &      4.66 \\
 \ion{Mg}{1} &  5183.604  &     2.72 &  $-$0.239 &    151.64 &      4.62 \\
 \ion{Mg}{1} &  5528.405  &     4.34 &  $-$0.498 &     30.19 &      4.62 \\
 \ion{Al}{1} &  3961.520  &     0.01 &  $-$0.340 &       syn &      2.70 \\
 \ion{Si}{1} &  4102.936  &     1.91 &  $-$3.140 &       syn &      4.43 \\
 \ion{K}{1}  &  7664.900  &     0.00 &     0.135 &     15.82 &      2.04 \\
 \ion{K}{1}  &  7698.960  &     0.00 &  $-$0.168 &     10.57 &      2.15 \\
 \ion{Ca}{1} &  4454.780  &     1.90 &     0.260 &     42.23 &      3.15 \\
 \ion{Ca}{1} &  4455.890  &     1.90 &  $-$0.530 &      9.73 &      3.10 \\
 \ion{Ca}{1} &  5588.760  &     2.52 &     0.210 &     14.42 &      3.22 \\
 \ion{Ca}{1} &  5594.468  &     2.52 &     0.097 &      8.48 &      3.07 \\
 \ion{Ca}{1} &  5598.487  &     2.52 &  $-$0.087 &      6.78 &      3.16 \\
 \ion{Ca}{1} &  6102.720  &     1.88 &  $-$0.790 &      7.03 &      3.11 \\
 \ion{Ca}{1} &  6122.220  &     1.89 &  $-$0.315 &     21.52 &      3.21 \\
 \ion{Ca}{1} &  6162.170  &     1.90 &  $-$0.089 &     29.02 &      3.16 \\
 \ion{Ca}{1} &  6439.070  &     2.52 &     0.470 &     18.79 &      3.07 \\
 \ion{Sc}{2} &  4415.544  &     0.59 &  $-$0.670 &     42.74 &   $-$0.08 \\
 \ion{Sc}{2} &  5526.785  &     1.77 &     0.020 &     14.79 &   $-$0.06 \\
 \ion{Sc}{2} &  5657.907  &     1.51 &  $-$0.600 &      7.95 &   $-$0.06 \\
 \ion{Ti}{1} &  3989.760  &     0.02 &  $-$0.062 &     27.51 &      1.56 \\
 \ion{Ti}{1} &  3998.640  &     0.05 &     0.010 &     30.30 &      1.58 \\
 \ion{Ti}{1} &  4533.249  &     0.85 &     0.532 &     18.55 &      1.68 \\
 \ion{Ti}{1} &  4981.730  &     0.84 &     0.560 &     18.31 &      1.60 \\
 \ion{Ti}{1} &  4991.070  &     0.84 &     0.436 &     13.27 &      1.56 \\
 \ion{Ti}{2} &  3380.276  &     0.05 &  $-$0.630 &    106.63 &      1.66 \\
 \ion{Ti}{2} &  3383.759  &     0.00 &     0.160 &    135.11 &      1.70 \\
 \ion{Ti}{2} &  3489.736  &     0.14 &  $-$1.980 &     63.11 &      1.70 \\
 \ion{Ti}{2} &  3759.291  &     0.61 &     0.280 &    126.62 &      1.65 \\
 \ion{Ti}{2} &  3761.320  &     0.57 &     0.180 &    122.07 &      1.57 \\
 \ion{Ti}{2} &  3913.461  &     1.12 &  $-$0.420 &     82.78 &      1.58 \\
 \ion{Ti}{2} &  4012.396  &     0.57 &  $-$1.750 &     52.68 &      1.59 \\
 \ion{Ti}{2} &  4417.714  &     1.17 &  $-$1.190 &     47.45 &      1.60 \\
 \ion{Ti}{2} &  4418.331  &     1.24 &  $-$1.970 &     10.35 &      1.58 \\
 \ion{Ti}{2} &  4443.801  &     1.08 &  $-$0.720 &     76.06 &      1.56 \\
 \ion{Ti}{2} &  4450.482  &     1.08 &  $-$1.520 &     33.21 &      1.57 \\
 \ion{Ti}{2} &  4464.448  &     1.16 &  $-$1.810 &     22.62 &      1.72 \\
 \ion{Ti}{2} &  4468.517  &     1.13 &  $-$0.600 &     83.05 &      1.64 \\
 \ion{Ti}{2} &  4470.853  &     1.17 &  $-$2.020 &     11.60 &      1.60 \\
 \ion{Ti}{2} &  4501.270  &     1.12 &  $-$0.770 &     72.62 &      1.56 \\
 \ion{Ti}{2} &  4533.960  &     1.24 &  $-$0.530 &     82.19 &      1.67 \\
 \ion{Ti}{2} &  4563.770  &     1.22 &  $-$0.960 &     61.57 &      1.66 \\
 \ion{Ti}{2} &  4571.971  &     1.57 &  $-$0.320 &     73.45 &      1.67 \\
 \ion{Ti}{2} &  4589.915  &     1.24 &  $-$1.790 &     19.49 &      1.71 \\
 \ion{Ti}{2} &  5129.156  &     1.89 &  $-$1.240 &     12.80 &      1.68 \\
 \ion{Ti}{2} &  5188.687  &     1.58 &  $-$1.050 &     30.24 &      1.57 \\
 \ion{Ti}{2} &  5226.538  &     1.57 &  $-$1.260 &     22.88 &      1.61 \\
 \ion{Ti}{2} &  5336.786  &     1.58 &  $-$1.590 &     13.81 &      1.68 \\
 \ion{Cr}{1} &  3578.680  &     0.00 &     0.420 &     84.88 &      1.75 \\
 \ion{Cr}{1} &  5206.040  &     0.94 &     0.020 &     31.85 &      1.75 \\
 \ion{Cr}{1} &  5208.419  &     0.94 &     0.160 &     38.08 &      1.72 \\
 \ion{Mn}{1} &  4041.380  &     2.11 &  $-$0.350 &       syn &      1.43 \\
 \ion{Mn}{1} &  4754.021  &     2.28 &  $-$0.647 &       syn &      1.48 \\
 \ion{Mn}{1} &  4783.424  &     2.30 &  $-$0.736 &       syn &      1.48 \\
 \ion{Mn}{1} &  4823.514  &     2.32 &  $-$0.466 &       syn &      1.48 \\
 \ion{Fe}{1} &  3476.702  &     0.12 &  $-$1.506 &    103.14 &      3.97 \\
 \ion{Fe}{1} &  3490.574  &     0.05 &  $-$1.105 &    120.06 &      3.99 \\
 \ion{Fe}{1} &  3565.379  &     0.96 &  $-$0.133 &    118.17 &      3.91 \\
 \ion{Fe}{1} &  3608.859  &     1.01 &  $-$0.090 &    120.76 &      3.95 \\
 \ion{Fe}{1} &  3618.768  &     0.99 &  $-$0.003 &    123.86 &      3.90 \\
 \ion{Fe}{1} &  3727.619  &     0.96 &  $-$0.609 &    110.58 &      3.93 \\
 \ion{Fe}{1} &  3743.362  &     0.99 &  $-$0.790 &    103.77 &      3.95 \\
 \ion{Fe}{1} &  3753.611  &     2.18 &  $-$0.890 &     37.23 &      3.81 \\
 \ion{Fe}{1} &  3758.233  &     0.96 &  $-$0.005 &    131.64 &      3.81 \\
 \ion{Fe}{1} &  3763.789  &     0.99 &  $-$0.221 &    124.07 &      3.90 \\
 \ion{Fe}{1} &  3765.539  &     3.24 &     0.482 &     46.35 &      3.85 \\
 \ion{Fe}{1} &  3767.192  &     1.01 &  $-$0.390 &    113.22 &      3.82 \\
 \ion{Fe}{1} &  3786.677  &     1.01 &  $-$2.185 &     47.17 &      3.92 \\
 \ion{Fe}{1} &  3787.880  &     1.01 &  $-$0.838 &     98.39 &      3.84 \\
 \ion{Fe}{1} &  3805.343  &     3.30 &     0.313 &     37.77 &      3.91 \\
 \ion{Fe}{1} &  3815.840  &     1.48 &     0.237 &    124.21 &      3.96 \\
 \ion{Fe}{1} &  3820.425  &     0.86 &     0.157 &    159.54 &      3.87 \\
 \ion{Fe}{1} &  3825.881  &     0.91 &  $-$0.024 &    143.00 &      3.88 \\
 \ion{Fe}{1} &  3827.823  &     1.56 &     0.094 &    110.35 &      3.87 \\
 \ion{Fe}{1} &  3840.438  &     0.99 &  $-$0.497 &    115.64 &      3.90 \\
 \ion{Fe}{1} &  3865.523  &     1.01 &  $-$0.950 &     95.35 &      3.82 \\
 \ion{Fe}{1} &  3887.048  &     0.91 &  $-$1.140 &     98.55 &      3.97 \\
 \ion{Fe}{1} &  3899.707  &     0.09 &  $-$1.515 &    118.38 &      4.00 \\
 \ion{Fe}{1} &  3917.181  &     0.99 &  $-$2.155 &     56.79 &      4.02 \\
 \ion{Fe}{1} &  3922.912  &     0.05 &  $-$1.626 &    114.89 &      3.96 \\
 \ion{Fe}{1} &  3940.878  &     0.96 &  $-$2.600 &     32.77 &      3.97 \\
 \ion{Fe}{1} &  3949.953  &     2.18 &  $-$1.251 &     25.93 &      3.91 \\
 \ion{Fe}{1} &  3977.741  &     2.20 &  $-$1.120 &     30.42 &      3.90 \\
 \ion{Fe}{1} &  4005.242  &     1.56 &  $-$0.583 &     93.96 &      4.02 \\
 \ion{Fe}{1} &  4021.866  &     2.76 &  $-$0.730 &     19.75 &      3.90 \\
 \ion{Fe}{1} &  4045.812  &     1.49 &     0.284 &    121.58 &      3.80 \\
 \ion{Fe}{1} &  4062.441  &     2.85 &  $-$0.860 &     16.32 &      4.03 \\
 \ion{Fe}{1} &  4063.594  &     1.56 &     0.062 &    112.62 &      3.87 \\
 \ion{Fe}{1} &  4067.978  &     3.21 &  $-$0.470 &      9.91 &      3.81 \\
 \ion{Fe}{1} &  4071.738  &     1.61 &  $-$0.008 &    109.82 &      3.92 \\
 \ion{Fe}{1} &  4076.629  &     3.21 &  $-$0.370 &     15.60 &      3.93 \\
 \ion{Fe}{1} &  4132.058  &     1.61 &  $-$0.675 &     81.69 &      3.81 \\
 \ion{Fe}{1} &  4134.678  &     2.83 &  $-$0.649 &     18.45 &      3.86 \\
 \ion{Fe}{1} &  4143.414  &     3.05 &  $-$0.200 &     22.46 &      3.77 \\
 \ion{Fe}{1} &  4143.868  &     1.56 &  $-$0.511 &     95.27 &      3.94 \\
 \ion{Fe}{1} &  4147.669  &     1.48 &  $-$2.071 &     22.83 &      3.81 \\
 \ion{Fe}{1} &  4152.169  &     0.96 &  $-$3.232 &      9.20 &      3.89 \\
 \ion{Fe}{1} &  4153.899  &     3.40 &  $-$0.320 &     11.21 &      3.93 \\
 \ion{Fe}{1} &  4156.799  &     2.83 &  $-$0.808 &     17.10 &      3.98 \\
 \ion{Fe}{1} &  4187.039  &     2.45 &  $-$0.514 &     43.23 &      3.81 \\
 \ion{Fe}{1} &  4191.430  &     2.47 &  $-$0.666 &     38.82 &      3.91 \\
 \ion{Fe}{1} &  4202.029  &     1.49 &  $-$0.689 &     86.91 &      3.79 \\
 \ion{Fe}{1} &  4216.184  &     0.00 &  $-$3.357 &     53.33 &      3.93 \\
 \ion{Fe}{1} &  4222.213  &     2.45 &  $-$0.914 &     28.52 &      3.92 \\
 \ion{Fe}{1} &  4227.427  &     3.33 &     0.266 &     34.77 &      3.89 \\
 \ion{Fe}{1} &  4233.603  &     2.48 &  $-$0.579 &     44.10 &      3.93 \\
 \ion{Fe}{1} &  4250.787  &     1.56 &  $-$0.713 &     90.86 &      3.98 \\
 \ion{Fe}{1} &  4260.474  &     2.40 &     0.077 &     79.14 &      3.88 \\
 \ion{Fe}{1} &  4383.545  &     1.48 &     0.200 &    132.48 &      3.99 \\
 \ion{Fe}{1} &  4404.750  &     1.56 &  $-$0.147 &    107.01 &      3.81 \\
 \ion{Fe}{1} &  4415.122  &     1.61 &  $-$0.621 &     94.81 &      4.01 \\
 \ion{Fe}{1} &  4447.717  &     2.22 &  $-$1.339 &     24.49 &      3.97 \\
 \ion{Fe}{1} &  4459.118  &     2.18 &  $-$1.279 &     29.92 &      3.98 \\
 \ion{Fe}{1} &  4461.653  &     0.09 &  $-$3.194 &     66.31 &      4.08 \\
 \ion{Fe}{1} &  4466.552  &     2.83 &  $-$0.600 &     22.03 &      3.88 \\
 \ion{Fe}{1} &  4476.019  &     2.85 &  $-$0.820 &     15.62 &      3.94 \\
 \ion{Fe}{1} &  4489.739  &     0.12 &  $-$3.899 &     24.23 &      4.03 \\
 \ion{Fe}{1} &  4494.563  &     2.20 &  $-$1.143 &     32.40 &      3.92 \\
 \ion{Fe}{1} &  4528.614  &     2.18 &  $-$0.822 &     48.24 &      3.86 \\
 \ion{Fe}{1} &  4531.148  &     1.48 &  $-$2.101 &     29.24 &      3.96 \\
 \ion{Fe}{1} &  4592.651  &     1.56 &  $-$2.462 &     11.74 &      3.92 \\
 \ion{Fe}{1} &  4602.941  &     1.49 &  $-$2.208 &     28.74 &      4.06 \\
 \ion{Fe}{1} &  4871.318  &     2.87 &  $-$0.362 &     35.31 &      3.95 \\
 \ion{Fe}{1} &  4872.137  &     2.88 &  $-$0.567 &     25.69 &      3.97 \\
 \ion{Fe}{1} &  4890.755  &     2.88 &  $-$0.394 &     27.67 &      3.84 \\
 \ion{Fe}{1} &  4891.492  &     2.85 &  $-$0.111 &     44.87 &      3.85 \\
 \ion{Fe}{1} &  4903.310  &     2.88 &  $-$0.926 &      8.93 &      3.78 \\
 \ion{Fe}{1} &  4918.994  &     2.85 &  $-$0.342 &     31.74 &      3.83 \\
 \ion{Fe}{1} &  4920.503  &     2.83 &     0.068 &     51.65 &      3.77 \\
 \ion{Fe}{1} &  4994.130  &     0.92 &  $-$2.969 &     20.20 &      3.91 \\
 \ion{Fe}{1} &  5006.119  &     2.83 &  $-$0.615 &     24.99 &      3.93 \\
 \ion{Fe}{1} &  5012.068  &     0.86 &  $-$2.642 &     46.11 &      4.03 \\
 \ion{Fe}{1} &  5041.072  &     0.96 &  $-$3.090 &     17.32 &      4.00 \\
 \ion{Fe}{1} &  5041.756  &     1.49 &  $-$2.200 &     25.17 &      3.94 \\
 \ion{Fe}{1} &  5049.820  &     2.28 &  $-$1.355 &     21.12 &      3.93 \\
 \ion{Fe}{1} &  5051.634  &     0.92 &  $-$2.764 &     31.00 &      3.94 \\
 \ion{Fe}{1} &  5068.766  &     2.94 &  $-$1.041 &      8.17 &      3.92 \\
 \ion{Fe}{1} &  5083.339  &     0.96 &  $-$2.842 &     28.29 &      4.01 \\
 \ion{Fe}{1} &  5110.413  &     0.00 &  $-$3.760 &     42.15 &      4.04 \\
 \ion{Fe}{1} &  5127.360  &     0.92 &  $-$3.249 &     14.02 &      3.99 \\
 \ion{Fe}{1} &  5150.839  &     0.99 &  $-$3.037 &     17.36 &      3.98 \\
 \ion{Fe}{1} &  5166.282  &     0.00 &  $-$4.123 &     22.79 &      4.01 \\
 \ion{Fe}{1} &  5171.596  &     1.49 &  $-$1.721 &     50.65 &      3.93 \\
 \ion{Fe}{1} &  5191.455  &     3.04 &  $-$0.551 &     16.60 &      3.88 \\
 \ion{Fe}{1} &  5192.344  &     3.00 &  $-$0.421 &     23.33 &      3.89 \\
 \ion{Fe}{1} &  5194.942  &     1.56 &  $-$2.021 &     30.26 &      3.94 \\
 \ion{Fe}{1} &  5202.336  &     2.18 &  $-$1.871 &     10.87 &      3.98 \\
 \ion{Fe}{1} &  5216.274  &     1.61 &  $-$2.082 &     21.60 &      3.87 \\
 \ion{Fe}{1} &  5232.940  &     2.94 &  $-$0.057 &     42.19 &      3.84 \\
 \ion{Fe}{1} &  5266.555  &     3.00 &  $-$0.385 &     22.17 &      3.83 \\
 \ion{Fe}{1} &  5269.537  &     0.86 &  $-$1.333 &    108.24 &      3.97 \\
 \ion{Fe}{1} &  5281.790  &     3.04 &  $-$0.833 &      8.36 &      3.82 \\
 \ion{Fe}{1} &  5283.621  &     3.24 &  $-$0.524 &     14.25 &      4.01 \\
 \ion{Fe}{1} &  5302.300  &     3.28 &  $-$0.720 &      7.06 &      3.91 \\
 \ion{Fe}{1} &  5324.179  &     3.21 &  $-$0.103 &     25.16 &      3.85 \\
 \ion{Fe}{1} &  5328.039  &     0.92 &  $-$1.466 &     97.68 &      3.92 \\
 \ion{Fe}{1} &  5328.531  &     1.56 &  $-$1.850 &     41.23 &      3.97 \\
 \ion{Fe}{1} &  5332.900  &     1.55 &  $-$2.776 &      7.85 &      3.98 \\
 \ion{Fe}{1} &  5339.930  &     3.27 &  $-$0.720 &      8.47 &      3.98 \\
 \ion{Fe}{1} &  5371.489  &     0.96 &  $-$1.644 &     94.91 &      4.07 \\
 \ion{Fe}{1} &  5397.128  &     0.92 &  $-$1.982 &     82.11 &      4.07 \\
 \ion{Fe}{1} &  5405.775  &     0.99 &  $-$1.852 &     83.50 &      4.05 \\
 \ion{Fe}{1} &  5429.696  &     0.96 &  $-$1.881 &     85.76 &      4.09 \\
 \ion{Fe}{1} &  5434.524  &     1.01 &  $-$2.126 &     65.62 &      4.00 \\
 \ion{Fe}{1} &  5446.917  &     0.99 &  $-$1.910 &     81.57 &      4.07 \\
 \ion{Fe}{1} &  5455.609  &     1.01 &  $-$2.090 &     71.07 &      4.06 \\
 \ion{Fe}{1} &  5497.516  &     1.01 &  $-$2.825 &     24.49 &      3.95 \\
 \ion{Fe}{1} &  5506.779  &     0.99 &  $-$2.789 &     30.32 &      4.01 \\
 \ion{Fe}{1} &  5572.842  &     3.40 &  $-$0.275 &     12.52 &      3.87 \\
 \ion{Fe}{1} &  5586.756  &     3.37 &  $-$0.144 &     16.42 &      3.83 \\
 \ion{Fe}{1} &  5615.644  &     3.33 &     0.050 &     26.93 &      3.87 \\
 \ion{Fe}{1} &  6065.481  &     2.61 &  $-$1.410 &      8.79 &      3.88 \\
 \ion{Fe}{1} &  6136.615  &     2.45 &  $-$1.410 &     17.74 &      4.03 \\
 \ion{Fe}{1} &  6137.691  &     2.59 &  $-$1.346 &     12.51 &      3.96 \\
 \ion{Fe}{1} &  6191.558  &     2.43 &  $-$1.416 &     14.18 &      3.90 \\
 \ion{Fe}{1} &  6230.723  &     2.56 &  $-$1.276 &     16.39 &      3.99 \\
 \ion{Fe}{1} &  6252.555  &     2.40 &  $-$1.687 &      8.74 &      3.90 \\
 \ion{Fe}{1} &  6393.601  &     2.43 &  $-$1.576 &     12.79 &      4.00 \\
 \ion{Fe}{1} &  6400.000  &     3.60 &  $-$0.290 &      7.57 &      3.84 \\
 \ion{Fe}{1} &  6430.846  &     2.18 &  $-$1.946 &     11.45 &      4.02 \\
 \ion{Fe}{1} &  6494.980  &     2.40 &  $-$1.239 &     23.26 &      3.93 \\
 \ion{Fe}{1} &  6677.986  &     2.69 &  $-$1.418 &     10.53 &      4.04 \\
 \ion{Fe}{2} &  4520.224  &     2.81 &  $-$2.600 &      9.38 &      3.92 \\
 \ion{Fe}{2} &  4555.890  &     2.83 &  $-$2.400 &     12.72 &      3.88 \\
 \ion{Fe}{2} &  4583.840  &     2.81 &  $-$1.930 &     35.22 &      3.96 \\
 \ion{Fe}{2} &  4923.930  &     2.89 &  $-$1.320 &     61.88 &      3.90 \\
 \ion{Fe}{2} &  5018.450  &     2.89 &  $-$1.220 &     69.87 &      3.94 \\
 \ion{Fe}{2} &  5197.580  &     3.23 &  $-$2.220 &      8.21 &      3.92 \\
 \ion{Fe}{2} &  5276.000  &     3.20 &  $-$2.010 &     15.34 &      3.98 \\
 \ion{Co}{1} &  3845.468  &     0.92 &     0.010 &     51.86 &      1.55 \\
 \ion{Co}{1} &  3995.306  &     0.92 &  $-$0.220 &     42.58 &      1.58 \\
 \ion{Co}{1} &  4121.318  &     0.92 &  $-$0.320 &     37.58 &      1.57 \\
 \ion{Ni}{1} &  3452.880  &     0.11 &  $-$0.900 &     85.60 &      2.48 \\
 \ion{Ni}{1} &  3483.770  &     0.28 &  $-$1.120 &     72.72 &      2.46 \\
 \ion{Ni}{1} &  3492.960  &     0.11 &  $-$0.265 &    107.58 &      2.50 \\
 \ion{Ni}{1} &  3500.850  &     0.17 &  $-$1.294 &     73.79 &      2.53 \\
 \ion{Ni}{1} &  3519.770  &     0.28 &  $-$1.422 &     64.22 &      2.50 \\
 \ion{Ni}{1} &  3524.540  &     0.03 &     0.007 &    122.09 &      2.47 \\
 \ion{Ni}{1} &  3597.710  &     0.21 &  $-$1.115 &     78.75 &      2.50 \\
 \ion{Ni}{1} &  3783.520  &     0.42 &  $-$1.420 &     65.82 &      2.55 \\
 \ion{Ni}{1} &  3807.140  &     0.42 &  $-$1.220 &     69.61 &      2.43 \\
 \ion{Ni}{1} &  5476.900  &     1.83 &  $-$0.890 &     17.68 &      2.52 \\
 \ion{Sr}{2} &  4077.714  &     0.00 &     0.150 &       syn &   $-$1.23 \\
 \ion{Sr}{2} &  4215.524  &     0.00 &  $-$0.180 &       syn &   $-$1.28 \\
 \ion{Y}{2}  &  5205.731  &     1.03 &  $-$0.340 &       syn &   $-$1.64 \\
 \ion{Zr}{2} &  3998.965  &     0.56 &  $-$0.520 &       syn &   $-$1.20 \\
 \ion{Zr}{2} &  4045.613  &     0.71 &  $-$0.860 &       syn &   $-$0.97 \\
 \ion{Ru}{1} &  3728.025  &     0.00 &     0.260 &       syn &   $-$1.20 \\
 \ion{Pd}{1} &  3404.579  &     0.81 &     0.320 &       syn &   $-$1.43 \\
 \ion{Ba}{2} &  4554.033  &     0.00 &     0.163 &       syn &   $-$0.97 \\
 \ion{Ba}{2} &  4934.086  &     0.00 &  $-$0.160 &       syn &   $-$1.02 \\
 \ion{Ba}{2} &  5853.680  &     0.60 &  $-$2.560 &       syn &   $-$1.12 \\
 \ion{Ba}{2} &  6141.710  &     0.70 &  $-$0.008 &       syn &   $-$1.04 \\
 \ion{Ba}{2} &  6496.896  &     0.60 &  $-$0.369 &       syn &   $-$1.07 \\
 \ion{La}{2} &  3995.740  &     0.17 &  $-$0.686 &       syn &   $-$2.10 \\
 \ion{La}{2} &  4086.710  &     0.00 &  $-$0.696 &       syn &   $-$2.15 \\
 \ion{Ce}{2} &  4053.503  &     0.00 &  $-$0.610 &       syn &   $-$1.57 \\
 \ion{Pr}{2} &  4179.475  &     0.20 &  $-$0.194 &       syn &   $-$2.28 \\
 \ion{Pr}{2} &  4222.934  &     0.05 &  $-$0.557 &       syn &   $-$2.43 \\
 \ion{Nd}{2} &  4012.700  &     0.00 &  $-$0.600 &       syn &   $-$1.88 \\
 \ion{Nd}{2} &  4043.590  &     0.32 &  $-$0.710 &       syn &   $-$1.88 \\
 \ion{Nd}{2} &  4061.080  &     0.47 &     0.550 &       syn &   $-$1.98 \\
 \ion{Sm}{2} &  4318.930  &     0.28 &  $-$0.250 &       syn &   $-$2.04 \\
 \ion{Sm}{2} &  4424.334  &     0.48 &     0.140 &       syn &   $-$2.34 \\
 \ion{Sm}{2} &  4642.230  &     0.38 &  $-$0.460 &       syn &   $-$2.09 \\
 \ion{Eu}{2} &  3724.934  &     0.00 &  $-$0.855 &       syn &   $-$2.38 \\
 \ion{Eu}{2} &  4435.457  &     0.21 &  $-$0.696 &       syn &   $-$2.33 \\
 \ion{Gd}{2} &  3549.360  &     0.24 &     0.290 &       syn &   $-$1.93 \\
 \ion{Gd}{2} &  4251.730  &     0.38 &  $-$0.220 &       syn &   $-$2.08 \\
 \ion{Dy}{2} &  3536.020  &     0.54 &      0.53 &       syn &   $-$1.80 \\
 \ion{Dy}{2} &  4077.970  &     0.10 &   $-$0.04 &       syn &   $-$2.10 \\
 \ion{Ho}{2} &  3456.010  &     0.00 &      0.76 &       syn &   $-$2.52 \\
 \ion{Ho}{2} &  3890.970  &     0.08 &      0.46 &       syn &   $-$2.67 \\
 \ion{Er}{2} &  3692.650  &     0.06 &      0.28 &       syn &   $-$2.18 \\
 \ion{Er}{2} &  3729.520  &     0.00 &   $-$0.59 &       syn &   $-$2.08 \\
 \ion{Er}{2} &  3906.310  &     0.00 &      0.12 &       syn &   $-$2.18 \\
 \ion{Tm}{2} &  3462.200  &     0.00 &      0.03 &       syn &   $-$2.85 \\
 \ion{Yb}{2} &  3694.195  &     0.00 &  $-$0.300 &       syn &   $-$2.46 \\
 \ion{Pb}{1} &  4057.814  &     1.32 &  $-$0.220 &       syn &  $<$0.30 \\
 \ion{Th}{2} &  4019.129  &     0.00 &  $-$0.650 &       syn &  $<-$2.20 \\

%% file: abund.tex
\ion{C}{0}     & 8.43 &    6.76   & $-$1.67   & $+$1.90  & 0.10    &   6 \\
\ion{C}{0}\xx  & 8.43 &    7.20   & $-$1.23   & $+$2.34  & 0.10    &   6 \\
\ion{N}{0}     & 7.83 &    5.35   & $-$2.48   & $+$1.09  & 0.15    &   3 \\
\ion{O}{1}     & 8.69 &    7.36   & $-$1.34   & $+$2.24  & 0.10    &   1 \\
\ion{Na}{1}    & 6.24 &    3.32   & $-$2.92   & $+$0.65  & 0.15    &   2 \\
\ion{Mg}{1}    & 7.60 &    4.60   & $-$3.00   & $+$0.58  & 0.10    &   6 \\
\ion{Al}{1}    & 6.45 &    2.70   & $-$3.75   & $-$0.18  & 0.15    &   1 \\
\ion{Si}{1}    & 7.51 &    4.43   & $-$3.08   & $+$0.49  & 0.15    &   1 \\
\ion{K}{1}     & 5.03 &    2.10   & $-$2.93   & $+$0.64  & 0.10    &   2 \\
\ion{Ca}{1}    & 6.34 &    3.14   & $-$3.20   & $+$0.37  & 0.10    &   9 \\
\ion{Sc}{2}    & 3.15 & $-$0.07   & $-$3.22   & $+$0.36  & 0.10    &   3 \\
\ion{Ti}{1}    & 4.95 &    1.60   & $-$3.35   & $+$0.22  & 0.10    &   5 \\
\ion{Ti}{2}    & 4.95 &    1.63   & $-$3.32   & $+$0.25  & 0.10    &  23 \\
\ion{Cr}{1}    & 5.64 &    1.74   & $-$3.90   & $-$0.33  & 0.10    &   3 \\
\ion{Mn}{1}    & 5.43 &    1.47   & $-$3.96   & $-$0.39  & 0.10    &   4 \\
\ion{Fe}{1}    & 7.50 &    3.93   & $-$3.57   &    0.00  & 0.10    & 127 \\
\ion{Fe}{2}    & 7.50 &    3.93   & $-$3.57   &    0.00  & 0.10    &   7 \\
\ion{Co}{1}    & 4.99 &    1.57   & $-$3.42   & $+$0.15  & 0.10    &   3 \\
\ion{Ni}{1}    & 6.22 &    2.49   & $-$3.73   & $-$0.16  & 0.10    &  10 \\
\ion{Sr}{2}    & 2.87 & $-$1.26   & $-$4.13   & $-$0.56  & 0.20    &   2 \\
\ion{Y}{2}     & 2.21 & $-$1.64   & $-$3.85   & $-$0.28  & 0.20    &   1 \\
\ion{Zr}{2}    & 2.58 & $-$1.09   & $-$3.67   & $-$0.10  & 0.25    &   2 \\
\ion{Ru}{1}    & 1.75 & $-$1.20   & $-$2.95   & $+$0.62  & 0.25    &   1 \\
\ion{Pd}{1}    & 1.57 & $-$1.43   & $-$3.00   & $+$0.57  & 0.25    &   1 \\
\ion{Ba}{2}    & 2.18 & $-$1.04   & $-$3.22   & $+$0.35  & 0.15    &   5 \\
\ion{La}{2}    & 1.10 & $-$2.13   & $-$3.23   & $+$0.34  & 0.20    &   2 \\
\ion{Ce}{2}    & 1.58 & $-$1.57   & $-$3.15   & $+$0.42  & 0.20    &   1 \\
\ion{Pr}{2}    & 0.72 & $-$2.36   & $-$3.08   & $+$0.49  & 0.30    &   2 \\
\ion{Nd}{2}    & 1.42 & $-$1.91   & $-$3.33   & $+$0.24  & 0.20    &   3 \\
\ion{Sm}{2}    & 0.96 & $-$2.16   & $-$3.12   & $+$0.45  & 0.30    &   3 \\
\ion{Eu}{2}    & 0.52 & $-$2.36   & $-$2.88   & $+$0.69  & 0.20    &   2 \\
\ion{Gd}{2}    & 1.07 & $-$2.00   & $-$3.07   & $+$0.50  & 0.25    &   2 \\
\ion{Dy}{2}    & 1.10 & $-$1.95   & $-$3.05   & $+$0.52  & 0.20    &   3 \\
\ion{Ho}{2}    & 0.48 & $-$2.60   & $-$3.08   & $+$0.49  & 0.20    &   2 \\
\ion{Er}{2}    & 0.92 & $-$2.15   & $-$3.07   & $+$0.50  & 0.20    &   3 \\
\ion{Tm}{2}    & 0.10 & $-$2.85   & $-$2.95   & $+$0.62  & 0.25    &   1 \\
\ion{Yb}{2}    & 0.84 & $-$2.46   & $-$3.30   & $+$0.27  & 0.25    &   1 \\
\ion{Pb}{1}    & 1.75 & $<$0.30   & $<-$1.45  & $<+$2.12 & \nodata &   1 \\
\ion{Th}{2}    & 0.02 & $<-$2.20  & $<-$2.22  & $<+$1.35 & \nodata &   1 \\

%% file: error.tex
\ion{O}{1}  & 0.12 &    0.05 & $-$0.00 & 0.10 & 0.17 \\
\ion{Na}{1} & 0.18 & $-$0.04 & $-$0.11 & 0.07 & 0.23 \\
\ion{Mg}{1} & 0.14 & $-$0.05 & $-$0.07 & 0.04 & 0.17 \\
\ion{Al}{1} & 0.17 & $-$0.05 & $-$0.12 & 0.10 & 0.24 \\
\ion{Si}{1} & 0.16 & $-$0.01 & $-$0.01 & 0.10 & 0.19 \\
\ion{K}{1}  & 0.12 & $-$0.01 & $-$0.00 & 0.07 & 0.14 \\
\ion{Ca}{1} & 0.10 & $-$0.01 & $-$0.01 & 0.03 & 0.11 \\
\ion{Sc}{2} & 0.08 &    0.05 & $-$0.01 & 0.06 & 0.11 \\
\ion{Ti}{1} & 0.17 & $-$0.01 & $-$0.01 & 0.04 & 0.18 \\
\ion{Ti}{2} & 0.10 &    0.04 & $-$0.07 & 0.02 & 0.13 \\
\ion{Cr}{1} & 0.19 & $-$0.03 & $-$0.06 & 0.06 & 0.21 \\
\ion{Mn}{1} & 0.21 & $-$0.02 & $-$0.05 & 0.07 & 0.23 \\
\ion{Fe}{1} & 0.17 & $-$0.02 & $-$0.06 & 0.01 & 0.18 \\
\ion{Fe}{2} & 0.03 &    0.06 & $-$0.02 & 0.04 & 0.08 \\
\ion{Co}{1} & 0.19 & $-$0.01 & $-$0.03 & 0.06 & 0.20 \\
\ion{Ni}{1} & 0.24 & $-$0.05 & $-$0.12 & 0.03 & 0.27 \\
\ion{Sr}{2} & 0.13 &    0.04 & $-$0.13 & 0.07 & 0.20 \\
\ion{Ba}{2} & 0.16 &    0.03 & $-$0.14 & 0.10 & 0.24 \\

%% file: placco.bbl
\begin{thebibliography}{}
\expandafter\ifx\csname natexlab\endcsname\relax\def\natexlab#1{#1}\fi
\providecommand{\url}[1]{\href{#1}{#1}}
\providecommand{\dodoi}[1]{doi:~\href{http://doi.org/#1}{\nolinkurl{#1}}}
\providecommand{\doeprint}[1]{\href{http://ascl.net/#1}{\nolinkurl{http://ascl.net/#1}}}
\providecommand{\doarXiv}[1]{\href{https://arxiv.org/abs/#1}{\nolinkurl{https://arxiv.org/abs/#1}}}

\bibitem[{{Abbott} {et~al.}(2017){Abbott}, {Abbott}, {Abbott}, {Acernese},
  {Ackley}, {Adams}, {Adams}, {Addesso}, {Adhikari}, {Adya}, \&
  et~al.}]{abbott2017}
{Abbott}, B.~P., {Abbott}, R., {Abbott}, T.~D., {et~al.} 2017, \apjl, 848, L12,
  \dodoi{10.3847/2041-8213/aa91c9}

\bibitem[{{Abohalima} \& {Frebel}(2018)}]{jinabase}
{Abohalima}, A., \& {Frebel}, A. 2018, The Astrophysical Journal Supplement
  Series, 238, 36, \dodoi{10.3847/1538-4365/aadfe9}

\bibitem[{{Aguado} {et~al.}(2018){Aguado}, {Allende Prieto}, {Gonz{\'a}lez
  Hern{\'a}ndez}, \& {Rebolo}}]{aguado2018}
{Aguado}, D.~S., {Allende Prieto}, C., {Gonz{\'a}lez Hern{\'a}ndez}, J.~I., \&
  {Rebolo}, R. 2018, \apjl, 854, L34, \dodoi{10.3847/2041-8213/aaadb8}

\bibitem[{Aho {et~al.}(1987)Aho, Kernighan, \& Weinberger}]{awk}
Aho, A.~V., Kernighan, B.~W., \& Weinberger, P.~J. 1987, The AWK Programming
  Language (Boston, MA, USA: Addison-Wesley Longman Publishing Co., Inc.)

\bibitem[{{Anders} {et~al.}(2019){Anders}, {Khalatyan}, {Chiappini}, {Queiroz},
  {Santiago}, {Jordi}, {Girardi}, {Brown}, {Matijevi{\v{c}}}, {Monari},
  {Cantat-Gaudin}, {Weiler}, {Khan}, {Miglio}, {Carrillo}, {Romero-G{\'o}mez},
  {Minchev}, {de Jong}, {Antoja}, {Ramos}, {Steinmetz}, \&
  {Enke}}]{starhorse19}
{Anders}, F., {Khalatyan}, A., {Chiappini}, C., {et~al.} 2019, \aap, 628, A94,
  \dodoi{10.1051/0004-6361/201935765}

\bibitem[{{Aoki} {et~al.}(2007){Aoki}, {Beers}, {Christlieb}, {Norris}, {Ryan},
  \& {Tsangarides}}]{aoki2007}
{Aoki}, W., {Beers}, T.~C., {Christlieb}, N., {et~al.} 2007, \apj, 655, 492,
  \dodoi{10.1086/509817}

\bibitem[{{Arnett}(1996)}]{arnett1996}
{Arnett}, D. 1996, {Supernovae and Nucleosynthesis: An Investigation of the
  History of Matter from the Big Bang to the Present} (Princeton,N.J.)

\bibitem[{{Asplund} {et~al.}(2009){Asplund}, {Grevesse}, {Sauval}, \&
  {Scott}}]{asplund2009}
{Asplund}, M., {Grevesse}, N., {Sauval}, A.~J., \& {Scott}, P. 2009, \araa, 47,
  481, \dodoi{10.1146/annurev.astro.46.060407.145222}

\bibitem[{{Bailer-Jones} {et~al.}(2018){Bailer-Jones}, {Rybizki}, {Fouesneau},
  {Mantelet}, \& {Andrae}}]{bailer-jones2018}
{Bailer-Jones}, C.~A.~L., {Rybizki}, J., {Fouesneau}, M., {Mantelet}, G., \&
  {Andrae}, R. 2018, \aj, 156, 58, \dodoi{10.3847/1538-3881/aacb21}

\bibitem[{{Banerjee} {et~al.}(2019){Banerjee}, {Heger}, \&
  {Qian}}]{banerjee2019}
{Banerjee}, P., {Heger}, A., \& {Qian}, Y.-Z. 2019, \apj, 887, 187,
  \dodoi{10.3847/1538-4357/ab517a}

\bibitem[{{Beers} \& {Christlieb}(2005)}]{beers2005}
{Beers}, T.~C., \& {Christlieb}, N. 2005, \araa, 43, 531

\bibitem[{{Beers} {et~al.}(2014){Beers}, {Norris}, {Placco}, {Lee}, {Rossi},
  {Carollo}, \& {Masseron}}]{beers2014}
{Beers}, T.~C., {Norris}, J.~E., {Placco}, V.~M., {et~al.} 2014, \apj, 794, 58,
  \dodoi{10.1088/0004-637X/794/1/58}

\bibitem[{{Beers} {et~al.}(2017){Beers}, {Placco}, {Carollo}, {Rossi}, {Lee},
  {Frebel}, {Norris}, {Dietz}, \& {Masseron}}]{beers2017}
{Beers}, T.~C., {Placco}, V.~M., {Carollo}, D., {et~al.} 2017, \apj, 835, 81,
  \dodoi{10.3847/1538-4357/835/1/81}

\bibitem[{{Bernstein} {et~al.}(2003){Bernstein}, {Shectman}, {Gunnels},
  {Mochnacki}, \& {Athey}}]{mike}
{Bernstein}, R., {Shectman}, S.~A., {Gunnels}, S.~M., {Mochnacki}, S., \&
  {Athey}, A.~E. 2003, in Society of Photo-Optical Instrumentation Engineers
  (SPIE) Conference Series, Vol. 4841, Society of Photo-Optical Instrumentation
  Engineers (SPIE) Conference Series, ed. {M.~Iye \& A.~F.~M.~Moorwood}, 1694,
  \dodoi{10.1117/12.461502}

\bibitem[{{Bovard} {et~al.}(2017){Bovard}, {Martin}, {Guercilena}, {Arcones},
  {Rezzolla}, \& {Korobkin}}]{bovard2017}
{Bovard}, L., {Martin}, D., {Guercilena}, F., {et~al.} 2017, \prd, 96, 124005,
  \dodoi{10.1103/PhysRevD.96.124005}

\bibitem[{{Bromm} \& {Larson}(2004)}]{bromm2004}
{Bromm}, V., \& {Larson}, R.~B. 2004, \araa, 42, 79,
  \dodoi{10.1146/annurev.astro.42.053102.134034}

\bibitem[{{Bromm} {et~al.}(2009){Bromm}, {Yoshida}, {Hernquist}, \&
  {McKee}}]{bromm2009}
{Bromm}, V., {Yoshida}, N., {Hernquist}, L., \& {McKee}, C.~F. 2009, \nat, 459,
  49, \dodoi{10.1038/nature07990}

\bibitem[{{Burbidge} {et~al.}(1957){Burbidge}, {Burbidge}, {Fowler}, \&
  {Hoyle}}]{b2fh}
{Burbidge}, E.~M., {Burbidge}, G.~R., {Fowler}, W.~A., \& {Hoyle}, F. 1957,
  Reviews of Modern Physics, 29, 547, \dodoi{10.1103/RevModPhys.29.547}

\bibitem[{{Burris} {et~al.}(2000){Burris}, {Pilachowski}, {Armandroff},
  {Sneden}, {Cowan}, \& {Roe}}]{burris2000}
{Burris}, D.~L., {Pilachowski}, C.~A., {Armandroff}, T.~E., {et~al.} 2000,
  \apj, 544, 302, \dodoi{10.1086/317172}

\bibitem[{{Buzzoni} {et~al.}(1984){Buzzoni}, {Delabre}, {Dekker}, {Dodorico},
  {Enard}, {Focardi}, {Gustafsson}, {Nees}, {Paureau}, \& {Reiss}}]{efosc}
{Buzzoni}, B., {Delabre}, B., {Dekker}, H., {et~al.} 1984, The Messenger, 38, 9

\bibitem[{{Caffau} {et~al.}(2011){Caffau}, {Bonifacio}, {Fran{\c c}ois},
  {Sbordone}, {Monaco}, {Spite}, {Spite}, {Ludwig}, {Cayrel}, {Zaggia},
  {Hammer}, {Randich}, {Molaro}, \& {Hill}}]{caffau2011b}
{Caffau}, E., {Bonifacio}, P., {Fran{\c c}ois}, P., {et~al.} 2011, \nat, 477,
  67, \dodoi{10.1038/nature10377}

\bibitem[{{Caffau} {et~al.}(2016){Caffau}, {Bonifacio}, {Spite}, {Spite},
  {Monaco}, {Sbordone}, {Fran{\c{c}}ois}, {Gallagher}, {Plez}, {Zaggia},
  {Ludwig}, {Cayrel}, {Koch}, {Steffen}, {Salvadori}, {Klessen}, {Glover}, \&
  {Christlieb}}]{caffau2016}
{Caffau}, E., {Bonifacio}, P., {Spite}, M., {et~al.} 2016, \aap, 595, L6,
  \dodoi{10.1051/0004-6361/201629776}

\bibitem[{{Cain} {et~al.}(2018){Cain}, {Frebel}, {Gull}, {Ji}, {Placco},
  {Beers}, {Mel{\'e}ndez}, {Ezzeddine}, {Casey}, {Hansen}, {Roederer}, \&
  {Sakari}}]{cain2018}
{Cain}, M., {Frebel}, A., {Gull}, M., {et~al.} 2018, \apj, 864, 43,
  \dodoi{10.3847/1538-4357/aad37d}

\bibitem[{{Cameron}(1957)}]{cameron1957}
{Cameron}, A.~G.~W. 1957, \pasp, 69, 201, \dodoi{10.1086/127051}

\bibitem[{{Castelli} \& {Kurucz}(2004)}]{castelli2004}
{Castelli}, F., \& {Kurucz}, R.~L. 2004, ArXiv Astrophysics e-prints

\bibitem[{{Cescutti} \& {Chiappini}(2014)}]{cescutti2014}
{Cescutti}, G., \& {Chiappini}, C. 2014, \aap, 565, A51,
  \dodoi{10.1051/0004-6361/201423432}

\bibitem[{{Cescutti} {et~al.}(2013){Cescutti}, {Chiappini}, {Hirschi},
  {Meynet}, \& {Frischknecht}}]{cescutti2013}
{Cescutti}, G., {Chiappini}, C., {Hirschi}, R., {Meynet}, G., \&
  {Frischknecht}, U. 2013, \aap, 553, A51, \dodoi{10.1051/0004-6361/201220809}

\bibitem[{{Cescutti} {et~al.}(2015){Cescutti}, {Romano}, {Matteucci},
  {Chiappini}, \& {Hirschi}}]{cescutti2015}
{Cescutti}, G., {Romano}, D., {Matteucci}, F., {Chiappini}, C., \& {Hirschi},
  R. 2015, \aap, 577, A139, \dodoi{10.1051/0004-6361/201525698}

\bibitem[{{Chiappini}(2013)}]{chiappini2013}
{Chiappini}, C. 2013, Astronomische Nachrichten, 334, 595,
  \dodoi{10.1002/asna.201311902}

\bibitem[{{Choplin} {et~al.}(2017){Choplin}, {Hirschi}, {Meynet}, \&
  {Ekstr{\"o}m}}]{choplin2017}
{Choplin}, A., {Hirschi}, R., {Meynet}, G., \& {Ekstr{\"o}m}, S. 2017, \aap,
  607, L3, \dodoi{10.1051/0004-6361/201731948}

\bibitem[{{Choplin} {et~al.}(2018){Choplin}, {Hirschi}, {Meynet},
  {Ekstr{\"o}m}, {Chiappini}, \& {Laird}}]{choplin2018}
{Choplin}, A., {Hirschi}, R., {Meynet}, G., {et~al.} 2018, \aap, 618, A133,
  \dodoi{10.1051/0004-6361/201833283}

\bibitem[{{Choplin} {et~al.}(2016){Choplin}, {Maeder}, {Meynet}, \&
  {Chiappini}}]{choplin2016}
{Choplin}, A., {Maeder}, A., {Meynet}, G., \& {Chiappini}, C. 2016, \aap, 593,
  A36, \dodoi{10.1051/0004-6361/201628083}

\bibitem[{{Christlieb} {et~al.}(2002){Christlieb}, {Bessell}, {Beers},
  {Gustafsson}, {Korn}, {Barklem}, {Karlsson}, {Mizuno-Wiedner}, \&
  {Rossi}}]{christlieb2002}
{Christlieb}, N., {Bessell}, M.~S., {Beers}, T.~C., {et~al.} 2002, \nat, 419,
  904, \dodoi{10.1038/nature01142}

\bibitem[{{Cohen} {et~al.}(2008){Cohen}, {Christlieb}, {McWilliam}, {Shectman},
  {Thompson}, {Melendez}, {Wisotzki}, \& {Reimers}}]{cohen2008}
{Cohen}, J.~G., {Christlieb}, N., {McWilliam}, A., {et~al.} 2008, \apj, 672,
  320, \dodoi{10.1086/523638}

\bibitem[{{Cohen} {et~al.}(2013){Cohen}, {Christlieb}, {Thompson}, {McWilliam},
  {Shectman}, {Reimers}, {Wisotzki}, \& {Kirby}}]{cohen2013}
{Cohen}, J.~G., {Christlieb}, N., {Thompson}, I., {et~al.} 2013, \apj, 778, 56,
  \dodoi{10.1088/0004-637X/778/1/56}

\bibitem[{{Cooke} \& {Madau}(2014)}]{cooke2014}
{Cooke}, R., \& {Madau}, P. 2014, ArXiv e-prints.
\newblock \doarXiv{1405.7369}

\bibitem[{{Cooke} {et~al.}(2011){Cooke}, {Pettini}, {Steidel}, {Rudie}, \&
  {Nissen}}]{cooke2011}
{Cooke}, R., {Pettini}, M., {Steidel}, C.~C., {Rudie}, G.~C., \& {Nissen},
  P.~E. 2011, \mnras, 417, 1534, \dodoi{10.1111/j.1365-2966.2011.19365.x}

\bibitem[{{C{\^o}t{\'e}} {et~al.}(2019){C{\^o}t{\'e}}, {Eichler}, {Arcones},
  {Hansen}, {Simonetti}, {Frebel}, {Fryer}, {Pignatari}, {Reichert},
  {Belczynski}, \& {Matteucci}}]{cote2019}
{C{\^o}t{\'e}}, B., {Eichler}, M., {Arcones}, A., {et~al.} 2019, \apj, 875,
  106, \dodoi{10.3847/1538-4357/ab10db}

\bibitem[{{Cseh} {et~al.}(2018){Cseh}, {Lugaro}, {D'Orazi}, {de Castro},
  {Pereira}, {Karakas}, {Moln{\'a}r}, {Plachy}, {Szab{\'o}}, {Pignatari}, \&
  {Cristallo}}]{cseh2018}
{Cseh}, B., {Lugaro}, M., {D'Orazi}, V., {et~al.} 2018, \aap, 620, A146,
  \dodoi{10.1051/0004-6361/201834079}

\bibitem[{{Drout} {et~al.}(2017){Drout}, {Piro}, {Shappee}, {Kilpatrick},
  {Simon}, {Contreras}, {Coulter}, {Foley}, {Siebert}, {Morrell}, {Boutsia},
  {Di Mille}, {Holoien}, {Kasen}, {Kollmeier}, {Madore}, {Monson},
  {Murguia-Berthier}, {Pan}, {Prochaska}, {Ramirez-Ruiz}, {Rest}, {Adams},
  {Alatalo}, {Ba{\~n}ados}, {Baughman}, {Beers}, {Bernstein}, {Bitsakis},
  {Campillay}, {Hansen}, {Higgs}, {Ji}, {Maravelias}, {Marshall}, {Bidin},
  {Prieto}, {Rasmussen}, {Rojas-Bravo}, {Strom}, {Ulloa},
  {Vargas-Gonz{\'a}lez}, {Wan}, \& {Whitten}}]{drout2017}
{Drout}, M.~R., {Piro}, A.~L., {Shappee}, B.~J., {et~al.} 2017, Science, 358,
  1570, \dodoi{10.1126/science.aaq0049}

\bibitem[{{Ezzeddine} {et~al.}(2019{\natexlab{a}}){Ezzeddine}, {Rasmussen},
  {Frebel}, {Placco}, {Roederer}, \& {Beers}}]{ezzeddine2019b}
{Ezzeddine}, R., {Rasmussen}, K., {Frebel}, A., {et~al.} 2019{\natexlab{a}},
  submitted

\bibitem[{{Ezzeddine} {et~al.}(2019{\natexlab{b}}){Ezzeddine}, {Frebel},
  {Roederer}, {Tominaga}, {Tumlinson}, {Ishigaki}, {Nomoto}, {Placco}, \&
  {Aoki}}]{ezzeddine2019}
{Ezzeddine}, R., {Frebel}, A., {Roederer}, I.~U., {et~al.} 2019{\natexlab{b}},
  \apj, 876, 97, \dodoi{10.3847/1538-4357/ab14e7}

\bibitem[{Frebel(2018)}]{frebel2018}
Frebel, A. 2018, Annual Review of Nuclear and Particle Science, 68, null,
  \dodoi{10.1146/annurev-nucl-101917-021141}

\bibitem[{{Frebel} {et~al.}(2013){Frebel}, {Casey}, {Jacobson}, \&
  {Yu}}]{frebel2013}
{Frebel}, A., {Casey}, A.~R., {Jacobson}, H.~R., \& {Yu}, Q. 2013, \apj, 769,
  57, \dodoi{10.1088/0004-637X/769/1/57}

\bibitem[{{Frebel} {et~al.}(2015){Frebel}, {Chiti}, {Ji}, {Jacobson}, \&
  {Placco}}]{frebel2015b}
{Frebel}, A., {Chiti}, A., {Ji}, A.~P., {Jacobson}, H.~R., \& {Placco}, V.~M.
  2015, \apjl, 810, L27, \dodoi{10.1088/2041-8205/810/2/L27}

\bibitem[{{Frebel} \& {Norris}(2015)}]{frebel2015}
{Frebel}, A., \& {Norris}, J.~E. 2015, \araa, 53, 631,
  \dodoi{10.1146/annurev-astro-082214-122423}

\bibitem[{{Frebel} {et~al.}(2006){Frebel}, {Christlieb}, {Norris}, {Beers},
  {Bessell}, {Rhee}, {Fechner}, {Marsteller}, {Rossi}, {Thom}, {Wisotzki}, \&
  {Reimers}}]{frebel2006}
{Frebel}, A., {Christlieb}, N., {Norris}, J.~E., {et~al.} 2006, \apj, 652,
  1585, \dodoi{10.1086/508506}

\bibitem[{{Frischknecht} {et~al.}(2016){Frischknecht}, {Hirschi}, {Pignatari},
  {Maeder}, {Meynet}, {Chiappini}, {Thielemann}, {Rauscher}, {Georgy}, \&
  {Ekstr{\"o}m}}]{frischknecht2016}
{Frischknecht}, U., {Hirschi}, R., {Pignatari}, M., {et~al.} 2016, \mnras, 456,
  1803, \dodoi{10.1093/mnras/stv2723}

\bibitem[{{Gaia Collaboration} {et~al.}(2018){Gaia Collaboration}, {Brown},
  {Vallenari}, {Prusti}, {de Bruijne}, {Babusiaux}, {Bailer-Jones}, {Biermann},
  {Evans}, {Eyer}, \& et~al.}]{gaia2018}
{Gaia Collaboration}, {Brown}, A.~G.~A., {Vallenari}, A., {et~al.} 2018, \aap,
  616, A1, \dodoi{10.1051/0004-6361/201833051}

\bibitem[{{Grichener} \& {Soker}(2019)}]{grichener2019}
{Grichener}, A., \& {Soker}, N. 2019, \apj, 878, 24,
  \dodoi{10.3847/1538-4357/ab1d5d}

\bibitem[{{Gull} {et~al.}(2018){Gull}, {Frebel}, {Cain}, {Placco}, {Ji},
  {Abate}, {Ezzeddine}, {Karakas}, {Hansen}, {Sakari}, {Holmbeck}, {Santucci},
  {Casey}, \& {Beers}}]{gull2018}
{Gull}, M., {Frebel}, A., {Cain}, M.~G., {et~al.} 2018, \apj, 862, 174,
  \dodoi{10.3847/1538-4357/aacbc3}

\bibitem[{{Hampel} {et~al.}(2016){Hampel}, {Stancliffe}, {Lugaro}, \&
  {Meyer}}]{hampel2016}
{Hampel}, M., {Stancliffe}, R.~J., {Lugaro}, M., \& {Meyer}, B.~S. 2016, \apj,
  831, 171, \dodoi{10.3847/0004-637X/831/2/171}

\bibitem[{{Hansen} {et~al.}(2019){Hansen}, {Hansen}, {Koch}, {Beers},
  {Nordstr{\"o}m}, {Placco}, \& {Andersen}}]{hansen2019}
{Hansen}, C.~J., {Hansen}, T.~T., {Koch}, A., {et~al.} 2019, \aap, 623, A128,
  \dodoi{10.1051/0004-6361/201834601}

\bibitem[{{Hansen} {et~al.}(2011){Hansen}, {Andersen}, {Nordstr{\"o}m},
  {Buchhave}, \& {Beers}}]{hansen2011}
{Hansen}, T., {Andersen}, J., {Nordstr{\"o}m}, B., {Buchhave}, L.~A., \&
  {Beers}, T.~C. 2011, \apjl, 743, L1, \dodoi{10.1088/2041-8205/743/1/L1}

\bibitem[{{Hansen} {et~al.}(2014){Hansen}, {Hansen}, {Christlieb}, {Yong},
  {Bessell}, {Garc{\'{\i}}a P{\'e}rez}, {Beers}, {Placco}, {Frebel}, {Norris},
  \& {Asplund}}]{hansen2014}
{Hansen}, T., {Hansen}, C.~J., {Christlieb}, N., {et~al.} 2014, \apj, 787, 162,
  \dodoi{10.1088/0004-637X/787/2/162}

\bibitem[{{Hansen} {et~al.}(2016{\natexlab{a}}){Hansen}, {Andersen},
  {Nordstr{\"o}m}, {Beers}, {Placco}, {Yoon}, \& {Buchhave}}]{hansen2016c}
{Hansen}, T.~T., {Andersen}, J., {Nordstr{\"o}m}, B., {et~al.}
  2016{\natexlab{a}}, \aap, 588, A3, \dodoi{10.1051/0004-6361/201527409}

\bibitem[{{Hansen} {et~al.}(2016{\natexlab{b}}){Hansen}, {Andersen},
  {Nordstr{\"o}m}, {Beers}, {Placco}, {Yoon}, \& {Buchhave}}]{hansen2016}
---. 2016{\natexlab{b}}, \aap, 586, A160, \dodoi{10.1051/0004-6361/201527235}

\bibitem[{{Hansen} {et~al.}(2015){Hansen}, {Andersen}, {Nordstr{\"o}m},
  {Beers}, {Yoon}, \& {Buchhave}}]{hansen2015b}
---. 2015, \aap, 583, A49, \dodoi{10.1051/0004-6361/201526812}

\bibitem[{{Hansen} {et~al.}(2017){Hansen}, {Simon}, {Marshall}, {Li},
  {Carollo}, {DePoy}, {Nagasawa}, {Bernstein}, {Drlica-Wagner}, {Abdalla},
  {Allam}, {Annis}, {Bechtol}, {Benoit-L{\'e}vy}, {Brooks}, {Buckley-Geer},
  {Carnero Rosell}, {Carrasco Kind}, {Carretero}, {Cunha}, {da Costa}, {Desai},
  {Eifler}, {Fausti Neto}, {Flaugher}, {Frieman}, {Garc{\'\i}a-Bellido},
  {Gaztanaga}, {Gerdes}, {Gruen}, {Gruendl}, {Gschwend}, {Gutierrez}, {James},
  {Krause}, {Kuehn}, {Kuropatkin}, {Lahav}, {Miquel}, {Plazas}, {Romer},
  {Sanchez}, {Santiago}, {Scarpine}, {Smith}, {Soares-Santos}, {Sobreira},
  {Suchyta}, {Swanson}, {Tarle}, {Walker}, \& {DES Collaboration}}]{hansen2017}
{Hansen}, T.~T., {Simon}, J.~D., {Marshall}, J.~L., {et~al.} 2017, \apj, 838,
  44, \dodoi{10.3847/1538-4357/aa634a}

\bibitem[{{Hansen} {et~al.}(2018){Hansen}, {Holmbeck}, {Beers}, {Placco},
  {Roederer}, {Frebel}, {Sakari}, {Simon}, \& {Thompson}}]{hansen2018}
{Hansen}, T.~T., {Holmbeck}, E.~M., {Beers}, T.~C., {et~al.} 2018, \apj, 858,
  92, \dodoi{10.3847/1538-4357/aabacc}

\bibitem[{{Hartwig} {et~al.}(2018){Hartwig}, {Yoshida}, {Magg}, {Frebel},
  {Glover}, {G{\'o}mez}, {Griffen}, {Ishigaki}, {Ji}, {Klessen}, {O'Shea}, \&
  {Tominaga}}]{hartwig2018}
{Hartwig}, T., {Yoshida}, N., {Magg}, M., {et~al.} 2018, \mnras, 478, 1795,
  \dodoi{10.1093/mnras/sty1176}

\bibitem[{{Haynes} \& {Kobayashi}(2019)}]{haynes2019}
{Haynes}, C.~J., \& {Kobayashi}, C. 2019, \mnras, 483, 5123,
  \dodoi{10.1093/mnras/sty3389}

\bibitem[{{Heger} \& {Woosley}(2010)}]{heger2010}
{Heger}, A., \& {Woosley}, S.~E. 2010, \apj, 724, 341,
  \dodoi{10.1088/0004-637X/724/1/341}

\bibitem[{{Henden} \& {Munari}(2014)}]{henden2014}
{Henden}, A., \& {Munari}, U. 2014, Contributions of the Astronomical
  Observatory Skalnate Pleso, 43, 518

\bibitem[{{Herwig}(2005)}]{herwig2005}
{Herwig}, F. 2005, \araa, 43, 435,
  \dodoi{10.1146/annurev.astro.43.072103.150600}

\bibitem[{{Holmbeck} {et~al.}(2020){Holmbeck}, {Hansen}, {Beers}, {Placco}, \&
  {Frebel}}]{holmbeck2020}
{Holmbeck}, E., {Hansen}, T.~T., {Beers}, T.~C., {Placco}, V.~M., \& {Frebel},
  A.~L. 2020, \apj, 000, 23, \dodoi{10.3847/1538-4357/aaefef}

\bibitem[{{Holmbeck} {et~al.}(2019){Holmbeck}, {Sprouse}, {Mumpower}, {Vassh},
  {Surman}, {Beers}, \& {Kawano}}]{holmbeck2019}
{Holmbeck}, E.~M., {Sprouse}, T.~M., {Mumpower}, M.~R., {et~al.} 2019, \apj,
  870, 23, \dodoi{10.3847/1538-4357/aaefef}

\bibitem[{{Holmbeck} {et~al.}(2018){Holmbeck}, {Beers}, {Roederer}, {Placco},
  {Hansen}, {Sakari}, {Sneden}, {Liu}, {Lee}, {Cowan}, \&
  {Frebel}}]{holmbeck2018}
{Holmbeck}, E.~M., {Beers}, T.~C., {Roederer}, I.~U., {et~al.} 2018, \apjl,
  859, L24, \dodoi{10.3847/2041-8213/aac722}

\bibitem[{{Hoyle}(1954)}]{hoyle1954}
{Hoyle}, F. 1954, \apjs, 1, 121, \dodoi{10.1086/190005}

\bibitem[{Hunter(2007)}]{matplotlib}
Hunter, J.~D. 2007, Computing in Science \& Engineering, 9, 90,
  \dodoi{10.1109/MCSE.2007.55}

\bibitem[{{Ishimaru} {et~al.}(2015){Ishimaru}, {Wanajo}, \&
  {Prantzos}}]{ishimaru2015}
{Ishimaru}, Y., {Wanajo}, S., \& {Prantzos}, N. 2015, \apjl, 804, L35,
  \dodoi{10.1088/2041-8205/804/2/L35}

\bibitem[{{Ito} {et~al.}(2013){Ito}, {Aoki}, {Beers}, {Tominaga}, {Honda}, \&
  {Carollo}}]{ito2013}
{Ito}, H., {Aoki}, W., {Beers}, T.~C., {et~al.} 2013, \apj, 773, 33,
  \dodoi{10.1088/0004-637X/773/1/33}

\bibitem[{{Jacobson} {et~al.}(2015){Jacobson}, {Keller}, {Frebel}, {Casey},
  {Asplund}, {Bessell}, {Da Costa}, {Lind}, {Marino}, {Norris}, {Pe{\~n}a},
  {Schmidt}, {Tisserand}, {Walsh}, {Yong}, \& {Yu}}]{jacobson2015}
{Jacobson}, H.~R., {Keller}, S., {Frebel}, A., {et~al.} 2015, \apj, 807, 171,
  \dodoi{10.1088/0004-637X/807/2/171}

\bibitem[{{Ji} {et~al.}(2019){Ji}, {Drout}, \& {Hansen}}]{ji2019}
{Ji}, A.~P., {Drout}, M.~R., \& {Hansen}, T.~T. 2019, \apj, 882, 40,
  \dodoi{10.3847/1538-4357/ab3291}

\bibitem[{{Ji} {et~al.}(2016){Ji}, {Frebel}, {Chiti}, \& {Simon}}]{ji2016}
{Ji}, A.~P., {Frebel}, A., {Chiti}, A., \& {Simon}, J.~D. 2016, \nat, 531, 610,
  \dodoi{10.1038/nature17425}

\bibitem[{{Keller} {et~al.}(2014){Keller}, {Bessell}, {Frebel}, {Casey},
  {Asplund}, {Jacobson}, {Lind}, {Norris}, {Yong}, {Heger}, {Magic}, {da
  Costa}, {Schmidt}, \& {Tisserand}}]{keller2014}
{Keller}, S.~C., {Bessell}, M.~S., {Frebel}, A., {et~al.} 2014, \nat, 506, 463,
  \dodoi{10.1038/nature12990}

\bibitem[{{Kelson}(2003)}]{kelson2003}
{Kelson}, D.~D. 2003, \pasp, 115, 688, \dodoi{10.1086/375502}

\bibitem[{{Kunder} {et~al.}(2017){Kunder}, {Kordopatis}, {Steinmetz},
  {Zwitter}, {McMillan}, {Casagrande}, {Enke}, {Wojno}, {Valentini},
  {Chiappini}, {Matijevi{\v c}}, {Siviero}, {de Laverny}, {Recio-Blanco},
  {Bijaoui}, {Wyse}, {Binney}, {Grebel}, {Helmi}, {Jofre}, {Antoja}, {Gilmore},
  {Siebert}, {Famaey}, {Bienaym{\'e}}, {Gibson}, {Freeman}, {Navarro},
  {Munari}, {Seabroke}, {Anguiano}, {{\v Z}erjal}, {Minchev}, {Reid},
  {Bland-Hawthorn}, {Kos}, {Sharma}, {Watson}, {Parker}, {Scholz}, {Burton},
  {Cass}, {Hartley}, {Fiegert}, {Stupar}, {Ritter}, {Hawkins}, {Gerhard},
  {Chaplin}, {Davies}, {Elsworth}, {Lund}, {Miglio}, \& {Mosser}}]{kunder2017}
{Kunder}, A., {Kordopatis}, G., {Steinmetz}, M., {et~al.} 2017, \aj, 153, 75,
  \dodoi{10.3847/1538-3881/153/2/75}

\bibitem[{{Langer} {et~al.}(1989){Langer}, {Arcoragi}, \&
  {Arnould}}]{langer1989}
{Langer}, N., {Arcoragi}, J.-P., \& {Arnould}, M. 1989, \aap, 210, 187

\bibitem[{{Lasker} {et~al.}(1990){Lasker}, {Sturch}, {McLean}, {Russell},
  {Jenkner}, \& {Shara}}]{lasker1990}
{Lasker}, B.~M., {Sturch}, C.~R., {McLean}, B.~J., {et~al.} 1990, \aj, 99,
  2019, \dodoi{10.1086/115483}

\bibitem[{{Lee} {et~al.}(2008{\natexlab{a}}){Lee}, {Beers}, {Sivarani},
  {Allende Prieto}, {Koesterke}, {Wilhelm}, {Re Fiorentin}, {Bailer-Jones},
  {Norris}, {Rockosi}, {Yanny}, {Newberg}, {Covey}, {Zhang}, \&
  {Luo}}]{lee2008a}
{Lee}, Y.~S., {Beers}, T.~C., {Sivarani}, T., {et~al.} 2008{\natexlab{a}}, \aj,
  136, 2022, \dodoi{10.1088/0004-6256/136/5/2022}

\bibitem[{{Lee} {et~al.}(2008{\natexlab{b}}){Lee}, {Beers}, {Sivarani},
  {Johnson}, {An}, {Wilhelm}, {Allende Prieto}, {Koesterke}, {Re Fiorentin},
  {Bailer-Jones}, {Norris}, {Yanny}, {Rockosi}, {Newberg}, {Cudworth}, \&
  {Pan}}]{lee2008b}
---. 2008{\natexlab{b}}, \aj, 136, 2050, \dodoi{10.1088/0004-6256/136/5/2050}

\bibitem[{{Lee} {et~al.}(2013){Lee}, {Beers}, {Masseron}, {Plez}, {Rockosi},
  {Sobeck}, {Yanny}, {Lucatello}, {Sivarani}, {Placco}, \& {Carollo}}]{lee2013}
{Lee}, Y.~S., {Beers}, T.~C., {Masseron}, T., {et~al.} 2013, \aj, 146, 132,
  \dodoi{10.1088/0004-6256/146/5/132}

\bibitem[{{Limongi} \& {Chieffi}(2018)}]{limongi2018}
{Limongi}, M., \& {Chieffi}, A. 2018, \apjs, 237, 13,
  \dodoi{10.3847/1538-4365/aacb24}

\bibitem[{{Lindegren} {et~al.}(2018){Lindegren}, {Hern{\'a}ndez}, {Bombrun},
  {Klioner}, {Bastian}, {Ramos-Lerate}, \& {de Torres}}]{gaiadr2}
{Lindegren}, L., {Hern{\'a}ndez}, J., {Bombrun}, A., {et~al.} 2018, \aap, 616,
  A2, \dodoi{10.1051/0004-6361/201832727}

\bibitem[{{Longeard} {et~al.}(2018){Longeard}, {Martin}, {Starkenburg},
  {Ibata}, {Collins}, {Geha}, {Laevens}, {Rich}, {Aguado}, {Arentsen},
  {Carlberg}, {C{\^o}t{\'e}}, {Hill}, {Jablonka}, {Gonz{\'a}lez Hern{\'a}ndez},
  {Navarro}, {S{\'a}nchez-Janssen}, {Tolstoy}, {Venn}, \&
  {Youakim}}]{longeard2018}
{Longeard}, N., {Martin}, N., {Starkenburg}, E., {et~al.} 2018, \mnras, 480,
  2609, \dodoi{10.1093/mnras/sty1986}

\bibitem[{{Mardini} {et~al.}(2019{\natexlab{a}}){Mardini}, {Placco}, {Taani},
  {Li}, \& {Zhao}}]{mardini2019b}
{Mardini}, M.~K., {Placco}, V.~M., {Taani}, A., {Li}, H., \& {Zhao}, G.
  2019{\natexlab{a}}, \apj, 882, 27, \dodoi{10.3847/1538-4357/ab3047}

\bibitem[{{Mardini} {et~al.}(2019{\natexlab{b}}){Mardini}, {Li}, {Placco},
  {Alexeeva}, {Carollo}, {Taani}, {Ablimit}, {Wang}, \& {Zhao}}]{mardini2019}
{Mardini}, M.~K., {Li}, H., {Placco}, V.~M., {et~al.} 2019{\natexlab{b}}, \apj,
  875, 89, \dodoi{10.3847/1538-4357/ab0fa2}

\bibitem[{{Marshall} {et~al.}(2019){Marshall}, {Hansen}, {Simon}, {Li},
  {Bernstein}, {Kuehn}, {Pace}, {DePoy}, {Palmese}, {Pieres}, {Strigari},
  {Drlica-Wagner}, {Bechtol}, {Lidman}, {Nagasawa}, {Bertin}, {Brooks},
  {Buckley-Geer}, {Burke}, {Carnero Rosell}, {Carrasco Kind}, {Carretero},
  {Cunha}, {D'Andrea}, {da Costa}, {De Vicente}, {Desai}, {Doel}, {Eifler},
  {Flaugher}, {Fosalba}, {Frieman}, {Garc{\'\i}a-Bellido}, {Gaztanaga},
  {Gerdes}, {Gruendl}, {Gschwend}, {Gutierrez}, {Hartley}, {Hollowood},
  {Honscheid}, {Hoyle}, {James}, {Kuropatkin}, {Maia}, {Menanteau}, {Miller},
  {Miquel}, {Plazas}, {Sanchez}, {Santiago}, {Scarpine}, {Schubnell},
  {Serrano}, {Sevilla-Noarbe}, {Smith}, {Soares-Santos}, {Suchyta}, {Swanson},
  {Tarle}, {Wester}, \& {DES Collaboration}}]{marshall2019}
{Marshall}, J.~L., {Hansen}, T., {Simon}, J.~D., {et~al.} 2019, \apj, 882, 177,
  \dodoi{10.3847/1538-4357/ab3653}

\bibitem[{{Masseron} {et~al.}(2006){Masseron}, {van Eck}, {Famaey}, {Goriely},
  {Plez}, {Siess}, {Beers}, {Primas}, \& {Jorissen}}]{masseron2006}
{Masseron}, T., {van Eck}, S., {Famaey}, B., {et~al.} 2006, \aap, 455, 1059,
  \dodoi{10.1051/0004-6361:20064802}

\bibitem[{{Matteucci} {et~al.}(2014){Matteucci}, {Romano}, {Arcones},
  {Korobkin}, \& {Rosswog}}]{matteucci2014}
{Matteucci}, F., {Romano}, D., {Arcones}, A., {Korobkin}, O., \& {Rosswog}, S.
  2014, \mnras, 438, 2177, \dodoi{10.1093/mnras/stt2350}

\bibitem[{McKinney(2010)}]{pandas}
McKinney, W. 2010, in Proceedings of the 9th Python in Science Conference, ed.
  S.~van~der Walt \& J.~Millman, 51 -- 56

\bibitem[{Mcmahon(1979)}]{sed}
Mcmahon, L.~E. 1979, in UNIX Programmer’s Manual - 7th Edition, volume 2,
  Bell Telephone Laboratories (Murray Hill)

\bibitem[{{McMillan}(2011)}]{mc11}
{McMillan}, P.~J. 2011, \mnras, 414, 2446,
  \dodoi{10.1111/j.1365-2966.2011.18564.x}

\bibitem[{{McMillan}(2017)}]{mc17}
---. 2017, \mnras, 465, 76, \dodoi{10.1093/mnras/stw2759}

\bibitem[{{Mel{\'e}ndez} {et~al.}(2016){Mel{\'e}ndez}, {Placco}, {Tucci-Maia},
  {Ram{\'{\i}}rez}, {Li}, \& {Perez}}]{melendez2016}
{Mel{\'e}ndez}, J., {Placco}, V.~M., {Tucci-Maia}, M., {et~al.} 2016, \aap,
  585, L5, \dodoi{10.1051/0004-6361/201527456}

\bibitem[{{Merrill}(1952)}]{merrill1952}
{Merrill}, P.~W. 1952, \apj, 116, 21, \dodoi{10.1086/145589}

\bibitem[{{Meynet} {et~al.}(2010){Meynet}, {Hirschi}, {Ekstrom}, {Maeder},
  {Georgy}, {Eggenberger}, \& {Chiappini}}]{meynet2010}
{Meynet}, G., {Hirschi}, R., {Ekstrom}, S., {et~al.} 2010, \aap, 521, A30,
  \dodoi{10.1051/0004-6361/200913377}

\bibitem[{{Nagasawa} {et~al.}(2018){Nagasawa}, {Marshall}, {Li}, {Hansen},
  {Simon}, {Bernstein}, {Balbinot}, {Drlica-Wagner}, {Pace}, {Strigari},
  {Pellegrino}, {DePoy}, {Suntzeff}, {Bechtol}, {Walker}, {Abbott}, {Abdalla},
  {Allam}, {Annis}, {Benoit-L{\'e}vy}, {Bertin}, {Brooks}, {Carnero Rosell},
  {Carrasco Kind}, {Carretero}, {Cunha}, {D'Andrea}, {da Costa}, {Davis},
  {Desai}, {Doel}, {Eifler}, {Flaugher}, {Fosalba}, {Frieman},
  {Garc{\'\i}a-Bellido}, {Gaztanaga}, {Gerdes}, {Gruen}, {Gruendl}, {Gschwend},
  {Gutierrez}, {Hartley}, {Honscheid}, {James}, {Jeltema}, {Krause}, {Kuehn},
  {Kuhlmann}, {Kuropatkin}, {March}, {Miquel}, {Nord}, {Roodman}, {Sanchez},
  {Santiago}, {Scarpine}, {Schindler}, {Schubnell}, {Sevilla-Noarbe}, {Smith},
  {Smith}, {Soares-Santos}, {Sobreira}, {Suchyta}, {Tarle}, {Thomas}, {Tucker},
  {Wechsler}, {Wolf}, \& {Yanny}}]{nagasawa2018}
{Nagasawa}, D.~Q., {Marshall}, J.~L., {Li}, T.~S., {et~al.} 2018, \apj, 852,
  99, \dodoi{10.3847/1538-4357/aaa01d}

\bibitem[{{Nishimura} {et~al.}(2015){Nishimura}, {Takiwaki}, \&
  {Thielemann}}]{nishimura2015}
{Nishimura}, N., {Takiwaki}, T., \& {Thielemann}, F.-K. 2015, \apj, 810, 109,
  \dodoi{10.1088/0004-637X/810/2/109}

\bibitem[{{Nomoto} {et~al.}(2013){Nomoto}, {Kobayashi}, \&
  {Tominaga}}]{nomoto2013}
{Nomoto}, K., {Kobayashi}, C., \& {Tominaga}, N. 2013, \araa, 51, 457,
  \dodoi{10.1146/annurev-astro-082812-140956}

\bibitem[{{Nomoto} {et~al.}(2006){Nomoto}, {Tominaga}, {Umeda}, {Kobayashi}, \&
  {Maeda}}]{nomoto2006}
{Nomoto}, K., {Tominaga}, N., {Umeda}, H., {Kobayashi}, C., \& {Maeda}, K.
  2006, Nuclear Physics A, 777, 424, \dodoi{10.1016/j.nuclphysa.2006.05.008}

\bibitem[{{Norris} {et~al.}(2007){Norris}, {Christlieb}, {Korn}, {Eriksson},
  {Bessell}, {Beers}, {Wisotzki}, \& {Reimers}}]{norris2007}
{Norris}, J.~E., {Christlieb}, N., {Korn}, A.~J., {et~al.} 2007, \apj, 670,
  774, \dodoi{10.1086/521919}

\bibitem[{Oliphant(2006)}]{numpy}
Oliphant, T.~E. 2006, A guide to NumPy, Vol.~1 (Trelgol Publishing USA)

\bibitem[{{Pignatari} {et~al.}(2008){Pignatari}, {Gallino}, {Meynet},
  {Hirschi}, {Herwig}, \& {Wiescher}}]{pignatari2008}
{Pignatari}, M., {Gallino}, R., {Meynet}, G., {et~al.} 2008, \apjl, 687, L95,
  \dodoi{10.1086/593350}

\bibitem[{{Placco} {et~al.}(2016{\natexlab{a}}){Placco}, {Beers}, {Reggiani},
  \& {Mel{\'e}ndez}}]{placco2016}
{Placco}, V.~M., {Beers}, T.~C., {Reggiani}, H., \& {Mel{\'e}ndez}, J.
  2016{\natexlab{a}}, \apjl, 829, L24, \dodoi{10.3847/2041-8205/829/2/L24}

\bibitem[{{Placco} {et~al.}(2014{\natexlab{a}}){Placco}, {Frebel}, {Beers},
  {Christlieb}, {Lee}, {Kennedy}, {Rossi}, \& {Santucci}}]{placco2014}
{Placco}, V.~M., {Frebel}, A., {Beers}, T.~C., {et~al.} 2014{\natexlab{a}},
  \apj, 781, 40, \dodoi{10.1088/0004-637X/781/1/40}

\bibitem[{{Placco} {et~al.}(2013){Placco}, {Frebel}, {Beers}, {Karakas},
  {Kennedy}, {Rossi}, {Christlieb}, \& {Stancliffe}}]{placco2013}
---. 2013, \apj, 770, 104, \dodoi{10.1088/0004-637X/770/2/104}

\bibitem[{{Placco} {et~al.}(2014{\natexlab{b}}){Placco}, {Frebel}, {Beers}, \&
  {Stancliffe}}]{placco2014c}
{Placco}, V.~M., {Frebel}, A., {Beers}, T.~C., \& {Stancliffe}, R.~J.
  2014{\natexlab{b}}, \apj, 797, 21, \dodoi{10.1088/0004-637X/797/1/21}

\bibitem[{{Placco} {et~al.}(2014{\natexlab{c}}){Placco}, {Beers}, {Roederer},
  {Cowan}, {Frebel}, {Filler}, {Ivans}, {Lawler}, {Schatz}, {Sneden}, {Sobeck},
  {Aoki}, \& {Smith}}]{placco2014b}
{Placco}, V.~M., {Beers}, T.~C., {Roederer}, I.~U., {et~al.}
  2014{\natexlab{c}}, \apj, 790, 34, \dodoi{10.1088/0004-637X/790/1/34}

\bibitem[{{Placco} {et~al.}(2015){Placco}, {Beers}, {Ivans}, {Filler}, {Imig},
  {Roederer}, {Abate}, {Hansen}, {Cowan}, {Frebel}, {Lawler}, {Schatz},
  {Sneden}, {Sobeck}, {Aoki}, {Smith}, \& {Bolte}}]{placco2015b}
{Placco}, V.~M., {Beers}, T.~C., {Ivans}, I.~I., {et~al.} 2015, \apj, 812, 109,
  \dodoi{10.1088/0004-637X/812/2/109}

\bibitem[{{Placco} {et~al.}(2016{\natexlab{b}}){Placco}, {Frebel}, {Beers},
  {Yoon}, {Chiti}, {Heger}, {Chan}, {Casey}, \& {Christlieb}}]{placco2016b}
{Placco}, V.~M., {Frebel}, A., {Beers}, T.~C., {et~al.} 2016{\natexlab{b}},
  \apj, 833, 21, \dodoi{10.3847/0004-637X/833/1/21}

\bibitem[{{Placco} {et~al.}(2018){Placco}, {Beers}, {Santucci}, {Chanam{\'e}},
  {Sep{\'u}lveda}, {Coronado}, {Points}, {Kaleida}, {Rossi}, {Kordopatis},
  {Lee}, {Matijevi{\v c}}, {Frebel}, {Hansen}, {Holmbeck}, {Rasmussen},
  {Roederer}, {Sakari}, \& {Whitten}}]{placco2018}
{Placco}, V.~M., {Beers}, T.~C., {Santucci}, R.~M., {et~al.} 2018, \aj, 155,
  256, \dodoi{10.3847/1538-3881/aac20c}

\bibitem[{{Placco} {et~al.}(2019){Placco}, {Santucci}, {Beers}, {Chanam{\'e}},
  {Sep{\'u}lveda}, {Coronado}, {Rossi}, {Lee}, {Starkenburg}, {Youakim},
  {Barrientos}, {Ezzeddine}, {Frebel}, {Hansen}, {Holmbeck}, {Ji}, {Rasmussen},
  {Roederer}, {Sakari}, \& {Whitten}}]{placco2019}
{Placco}, V.~M., {Santucci}, R.~M., {Beers}, T.~C., {et~al.} 2019, \apj, 870,
  122, \dodoi{10.3847/1538-4357/aaf3b9}

\bibitem[{{Prantzos} {et~al.}(2020){Prantzos}, {Abia}, {Cristallo}, {Limongi},
  \& {Chieffi}}]{prantzos2020}
{Prantzos}, N., {Abia}, C., {Cristallo}, S., {Limongi}, M., \& {Chieffi}, A.
  2020, \mnras, 491, 1832, \dodoi{10.1093/mnras/stz3154}

\bibitem[{{Prantzos} {et~al.}(1990){Prantzos}, {Hashimoto}, \&
  {Nomoto}}]{prantzos1990}
{Prantzos}, N., {Hashimoto}, M., \& {Nomoto}, K. 1990, \aap, 234, 211

\bibitem[{{R Core Team}(2015)}]{rproject}
{R Core Team}. 2015, R: A Language and Environment for Statistical Computing, R
  Foundation for Statistical Computing, Vienna, Austria.
\newblock \url{https://www.R-project.org/}

\bibitem[{{Raiteri} {et~al.}(1991){Raiteri}, {Busso}, {Gallino}, {Picchio}, \&
  {Pulone}}]{raiteri1991a}
{Raiteri}, C.~M., {Busso}, M., {Gallino}, R., {Picchio}, G., \& {Pulone}, L.
  1991, \apj, 367, 228, \dodoi{10.1086/169622}

\bibitem[{{Roederer}(2017)}]{roederer2017}
{Roederer}, I.~U. 2017, \apj, 835, 23, \dodoi{10.3847/1538-4357/835/1/23}

\bibitem[{{Roederer} {et~al.}(2010){Roederer}, {Cowan}, {Karakas}, {Kratz},
  {Lugaro}, {Simmerer}, {Farouqi}, \& {Sneden}}]{roederer2010b}
{Roederer}, I.~U., {Cowan}, J.~J., {Karakas}, A.~I., {et~al.} 2010, \apj, 724,
  975, \dodoi{10.1088/0004-637X/724/2/975}

\bibitem[{{Roederer} {et~al.}(2018{\natexlab{a}}){Roederer}, {Hattori}, \&
  {Valluri}}]{roederer2018b}
{Roederer}, I.~U., {Hattori}, K., \& {Valluri}, M. 2018{\natexlab{a}}, \aj,
  156, 179, \dodoi{10.3847/1538-3881/aadd9c}

\bibitem[{{Roederer} {et~al.}(2016){Roederer}, {Placco}, \&
  {Beers}}]{roederer2016}
{Roederer}, I.~U., {Placco}, V.~M., \& {Beers}, T.~C. 2016, \apjl, 824, L19,
  \dodoi{10.3847/2041-8205/824/2/L19}

\bibitem[{{Roederer} {et~al.}(2014{\natexlab{a}}){Roederer}, {Preston},
  {Thompson}, {Shectman}, \& {Sneden}}]{roederer2014b}
{Roederer}, I.~U., {Preston}, G.~W., {Thompson}, I.~B., {Shectman}, S.~A., \&
  {Sneden}, C. 2014{\natexlab{a}}, \apj, 784, 158,
  \dodoi{10.1088/0004-637X/784/2/158}

\bibitem[{{Roederer} {et~al.}(2014{\natexlab{b}}){Roederer}, {Preston},
  {Thompson}, {Shectman}, {Sneden}, {Burley}, \& {Kelson}}]{roederer2014}
{Roederer}, I.~U., {Preston}, G.~W., {Thompson}, I.~B., {et~al.}
  2014{\natexlab{b}}, \aj, 147, 136, \dodoi{10.1088/0004-6256/147/6/136}

\bibitem[{{Roederer} {et~al.}(2018{\natexlab{b}}){Roederer}, {Sakari},
  {Placco}, {Beers}, {Ezzeddine}, {Frebel}, \& {Hansen}}]{roederer2018}
{Roederer}, I.~U., {Sakari}, C.~M., {Placco}, V.~M., {et~al.}
  2018{\natexlab{b}}, \apj, 865, 129, \dodoi{10.3847/1538-4357/aadd92}

\bibitem[{{Roederer} {et~al.}(2012){Roederer}, {Lawler}, {Sobeck}, {Beers},
  {Cowan}, {Frebel}, {Ivans}, {Schatz}, {Sneden}, \&
  {Thompson}}]{roederer2012d}
{Roederer}, I.~U., {Lawler}, J.~E., {Sobeck}, J.~S., {et~al.} 2012, \apjs, 203,
  27, \dodoi{10.1088/0067-0049/203/2/27}

\bibitem[{{Sakari} {et~al.}(2018{\natexlab{a}}){Sakari}, {Placco}, {Farrell},
  {Roederer}, {Wallerstein}, {Beers}, {Ezzeddine}, {Frebel}, {Hansen},
  {Holmbeck}, {Sneden}, {Cowan}, {Venn}, {Davis}, {Matijevi{\v c}}, {Wyse},
  {Bland-Hawthorn}, {Chiappini}, {Freeman}, {Gibson}, {Grebel}, {Helmi},
  {Kordopatis}, {Kunder}, {Navarro}, {Reid}, {Seabroke}, {Steinmetz}, \&
  {Watson}}]{sakari2018b}
{Sakari}, C.~M., {Placco}, V.~M., {Farrell}, E.~M., {et~al.}
  2018{\natexlab{a}}, \apj, 868, 110, \dodoi{10.3847/1538-4357/aae9df}

\bibitem[{{Sakari} {et~al.}(2018{\natexlab{b}}){Sakari}, {Placco}, {Hansen},
  {Holmbeck}, {Beers}, {Frebel}, {Roederer}, {Venn}, {Wallerstein}, {Davis},
  {Farrell}, \& {Yong}}]{sakari2018}
{Sakari}, C.~M., {Placco}, V.~M., {Hansen}, T., {et~al.} 2018{\natexlab{b}},
  \apjl, 854, L20, \dodoi{10.3847/2041-8213/aaa9b4}

\bibitem[{{Salvadori} {et~al.}(2015){Salvadori}, {Sk{\'u}lad{\'o}ttir}, \&
  {Tolstoy}}]{salvadori2015}
{Salvadori}, S., {Sk{\'u}lad{\'o}ttir}, {\'A}., \& {Tolstoy}, E. 2015, \mnras,
  454, 1320, \dodoi{10.1093/mnras/stv1969}

\bibitem[{{Schlafly} \& {Finkbeiner}(2011)}]{schlafly2011}
{Schlafly}, E.~F., \& {Finkbeiner}, D.~P. 2011, \apj, 737, 103,
  \dodoi{10.1088/0004-637X/737/2/103}

\bibitem[{{Sch{\"o}nrich} {et~al.}(2019){Sch{\"o}nrich}, {McMillan}, \&
  {Eyer}}]{sch19}
{Sch{\"o}nrich}, R., {McMillan}, P., \& {Eyer}, L. 2019, \mnras, 487, 3568,
  \dodoi{10.1093/mnras/stz1451}

\bibitem[{{Sestito} {et~al.}(2020){Sestito}, {Martin}, {Starkenburg},
  {Arentsen}, {Ibata}, {Longeard}, {Kielty}, {Youakim}, {Venn}, {Aguado},
  {Carlberg}, {Gonz{\'a}lez Hern{\'a}ndez}, {Hill}, {Jablonka}, {Kordopatis},
  {Malhan}, {Navarro}, {S{\'a}nchez-Janssen}, {Thomas}, {Tolstoy}, {Wilson},
  {Palicio}, {Bialek}, {Garcia-Dias}, {Lucchesi}, {North}, {Osorio}, {Patrick},
  \& {Peralta de Arriba}}]{sestito2020}
{Sestito}, F., {Martin}, N.~F., {Starkenburg}, E., {et~al.} 2020, \mnras,
  \dodoi{10.1093/mnrasl/slaa022}

\bibitem[{{Shappee} {et~al.}(2017){Shappee}, {Simon}, {Drout}, {Piro},
  {Morrell}, {Prieto}, {Kasen}, {Holoien}, {Kollmeier}, {Kelson}, {Coulter},
  {Foley}, {Kilpatrick}, {Siebert}, {Madore}, {Murguia-Berthier}, {Pan},
  {Prochaska}, {Ramirez-Ruiz}, {Rest}, {Adams}, {Alatalo}, {Ba{\~n}ados},
  {Baughman}, {Bernstein}, {Bitsakis}, {Boutsia}, {Bravo}, {Di Mille}, {Higgs},
  {Ji}, {Maravelias}, {Marshall}, {Placco}, {Prieto}, \& {Wan}}]{shappee2017}
{Shappee}, B.~J., {Simon}, J.~D., {Drout}, M.~R., {et~al.} 2017, Science, 358,
  1574, \dodoi{10.1126/science.aaq0186}

\bibitem[{{Shen} {et~al.}(2015){Shen}, {Cooke}, {Ramirez-Ruiz}, {Madau},
  {Mayer}, \& {Guedes}}]{shen2015}
{Shen}, S., {Cooke}, R.~J., {Ramirez-Ruiz}, E., {et~al.} 2015, \apj, 807, 115,
  \dodoi{10.1088/0004-637X/807/2/115}

\bibitem[{{Siegel} {et~al.}(2019){Siegel}, {Barnes}, \& {Metzger}}]{siegel2019}
{Siegel}, D.~M., {Barnes}, J., \& {Metzger}, B.~D. 2019, \nat, 569, 241,
  \dodoi{10.1038/s41586-019-1136-0}

\bibitem[{{Siqueira Mello} {et~al.}(2014){Siqueira Mello}, {Hill}, {Barbuy},
  {Spite}, {Spite}, {Beers}, {Caffau}, {Bonifacio}, {Cayrel}, {Fran{\c c}ois},
  {Schatz}, \& {Wanajo}}]{siqueira2014}
{Siqueira Mello}, C., {Hill}, V., {Barbuy}, B., {et~al.} 2014, \aap, 565, A93,
  \dodoi{10.1051/0004-6361/201423826}

\bibitem[{{Skrutskie} {et~al.}(2006){Skrutskie}, {Cutri}, {Stiening},
  {Weinberg}, {Schneider}, {Carpenter}, {Beichman}, {Capps}, {Chester}, \&
  {Elias}}]{skrutskie2006}
{Skrutskie}, M.~F., {Cutri}, R.~M., {Stiening}, R., {et~al.} 2006, \aj, 131,
  1163, \dodoi{10.1086/498708}

\bibitem[{{Sneden} {et~al.}(2008){Sneden}, {Cowan}, \& {Gallino}}]{sneden2008}
{Sneden}, C., {Cowan}, J.~J., \& {Gallino}, R. 2008, \araa, 46, 241,
  \dodoi{10.1146/annurev.astro.46.060407.145207}

\bibitem[{{Sneden}(1973)}]{sneden1973}
{Sneden}, C.~A. 1973, PhD thesis, The University of Texas at Austin.

\bibitem[{{Starkenburg} {et~al.}(2014){Starkenburg}, {Shetrone}, {McConnachie},
  \& {Venn}}]{starkenburg2014}
{Starkenburg}, E., {Shetrone}, M.~D., {McConnachie}, A.~W., \& {Venn}, K.~A.
  2014, \mnras, 441, 1217, \dodoi{10.1093/mnras/stu623}

\bibitem[{{Starkenburg} {et~al.}(2018){Starkenburg}, {Aguado}, {Bonifacio},
  {Caffau}, {Jablonka}, {Lardo}, {Martin}, {S{\'a}nchez-Janssen}, {Sestito},
  {Venn}, {Youakim}, {Allende Prieto}, {Arentsen}, {Gentile}, {Gonz{\'a}lez
  Hern{\'a}ndez}, {Kielty}, {Koppelman}, {Longeard}, {Tolstoy}, {Carlberg},
  {C{\^o}t{\'e}}, {Fouesneau}, {Hill}, {McConnachie}, \&
  {Navarro}}]{starkenburg2018}
{Starkenburg}, E., {Aguado}, D.~S., {Bonifacio}, P., {et~al.} 2018, \mnras,
  481, 3838, \dodoi{10.1093/mnras/sty2276}

\bibitem[{{Steinmetz} {et~al.}(2006){Steinmetz}, {Zwitter}, {Siebert},
  {Watson}, {Freeman}, {Munari}, {Campbell}, {Williams}, {Seabroke}, {Wyse},
  {Parker}, {Bienaym{\'e}}, {Roeser}, {Gibson}, {Gilmore}, {Grebel}, {Helmi},
  {Navarro}, {Burton}, {Cass}, {Dawe}, {Fiegert}, {Hartley}, {Russell},
  {Saunders}, {Enke}, {Bailin}, {Binney}, {Bland-Hawthorn}, {Boeche}, {Dehnen},
  {Eisenstein}, {Evans}, {Fiorucci}, {Fulbright}, {Gerhard}, {Jauregi}, {Kelz},
  {Mijovi{\'c}}, {Minchev}, {Parmentier}, {Pe{\~n}arrubia}, {Quillen}, {Read},
  {Ruchti}, {Scholz}, {Siviero}, {Smith}, {Sordo}, {Veltz}, {Vidrih}, {von
  Berlepsch}, {Boyle}, \& {Schilbach}}]{steinmetz2006}
{Steinmetz}, M., {Zwitter}, T., {Siebert}, A., {et~al.} 2006, \aj, 132, 1645,
  \dodoi{10.1086/506564}

\bibitem[{{Suda} {et~al.}(2008){Suda}, {Katsuta}, {Yamada}, {Suwa}, {Ishizuka},
  {Komiya}, {Sorai}, {Aikawa}, \& {Fujimoto}}]{saga2008}
{Suda}, T., {Katsuta}, Y., {Yamada}, S., {et~al.} 2008, \pasj, 60, 1159.
\newblock \doarXiv{0806.3697}

\bibitem[{{Tody}(1986)}]{tody1986}
{Tody}, D. 1986, in \procspie, Vol. 627, Instrumentation in astronomy VI, ed.
  D.~L. {Crawford}, 733, \dodoi{10.1117/12.968154}

\bibitem[{{Tody}(1993)}]{tody1993}
{Tody}, D. 1993, in Astronomical Society of the Pacific Conference Series,
  Vol.~52, Astronomical Data Analysis Software and Systems II, ed. R.~J.
  {Hanisch}, R.~J.~V. {Brissenden}, \& J.~{Barnes}, 173

\bibitem[{{Tominaga} {et~al.}(2014){Tominaga}, {Iwamoto}, \&
  {Nomoto}}]{tominaga2014}
{Tominaga}, N., {Iwamoto}, N., \& {Nomoto}, K. 2014, \apj, 785, 98,
  \dodoi{10.1088/0004-637X/785/2/98}

\bibitem[{{Umeda} \& {Nomoto}(2005)}]{umeda2005}
{Umeda}, H., \& {Nomoto}, K. 2005, \apj, 619, 427, \dodoi{10.1086/426097}

\bibitem[{{van de Voort} {et~al.}(2015){van de Voort}, {Quataert}, {Hopkins},
  {Kere{\v{s}}}, \& {Faucher-Gigu{\`e}re}}]{vandevoort2015}
{van de Voort}, F., {Quataert}, E., {Hopkins}, P.~F., {Kere{\v{s}}}, D., \&
  {Faucher-Gigu{\`e}re}, C.-A. 2015, \mnras, 447, 140,
  \dodoi{10.1093/mnras/stu2404}

\bibitem[{{Vasiliev}(2019)}]{agama}
{Vasiliev}, E. 2019, \mnras, 482, 1525, \dodoi{10.1093/mnras/sty2672}

\bibitem[{{Vincenzo} {et~al.}(2015){Vincenzo}, {Matteucci}, {Recchi}, {Calura},
  {McWilliam}, \& {Lanfranchi}}]{vincenzo2015}
{Vincenzo}, F., {Matteucci}, F., {Recchi}, S., {et~al.} 2015, \mnras, 449,
  1327, \dodoi{10.1093/mnras/stv357}

\bibitem[{{Wanajo} {et~al.}(2002){Wanajo}, {Itoh}, {Ishimaru}, {Nozawa}, \&
  {Beers}}]{wanajo2002}
{Wanajo}, S., {Itoh}, N., {Ishimaru}, Y., {Nozawa}, S., \& {Beers}, T.~C. 2002,
  \apj, 577, 853, \dodoi{10.1086/342230}

\bibitem[{{Wang} {et~al.}(2019){Wang}, {Fields}, {Mumpower}, {Sprouse},
  {Surman}, \& {Vassh}}]{wang2019}
{Wang}, X., {Fields}, B.~D., {Mumpower}, M., {et~al.} 2019, arXiv e-prints,
  arXiv:1909.12889.
\newblock \doarXiv{1909.12889}

\bibitem[{{Wehmeyer} {et~al.}(2015){Wehmeyer}, {Pignatari}, \&
  {Thielemann}}]{wehmeyer2015}
{Wehmeyer}, B., {Pignatari}, M., \& {Thielemann}, F.-K. 2015, \mnras, 452,
  1970, \dodoi{10.1093/mnras/stv1352}

\bibitem[{{Welsh} {et~al.}(2020){Welsh}, {Cooke}, {Fumagalli}, \&
  {Pettini}}]{welsh2020}
{Welsh}, L., {Cooke}, R., {Fumagalli}, M., \& {Pettini}, M. 2020, \mnras, 494,
  1411, \dodoi{10.1093/mnras/staa807}

\bibitem[{Williams \& Kelley(2015)}]{gnuplot}
Williams, T., \& Kelley, C. 2015, Gnuplot 5.0: an interactive plotting program,
  \href{http://www.gnuplot.info/}{http://www.gnuplot.info/}

\bibitem[{{Winteler} {et~al.}(2012){Winteler}, {K{\"a}ppeli}, {Perego},
  {Arcones}, {Vasset}, {Nishimura}, {Liebend{\"o}rfer}, \&
  {Thielemann}}]{winteler2012}
{Winteler}, C., {K{\"a}ppeli}, R., {Perego}, A., {et~al.} 2012, \apjl, 750,
  L22, \dodoi{10.1088/2041-8205/750/1/L22}

\bibitem[{{Yoon} {et~al.}(2019){Yoon}, {Whitten}, {Beers}, {Lee}, \&
  {Placco}}]{yoon2019}
{Yoon}, J., {Whitten}, D.~D., {Beers}, T.~C., {Lee}, Y.~S., \& {Placco}, V.~M.
  2019, arXiv e-prints, arXiv:1910.10038.
\newblock \doarXiv{1910.10038}

\bibitem[{{Yoon} {et~al.}(2016){Yoon}, {Beers}, {Placco}, {Rasmussen},
  {Carollo}, {He}, {Hansen}, {Roederer}, \& {Zeanah}}]{yoon2016}
{Yoon}, J., {Beers}, T.~C., {Placco}, V.~M., {et~al.} 2016, \apj, 833, 20,
  \dodoi{10.3847/0004-637X/833/1/20}

\end{thebibliography}
